\documentclass[PhysLectNotes]{SciPost}

\hypersetup{
    colorlinks,
    linkcolor={red!50!black},
    citecolor={blue!50!black},
    urlcolor={blue!80!black}
}

\usepackage[bitstream-charter]{mathdesign}
\DeclareSymbolFont{usualmathcal}{OMS}{cmsy}{m}{n}
\DeclareSymbolFontAlphabet{\mathcal}{usualmathcal}

\begin{document}

\begin{center}{\Large \textbf{Ambitions for theory in the physics of life}}\end{center}

\begin{center}
William Bialek
\end{center}
 
 \begin{center}
Joseph Henry Laboratories of Physics and Lewis--Sigler Institute for Integrative Genomics,   Princeton University, Princeton, NJ 08544 USA
\\
 {\small \sf wbialek@princeton.edu}
\end{center}

\begin{center}
\today
\end{center}

\section*{Abstract}
{\bf Theoretical physicists have been fascinated by the phenomena of life for more than a century.  As we engage with more realistic descriptions of living systems, however, things get complicated.   After reviewing different reactions to this complexity,  I explore the optimization of  information flow as a potentially general theoretical principle. The primary example is a genetic network guiding development of the fly embryo, but each idea also is illustrated by examples from neural systems.  In each case, optimization makes detailed, largely parameter--free predictions that connect quantitatively with experiment.}

\vspace{10pt}
\noindent\rule{\textwidth}{1pt}
\tableofcontents\thispagestyle{fancy}
\noindent\rule{\textwidth}{1pt}
\vspace{10pt}

\section{Introduction}

The history of physics teaches us that qualitatively striking phenomena have correspondingly deep theoretical explanations.  In some cases the relevant phenomena are quite mundane, and it takes time to appreciate just how surprised we should be.  The (literally) everyday observation that the sky gets dark at night turns out to be one of these familiar but profound facts \cite{harrison_87, peebles_93}, as does the rigidity of solid objects \cite{anderson_84}. Today, however, the search for  fundamentally new physics is concentrated in places very far from our immediate experience: looking back to the earliest times in our universe's history; at the shortest distances and highest energies; at the lowest temperatures; and in materials that do not occur in nature.   One might be tempted to conclude that everyday phenomena are understood, at least in outline.

For theoretical physicists, declaring something to be ``understood'' requires meeting a high standard.  We expect a wide range of phenomena to be explained using a small set of general principles; we expect these principles to be summarized in compact mathematical form; and we expect this framework to be tested in quantitative experiments, often with little room for adjusting parameters as we try to reach detailed numerical agreement between theory and experiment.  By these standards, the everyday phenomena of life are {\em not} understood.   

In living systems matter organizes itself with an intricacy that is unmatched in the inanimate world.  This organized state maintains  and even reproduces itself, with extraordinary fidelity.  Once organized, living systems behave in ways that are reasonably described as functional or even purposeful and intelligent.  While it seems silly to say that water is trying to flow downhill, it would seem equally silly not to admit that a predator is trying to catch its prey.  We should be careful not to anthropomorphize, and certainly we no longer believe in a vital force, but surely there is something different about life.   

A physicist's understanding would make the difference between animate and inanimate matter precise. This  should yield a classification of complexity in cellular and animal behavior, and predict quantitative connections between this macroscopic complexity and the richness of underlying microscopic mechanisms.  These are ambitious goals, but theoretical physics is not a modest enterprise.\footnote{I am hoping that these notes capture some of the fun and informality of the original lectures.  One advantage of the written version is that I can give references.  I may have been over--enthusiastic about this, but perhaps the long bibliography will provide guidance to a literature that sprawls across physics and multiple subfields of biology.}

\subsection{An example, and a problem}
\label{sec:introproblem}

Crucial facts about living systems provide ingredients for  sharpening our questions.  A famous example is the discovery in the 1930s that genes are the size of molecules, or more precisely that the targets for radiation to produce mutations are of molecular dimensions \cite{3man,sloan+fogel_11}, which was the foundation for the questions and conjectures in {\em What is Life?}~\cite{schrodinger_44}.  The explosive growth of molecular biology has given us a veritable encyclopedia of facts about life's microscopic mechanisms \cite{watson+al_14,alberts+al_02}, and it is natural to try and summarize these facts in mathematical terms, perhaps leading to something that we would recognize as a theory of the phenomena that first attracted our attention.  But such efforts lead to a forest of arbitrary parameters.

\begin{figure}
\centerline{\includegraphics[width=0.75\linewidth]{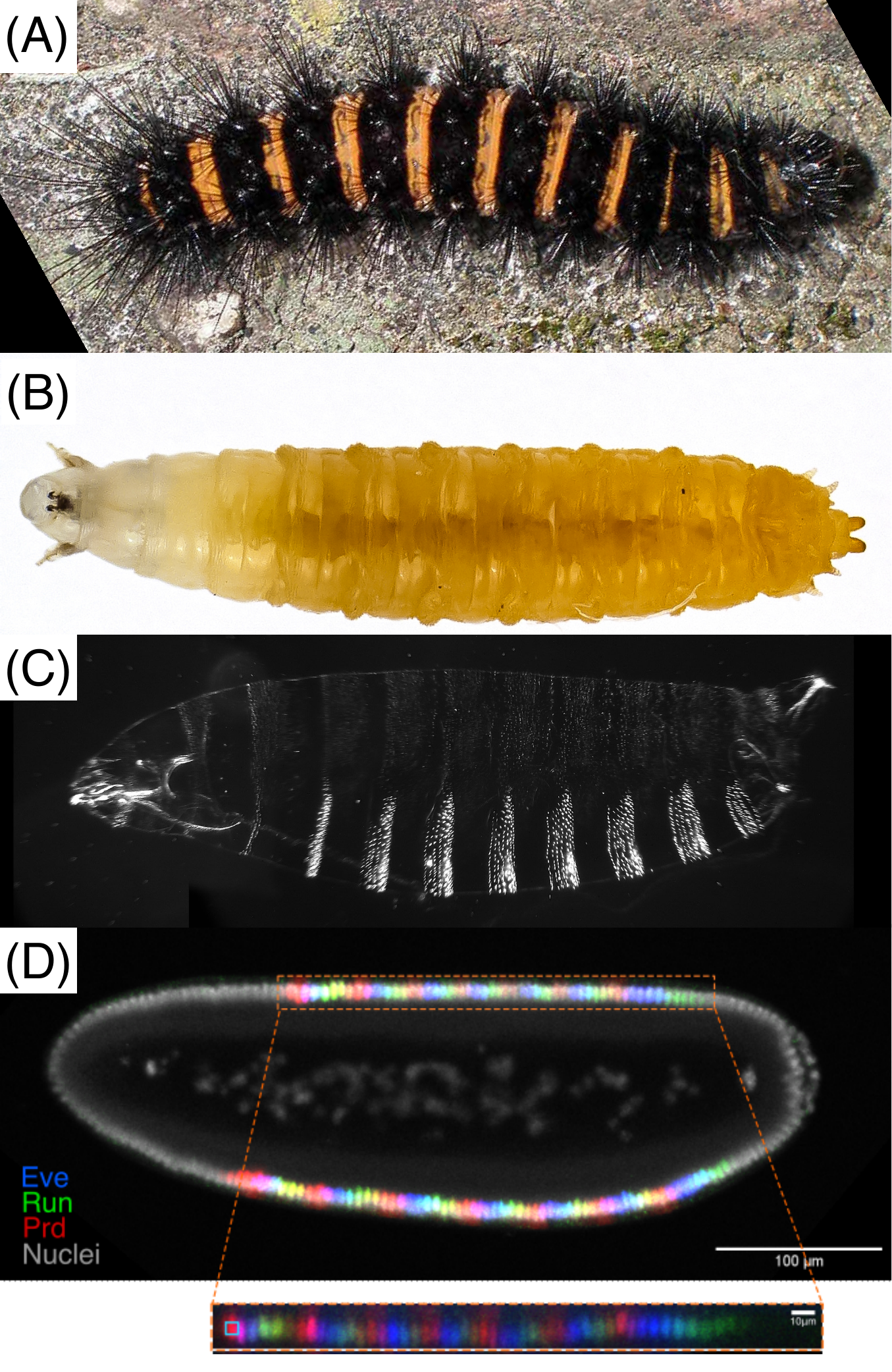}}
\caption{Segmented body plans of larval insects, and the underlying molecular blueprint.  (A) Caterpillar of the agreeable tiger moth, in which the segments are especially visible.  Image by Cyndy Syms Parr, from Wikipedia under the CC--SA 2.0 license. (B) Larva (maggot) of the fruit fly {\em Drosophila melanogaster}. Image by Salvatore Vitanza, with permission.  (C)  The ``cuticle preparation'' of the {\em Drosophila} maggot shortly after hatching, highlighting the segmented structure.  Thanks to Eric Wieschaus for the image, from experiments described in Ref   \cite{nusslein+wieschaus_80}. (D) An optical section through an embryo stained for three of the ``pair--rule'' proteins, showing striped patterns that align with the body segments; data from Ref \cite{petkova+al_19}, with thanks to M Nikoli\'c for the image.   \label{maggot+stripes}}
\end{figure}

Let us start with an unambiguously striking phenomenon, the development of a single cell into a complete multicellular organism.  Building on a century of foundational work by biologists, we will focus on the case of a fruit fly, {\em Drosophila melanogaster,} where it takes just twenty--four hours to go from one cell to a larva (maggot) that emerges from the egg shell and walks away, ready to navigate the world.  The maggot has a segmented body, and it is remarkable that if you know which molecules to look at then you can measure striped patterns in the concentration of these molecules  (Fig \ref{maggot+stripes}); these stripes provide a preview of the segmented structure.  It is perhaps even more striking that these stripes develop just {\em three hours} after the egg is laid, a time when almost all the $\sim 2^{14}$ cells are geometrically equivalent,  arrayed in a featureless lattice covering the embryo's surface.   Thus (in this case) Nature has divided the problem of development into laying out a blueprint---transmitting to each cell information about its position in the embryo, and hence its ultimate fate in the final structure---and then actually building the structure by changing the shape of the embryo.

\begin{figure}[t]
\centerline{\includegraphics[width=\linewidth]{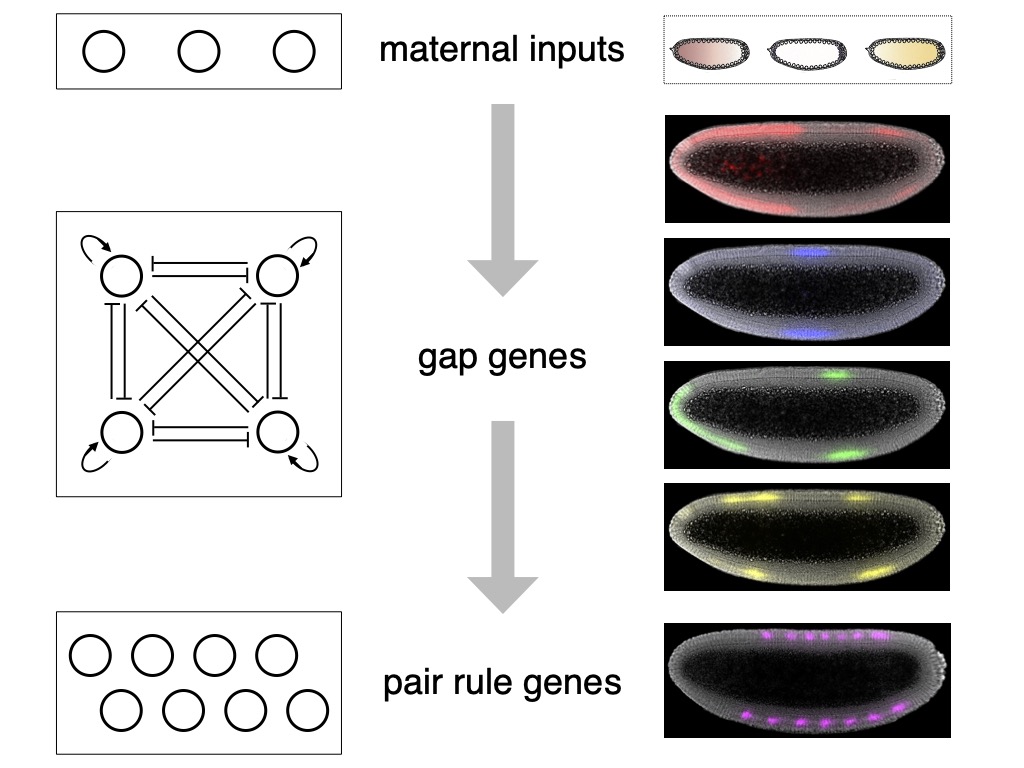}}
\caption{A schematic of information flow in the early fly embryo.  The maternal inputs  have simple spatial profiles: one with high concentration at the anterior end (left), one with high concentration at the posterior end (right), and one that is symmetrically high at the both ends (middle).  These molecules activate the expression of four gap genes, which also regulate one another.  Finally the gap genes modulate the expression of eight pair--rule genes, whose concentration profiles consist of striped patterns; one example is shown. Not illustrated are paths by which the maternal inputs can reach around the gap genes to regulate the pair--rule genes directly. Images reproduced from Ref \cite{petkova+al_19}, with permission.\label{iflow}}
\end{figure}

How does the fly embryo make stripes?  By the turn of this century, there was a clear outline of how this works.  To start, all the relevant molecules were identified, and this was one of the greatest triumphs of using genetic methods to dissect a complex phenomenon in living systems \cite{nusslein+wieschaus_80,efw+cnv_16,lawrence_92}.  The molecules with striped patterns of concentration are a group of eight proteins that are encoded by the ``pair--rule'' genes.  Whether these proteins are synthesized\footnote{When the information encoded in a particular gene is read out to make the corresponding protein we say that the gene is ``expressed.''  This occurs in two steps,  the synthesis of mRNA from the DNA template (transcription) and then the synthesis of protein from the mRNA template (translation).  Both steps are regulated, although our emphasis will be on the regulation of transcription.  When someone talks about the ``expression level of a gene'' it can be ambiguous whether this means the number of mRNA molecules or the number of protein molecules.  } is controlled largely by the concentrations of another set of four proteins that are encoded in the ``gap genes,'' so named because a mutation in one of these genes causes a large gap in the body plan.  The gap genes not only regulate the pair--rule genes, they also regulate one another, forming a network.   Finally, there are three inputs to the gap gene network that are provided by the mother when she makes the egg.  This flow of information through three layers of a molecular network is schematized in Fig \ref{iflow}.

\begin{figure}[t]
\centerline{\includegraphics[width=\linewidth]{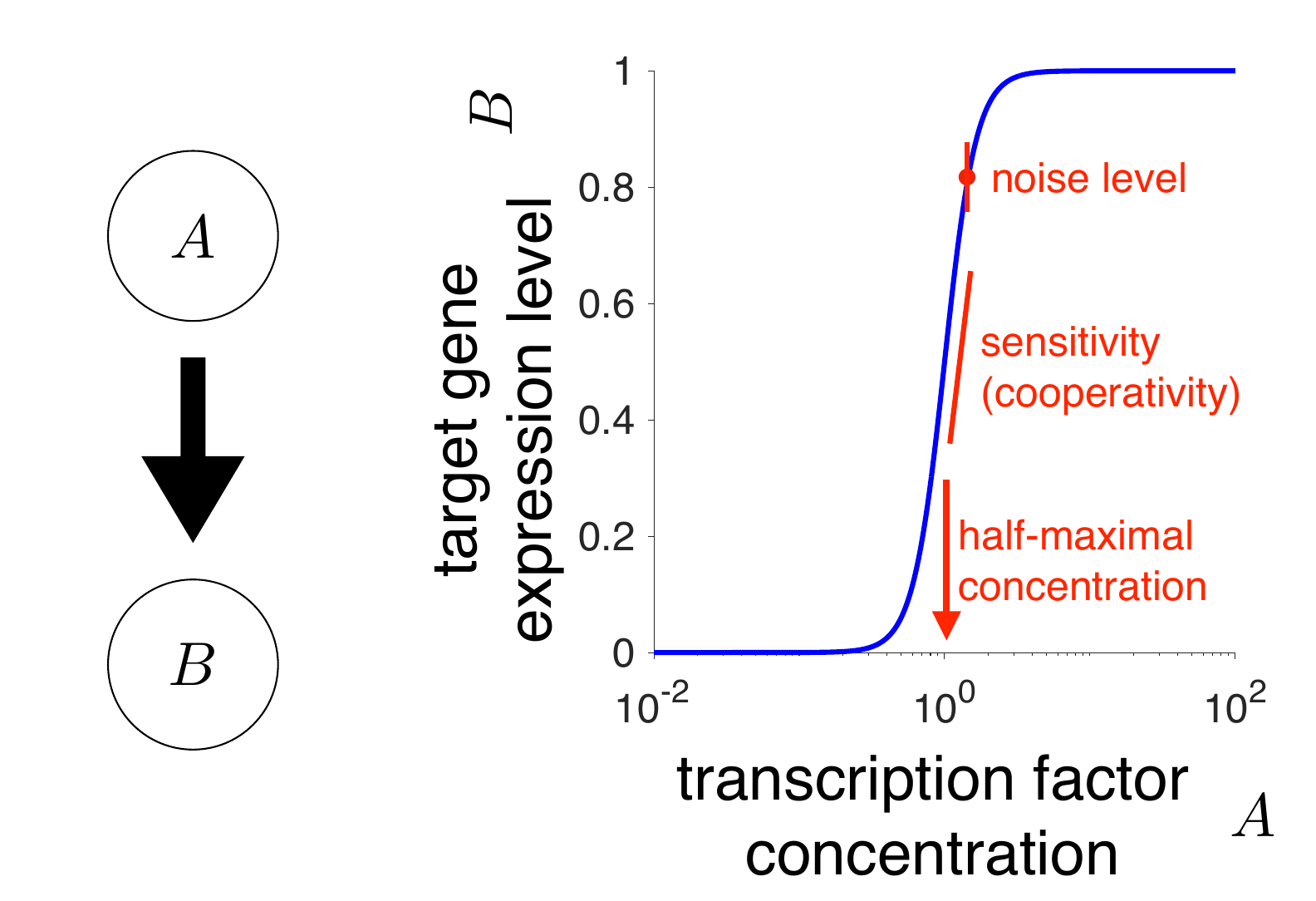}}
\caption{Parameters hiding under the arrow. $A$ is a transcription factor activating the expression of $B$, as in Eq (\ref{AregB}).  If this system comes to steady state then the expression level $B = r_{\rm max}\tau f(A)$, and we are free to choose units such that $r_{\rm max}\tau =1$, so that $B=f(A)$ as shown here.  The regulatory function $f(A)$ has at least two parameters, as shown.  The system also is  noisy, which we neglect in Eq (\ref{AregB}), but this will be important below. \label{regulation}}
\end{figure}

Schematics with nodes connected by arrows are common in descriptions of living systems.  How do we turn these schematics into equations?  There is no unique mapping but there are some common themes.  In a genetic network such as the one relevant for the fly embryo, drawing an arrow $A \rightarrow B$ means that the rate of synthesis of $B$ molecules depends on the concentration of $A$ molecules  (Fig \ref{regulation}), so at the very least we must have something like
\begin{equation}
{{dB}\over {dt}} = r_{\rm max} f(A) - {1\over \tau} B ,
\label{AregB}
\end{equation}
where $r_{\rm max}$ is the maximum synthesis rate, $1 < f(A) < 0$ is a normalized regulatory function, and $\tau$ is the lifetime of $B$ molecules; throughout these lectures I'll usually use the same symbol for the name of the molecular species and for its concentration.  We can use $\tau$ to set the units of time, and the combination $r_{\rm max}\tau$ to set the units of $B$,  but we still need to describe the regulatory function.  Plausibly it is monotonic, increasing if $A$ activates the expression of $B$ and decreasing if $A$ represses the expression of $B$.\footnote{In networks of neurons we speak conventionally of excitation and inhibition, rather than activation and repression as in genetic networks. I am not sure how these differences in jargon arose.}  A smooth function running monotonically between $0$ and $1$ has at least two parameters, roughly the concentration $A$ at which $f(A) = 1/2$ and the slope or sensitivity at this point.  We can imagine that the slope is controlled by the number of $A$ molecules that bind cooperatively to the relevant sites along the DNA and regulate the gene encoding $B$, although we should not take this too literally.   

So, to give a quantitative description we need to attach at least two numbers to every arrow in our schematic, and even this leaves things out:
\begin{itemize}
\item Something complicated could happen at places where multiple arrows converge.
\item We have associated one molecular species  with each gene, but  there could be at least two---the protein and the corresponding messenger RNA.  
\item We have assumed the dependence of synthesis rates on regulatory inputs is instantaneous, but switching among regulatory states could introduce relevant time scales.
\item In the embryo there are many cells,\footnote{In the fly, there are actually no membranes separating the cells until well into the fourteenth cycle.  It would be more precise to say ``many nuclei.''} so there is a separate copy of these dynamics at each of $\sim 2^{14}$ sites.  Exchange of molecules between sites can be described by an effective diffusion constant, but the dynamics could be more complicated.
\item During the relevant time period there are multiple rounds of nuclear division, and the regulation of gene expression could interact with this cycle.
\end{itemize}
Even with these simplifications, a network of four gap genes with three maternal inputs could have $(4\times4) + (3\times 4) = 28$ arrows, and in fact there is evidence that most of these exist; certainly there is no reason to exclude any of them {\em a priori} from a theory of this network.  But then there are at least 56 parameters needed to describe the first few hours of development along one axis of the embryo of one particular organism. It seems fair to say that this kind of complex model is uncomfortable for most theoretical physicists.

In the traditional core of theoretical physics our models of the world have parameters \cite{cahn_96}, but somehow fitting these parameters does not seem to be the central focus.  We assume that if there are too many parameters there must be something that unifies them, and if we need to set these parameters precisely to non--generic values we hope there is some extra dynamics that can make this happen more naturally.  More strongly, their are spectacular successes which are almost parameter independent, such as the BCS theory of superconductivity \cite{bcs}, the renormalization group theory of critical phenomena \cite{wilson_75,parisi_88,zinn-justin_89,cardy_96},  the theory of the fractional quantum Hall effect \cite{laughlin_83}, and more.  These theories make detailed quantitative predictions about the properties of real materials, despite the complexity of these materials on an atomic scale. In fact the periodic table of the elements, and hence most of the rules of chemical bonding, can be derived from quantum mechanics by knowing only one parameter ($\alpha \sim 1/137$), and even this isn't so important in the first rows of the table.  Perhaps ironically,  the standard model of elementary particles has many more parameters \cite{cahn_96}, but non--trivial predictions that are the foundation for our confidence in this model are nearly parameter--free, as with the connection of deep inelastic scattering experiments to the asymptotic freedom of QCD \cite{gross+wilczek_73,politzer_73,callan+gross_73,newman_04}.

Not all our efforts at theorizing reach the lofty heights of BCS or QCD.  But generations of theoretical physicists have developed a distaste for highly parameterized models, and by and large this bias has served the community well.\footnote{It will be interesting to see whether this view survives the current revolution in artificial intelligence.}  If we need 50+ parameters to describe one genetic network in one organism, and there are no principles that cut through the arbitrariness of these parameters, then we will be led to a different model for each of the many different genetic networks relevant in the life of complex organisms.  The same concern applies to other classes of processes.  The resulting collection of independent models for each of many different but related phenomena is almost the opposite of the physicist's search for unification.  

\subsection{Reacting to complexity}
\label{lec1-reactions}

In many ways, theoretical physicists’ engagement with the phenomena of life can be classified by their reaction to this proliferation of parameters.  At the risk of being cartoonish, let me list some of these broad classes:\footnote{References  are meant to be illustrative rather than exhaustive, and are a mix of original papers and reviews.}

\paragraph{0.~Give up, life really is that complicated.} A positive way of saying this is that living systems really must be described by models with many parameters, as in the example of the fly embryo, and so the interesting theoretical problems concern how we infer these parameters from data, or how prediction may be possible even when parameters are underdetermined.  This once seemed  pessimistic, but  it has received renewed attention in response to the dramatic successes of deep neural networks  \cite{lecun+al_15} and large language models \cite{vaswani+al_17,brown+al_20}.  These models grew, over a long period, out of efforts in the physics community to build theories of brain function \cite{block_62,block+al_62,hopfield_82,amit_89,hertz+al_91}, and many physicists now are interested in the question of why these networks `work' as well as they do \cite{mehta+schwab_14,mehta+al_19,carleo+al_19,kaplan+al_20,roberts+al_22}.  It is not unreasonable to think that a theory of deep networks will circle back to influence how we think about the physics of life.

\paragraph{1.~We should study theories that remind us of the real thing, and not try for quantitative comparison of theory with experiments on real living systems.} There was a period in which this style of work came under the heading of ``biologically inspired physics'' \cite{peliti_91}.   As an example, much of the work on soft  and active matter grew out of efforts to create simpler, better controlled examples of phenomena that we first encounter in the living world, from fluid membranes \cite{nelson+al_89} to flocks and swarms \cite{ramaswamy_17}.  In the same spirit, neural networks are a source of statistical physics problems that now are quite independent of efforts to understand how real brains process information and learn from their experience.  In the background of this approach is the worry that experiments on living systems are irreducibly messy, and so we will never have the kind of theory/experiment comparison in biological physics that is characteristic of physics more broadly.   This concern is addressed explicitly below.

\paragraph{2.~The only real theory is of how things are related to one another.} One sometimes hears the claim that biology is different from physics because biology is historical and physics is not.  This is a complicated claim,\footnote{In our modern understanding, the particular locations of stars or galaxies are accidents of history, but the distribution out of which these positions are drawn is not.  Thus one can see cosmology as historical.  But this absolutely does not preclude having theories in the same way as in other areas of physics---based on fundamental principles and tested by detailed comparison to (very!) quantitative experiments.}  but it suggests that even if we can't have a theory of life as we see it today, we could have a quantitative theory of the relationships among different life forms, over time---evolution.  Indeed, circa 2000, a number of physicists realized that population genetics and evolutionary dynamics could be seen as statistical physics problems, and this has been extraordinarily productive: even the simplest models of evolutionary change have subtle properties, the progress of a population of organisms over a fitness landscape is dominated by individuals in the tail of the distribution \cite{desai+al_07,fisher_11}, and more realistic contexts lead to fascinating interacting many--body problems \cite{neher+hallatschek_12,desai+al_13,fisher_13}.  This work has involved both sophisticated theory and quantitative connections to experiment, both in the laboratory \cite{levy+al_15,good+al_17,ba+al_19} and in the populations of viruses that infect humans all over the world \cite{luksza+lassig_14,neher+al_16,zanini+al_16}.\footnote{As a result there have been important practical consequences of this work, both in designing next season's flu vaccines and in the global response to covid--19. See, for example, \url{https://nextstrain.org}.}

\paragraph{3.~We are interested in (relatively) macroscopic behaviors, and these could be more universal than their microscopic mechanisms, in the spirit of the renormalization group.}  Biologists often complain that physicists oversimplify when we think about living systems, but we also simplify when we think about inanimate matter. These simplifications work not just because we are lucky.  The renormalization group teaches us that if we start with a detailed microscopic description of a system and coarse--grain to arrive at a model for behavior on long distances and long time scales, then in this process many of the microscopic details will be lost.  In technical language, there usually are only a small number of relevant operators \cite{wilson_75,parisi_88,zinn-justin_89,cardy_96}.  Thus, quantitative descriptions of macroscopic phenomena can be simpler and more universal that the underlying microscopic mechanisms.  This inspires us to think that essentially macroscopic functional phenomena in living systems could be similarly independent of microscopic details. Possibly related ideas of simplification arise from thinking about the broad spectrum of sensitivities to different parameters \cite{brown+al_04,transtrum+al_15}, and the universal behavior of dynamical systems near transition or bifurcation points \cite{rand+al_21}.

\paragraph{4.~The fact that living systems function (often quite well!) can be promoted to a principle that selects parameters or behaviors, circumventing details.}   On a dark night, our visual system can count single photons \cite{hecht+al_42}; in bright daylight, insect eyes reach a resolution close to the diffraction limit \cite{barlow_52};  bacteria navigate chemical gradients so reliably that they must be counting every molecule that arrives at the their surface \cite{berg+purcell_77}.  These and other examples suggest that organisms can reach levels of functional performance close to the limits of what is allowed by the laws of physics \cite{bialek_12}.  We can turn this around, promoting evidence of near optimal performance to a principle from which we can  derive aspects of the underlying mechanisms: rather than fitting highly parameterized models to data, we can hope that parameters have been driven to values that optimize performance.  This is in keeping with the formulation of many ideas in the core of physics as variational principles, such as least action or the minimization of free energy.

\bigskip

\noindent Exploring each of these reactions to complexity would make for a long review article, or perhaps a whole semester's worth of lectures.  With four lectures, it seems  best   to choose one approach and explore it more deeply, so these lectures will be about approach \#4.  We will keep coming back to the example of the fruit fly embryo,  but I also will try in each lecture to connect what we have been doing with the embryo to work on other examples, encouraging you to think about the generality of the principles involved.   You will have to decide how far we have come, but I hope to communicate my ambitions.
But first ...

\subsection{A few words about experiments}
\label{lec1-expts}

Theoretical physics aims at a  compact and compelling mathematical description of the world.  Because our theories are  mathematical, our predictions are  numerical.  Testing our theories thus involves measuring numbers, and this is such a central feature of physics that we can take it for granted.  But can we do this in living systems, with all their functional complexity?

Not so long ago, experiments on living systems seemed hopelessly noisy and messy.  Few things were measured quantitatively.  Some suggested that this absence of quantification was an essential difference between biology and physics, that crucial features of living systems were not reproducible as we expect in experiments on the inanimate world.  This view---which was surprisingly popular---always involved ignoring particular fields where experiments had reached physics--level precision, e.g. in studying the ability of the visual system to count single photons or the connection of neural dynamics to the properties of single ion channel molecules.  If correct, the view of biological systems as intrinsically messy would mean that there simply is no path to build an understanding of life that parallels the theoretical physicists' understanding of the inanimate world.  It is not just that we would need new principles, which would be welcome, but that we would have to retreat from what we mean by ``understand.''   This all has changed dramatically.  There has been an explosion of opportunities for physics experiments on the functional behavior of living systems, across all scales from molecules to ecosystems.  I hope we can banish forever the prejudice that life is a mess.\footnote{Certain aspects of life, of course, will remain messy, and delightfully so.}  
As data improve, we should ask more from our theories.

As a theorist I can be an unapologetic fan of what my experimental colleagues are doing. Since the early fly embryo will be our prime example,  let's focus on how measurements are made in this system, and then circle back to give  a quick survey of how things work in other contexts.   In the embryo we have a network of interacting genes, so we'd like to monitor the dynamical variables  at each node of the network.  How do we do this?  In Les Houches I only drew schematics on the blackboard, but here I take the liberty of showing real data, including some of the raw microscope images from which these data are extracted.   I find these very beautiful, and hope you will too.

The relevant variables in a genetic network are the concentrations of proteins and mRNAs.  There are ways of measuring both, and this can be done in live embryos and in fixed embryos that give us snapshots of the underlying dynamics.  Each method has pros and cons, and methods evolve with time.  I am writing this just after hearing about (admittedly preliminary) measurements that were not possible a few months before when I gave these lectures in Les Houches. As a theorist you will need to keep up with what can be done experimentally.

The oldest methods for exploring the genetic networks of the fly embryo involve measurements of protein concentrations in fixed samples.  One gently  cooks the egg, stopping all the action, and then does some chemistry to make the embryo permeable and cross--link the proteins so that they don't move around.  At this point we have plenty of time to make measurements.   Cells make many thousands of different proteins, and we want to know the concentrations of just a few of these, marked with the labels on the network nodes.  To do this we exploit the specificity of life's own mechanisms.  

We can purify the protein we are interested in and inject it into an animal which will then mount an immune response. To a good approximation, we can extract antibodies that bind to the protein of interest and nothing else.  If we can tag these antibodies with a fluorescent molecule,  then when they diffuse into the embryo they will stick to the target protein and the local fluorescence intensity will be proportional to the protein concentration; this is immunofluorescent staining.\footnote{This technique is a little more complicated.  Rather than tagging the (precious) antibodies against the protein of interest, one makes a large batch of general purpose anti--antibody antibodies, and tags these.  This depends on the fact that antibody molecules have two parts, one specific to their target and one common to all targets but varying from animal to animal.  There thus are two steps: exposing the embryo first to antibodies against the protein whose concentration we want to measure, and then to the fluorescently tagged secondary antibodies.}     With care in the choice of fluorophores, one can now measure, simultaneously, the concentrations of four proteins, sufficient to probe all the nodes of the gap gene network, for example.

\begin{figure}[t]
\centerline{\includegraphics[width = \linewidth]{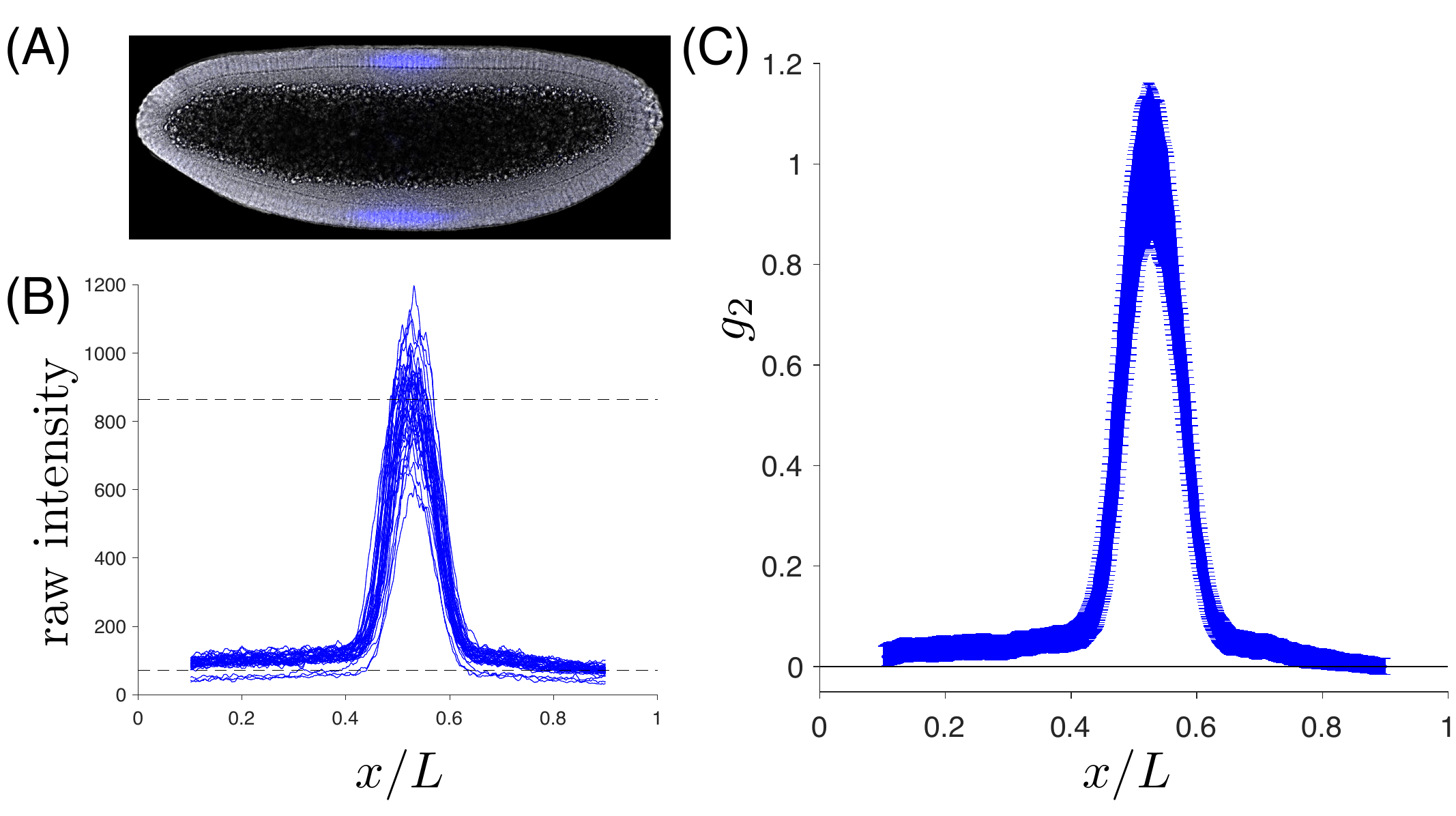}}
\caption{Measuring protein concentrations in the {\em Drosophila} embryo. (A) Image of fluorescence from labeled antibodies against one of the proteins encoded by the gap genes, corresponding to $g_2$ in Fig \ref{gapgenes} below; the gene is named {\em kr\"uppel}.  Embryo is $\sim500\,\mu{\rm m}$ long. Reproduced from Ref \cite{petkova+al_19}, with permission. (B) Raw fluorescence intensity from images of many such embryos, measured along the (straigher) upper edge.  Embryos are chosen to be a in narrow time window after the egg is laid, and in a narrow range of orientations.   We define a background (lower dashed line) and a scale (upper dashed line) so that the mean concentration is in the range $0 < \langle g_2(x) \rangle < 1$. (C) Mean and standard deviation across this large ensemble of embryos.  Note that fluctuations in the peak concentration are $\sim 10\%$, and the flanks around the peak are defined very sharply.  Data from Ref  \cite{petkova+al_19}, with thanks to T Gregor, MD Petkova, G Tka\v{c}ik, and EF Wieschaus. \label{KrAb}}
\end{figure}

Figure \ref{KrAb} shows an example of these measurements in the embryo, for one of the gap genes.
The embryo is intact, but the plane of focus is midway through its depth, so cells are arrayed along the rim of the optical slice.  One can resolve individual nuclei, but here we just plot the fluorescence intensity averaged over a window with diameter equal to that of the nuclei, sliding along the rim; we measure the position $x$ of the window by projecting onto the midline, so that $x=0$ is the anterior end (future head) and $x=L$ is the posterior end (future tail).  It is hard to convert the raw fluorescence intensity into absolute protein concentrations, so we choose units where the maximum mean concentration across embryos is $\max_x \langle g(x/L) \rangle = 1$, and we subtract a background so that $\min_x\langle g(x/L) \rangle = 0$, where $\langle g(x/L) \rangle$ is the average across the ensemble of embryos  at fixed $x/L$.\footnote{These experiments are done in inbred laboratory stocks of flies, minimizing genetic variation.  Nonetheless there are small variations in the embryo length $L$, even in eggs laid by the same mother in succession.  Our (very conventional) choice to plot always vs $x/L$ suggests that the underlying pattern formation dynamics have some mechanism that compensates for these variations, achieving a kind of scale--invariance.  This is a subtle problem that I stayed away from in my lectures, but we have come back to it since then \cite{nikolic+al_23}.}  To be clear, this single  normalization is applied to all the embryos in the sample.  Plausible alternatives, such as normalizing each image by the maximum intensity in that image, are unphysical and distort our estimates of noise, which will be important below.  The small fluctuations  from embryo to embryo that we see here provide a first glimpse of how the picture of biology as noisy can be conquered by careful experiments.

It would be nice to take a shortcut and have the proteins themselves be fluorescent.  Most of the fluorescence, and indeed most of the color that we see in living systems is generated by medium--sized organic pigment molecules that are synthesized through pathways that engage several enzymes (proteins that catalyze specific chemical reactions).  A major advance was the discovery that the fluorescence we see in some species of jellyfish and other sea creatures arises directly from a single protein molecule, with no accessory pigments, soon named the ``green fluorescent protein'' or GFP \cite{shinomura+al_62,johnson+al_62}.  It would take thirty years until it was possible to clone and sequence the gene that encodes this protein, and then insert this gene into other organisms \cite{prasher+al_92,chalfie+al_94}.  Importantly one can attach the gene for GFP to the gene for a protein you are interested in, with a short linker, so that the protein is synthesized with an intrinsic fluorescent tag; these are called GFP fusions.  Considerable effort has gone into engineering the fluorescent proteins so that they have a range of emission and absorption spectra \cite{tsien_09}.

All of the proteins we are interested in here are present at very low concentrations, so the proportionality of GFP fluorescence to its concentration is guaranteed.  With immunofluorescent staining it is a bit less obvious, because nonlinearities might creep in through the two steps of antibody binding.  By making a GFP fusion with a particular protein and then directing immunofluorescent stains against both  the protein and GFP itself, one can verify that immunofluorescence intensity really is proportional to protein concentration, and that nothing funny happens to break the 1--to--1 link between the protein and the GFP tag \cite{gregor+al_07a}.

The construction of GFP fusion proteins has the obvious advantage that one can measure protein concentrations in live cells or embryos, making the dynamics of these signals (literally) visible in real time.  The disadvantage is that GFP does not immediately fold into the structure that supports maximal fluorescence, but rather takes time to ``mature.''  There is a continuing stream of work to engineer GFP variants that have shorter maturation times, but we are not quite where we would like to be.  Thus, if we want to measure the concentration of a maternal morphogen during nuclear cycle fourteen, we are looking at proteins that have been synthesized more than two hours ago and everything is fine.\footnote{Dual immunofluorescence experiments (above) can detect small corrections due to the maturation time.}  On the other hand, if we make GFP fusions with the pair--rule genes we can see the stripes emerging during cycle fourteen but the dynamics we see probably are lagging the true dynamics.

The obvious disadvantage of immunofluorescent staining is that one has only a snapshot of a dynamic process.  We can do better by fixing and staining hundreds of embryos at once so that we get many snapshots, and unless we make a big effort to synchronize things these snapshots will be at different times after the eggs are laid.  Just by counting we can see whether we have stopped the action in nuclear cycle 12, 13, or 14, but we can do better.  As noted above, the first cycles of nuclear division are completed without pause to make membranes that define separate cells.  During nuclear cycle fourteen these membranes are constructed by infolding of the membrane that surrounds the whole embryo---the membrane of the initial fertilized egg cell.  If you watch this process in many live embryos, you see that the distance the membrane has progressed is such a reproducible function of time that you can take it as a clock accurate to one minute \cite{dubuis+al_13a}.  If you are not careful about this and mix together embryos fixed at different times, you vastly over estimate the noise in the expression levels.\footnote{More subtly, the embryo is not cylindrically symmetric, so you also have to be careful not to mix measurements from different orientations.  For a full discussion of all these concerns see Ref \cite{dubuis+al_13a}.}  As an example, in Fig \ref{KrAb} we look only at embryos in the window $40 < t < 44\,{\rm min}$ into nuclear cycle fourteen.

\begin{figure}[t]
\includegraphics[width = \linewidth]{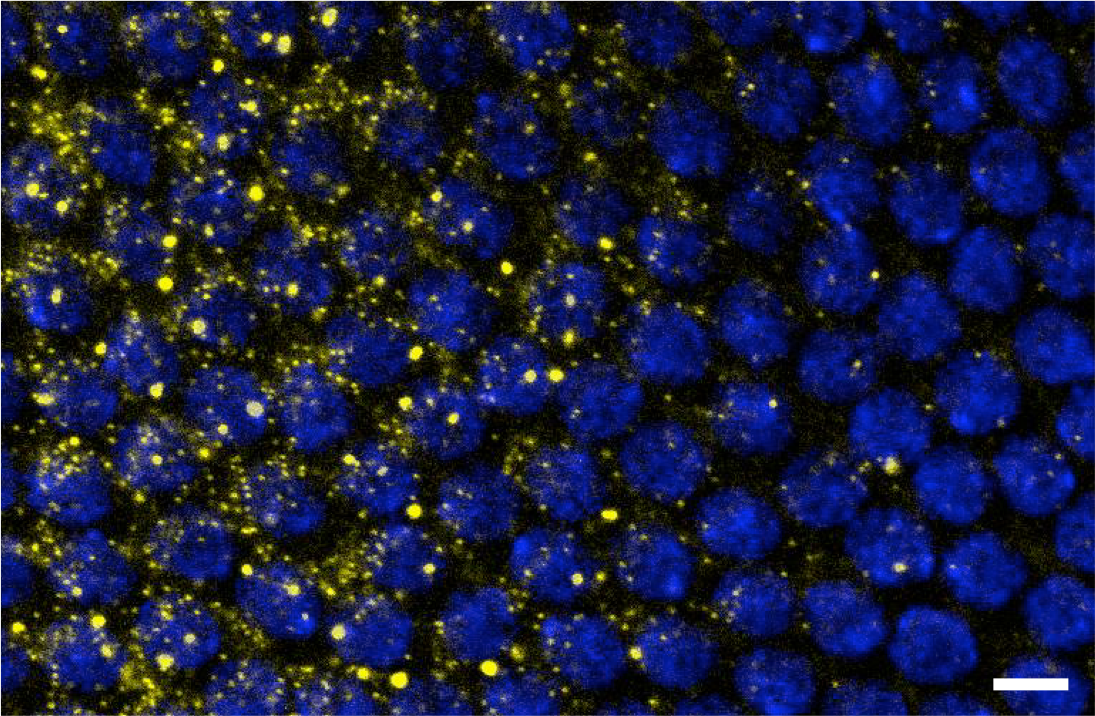}
\caption{Single molecule mRNA detection in a small region of the {\em Drosophila} embryo.  Yellow probes are complementary to the mRNA of the {\em hb} gene (protein concentration is shown as $g_1$ in Fig \ref{gapgenes}) and all nuclei are labeled in blue. Detailed analysis shows that all the spots outside the nuclei have brightness drawn from a narrow distribution, and a size consistent with the point spread function of the microscope; these and other observations indicate that they are single molecules.  The one or two brighter spots inside the nuclei are points at which mRNAs are being transcribed, as explained in the text. Expression is larger toward the left, corresponding to the anterior of the embryo. Scale bar $5\,\mu{\rm m}$.   Image courtesy of T Gregor, from the experiments of Ref  \cite{little+gregor_18}. \label{singlemRNA}}
\end{figure}

Rather than measuring protein concentrations one can observe the mRNA molecules.  Again we rely on the specificity of interactions among biological molecules to point accurately to the mRNAs that are transcribed from particular genes.  Short segments of DNA can be synthesized that are complementary to different pieces of an mRNA sequence, each labelled with a fluorescent molecule.  As with the antibodies directed at proteins, we can diffuse these molecules into a fixed and permeabilized embryo, in effect ``lighting up'' each individual  mRNA molecules with dozens of fluorophores, making it bright enough that we can count molecules one by one, as shown in Fig \ref{singlemRNA} \cite{little+gregor_18}.  This can be done in multiple colors, counting the mRNAs from multiple genes, and again this is sufficient to monitor all of the gap genes simultaneously.  These methods can be extended using combinations of fluorophores and cycles of washing and relabelling, so that eventually one can count hundreds of different mRNA species, each with single molecule resolution \cite{lubeck+cai_12,chen+al_15}.  

In Figure \ref{singlemRNA} you see not only the individual mRNA molecules in the cytoplasm, but also one or two exceedingly bright spots inside the nuclei.  These are the locations along the two chromosomes where the mRNA for this gene is being transcribed.  The gene is long enough that many copies of the transcriptional apparatus can operate simultaneously, with the result that many mRNA strands are ``in progress'' and still tethered to the DNA.  With care one can calibrate against the fluorescence intensity of the cytoplasmic spots and effectively count these nascent transcripts.  By tagging probes directed against the early and late parts of the sequence one gets a distribution of both colors and intensities across spots, and this can be used to test models of the underlying dynamics \cite{zoller+al_18}.  By interleaving the different colors one can check the precision of the measurement.

As with proteins, there is a strategy for live measurements of transcriptional activity.  Instead of attaching the DNA sequence encoding GFP to the gene of interest, one can add a sequence drawn from a virus that carries its genome as RNA rather than DNA.  To package the genome, proteins that form the coat of the virus bind to these specific RNA sequences, which fold into three--dimensional structures called ``stem loops.''  If the fly also has been genetically engineered to produce the coat protein tagged with GFP, then as the gene we are interested in gets transcribed these molecules, initially distributed throughout the nucleus, will bind to the nascent transcript \cite{larson+al_11,lucas+al_13,garcia+al_13}.  If there are many stem loops there will be many GFPs, enough to light up the mRNAs much as in the nuclear spots of Fig \ref{singlemRNA}.  After a decade of development the noise levels in these measurements now are at the point where one can almost count transcripts one by one \cite{chen+al_23}.

This strategy for visualizing transcriptional activity again exploits the specificity of interactions among biological molecules (here the coat protein and the stem loop), it makes use of sophisticated genetic engineering, and it depends on pushing the state of the art in optical microscopy.  These experiments get directly at the dynamics, showing for example that genes switch between near zero (silent) and near maximal (active) transcription rates, with the probability of being active responding to the input transcription factors.  These switching dynamics are essentially universal across all four gap genes.  Relatedly, the maximum number of mRNA molecules that one finds in cell--sized volumes surrounding a nucleus also is the same for all the gap genes.  These kinds of absolute statements would have been impossible not so long ago, and this is just a start. 

Before leaving the fly, let me note that one can adapt these live fluorescence experiments for other purposes, probing the nanoscale molecular events that underlie the control of concentrations and flow of information through gene networks.  As an example one can label a point along the DNA close to the start of transcription for one gene (the promoter), introduce stem loops into the sequence of that gene, and also label a point along the DNA where regulatory proteins bind (an enhancer).  With these three labels you can measure the distance between the promoter and enhancer, and find that these must be in close proximity in order for transcription to start \cite{chen+al_18}.  Higher resolution versions of these experiments show that proximity is not contact, so that even when transcription is active the promoter and enhancer are separated by $\sim 150\,{\rm nm}$ \cite{barinov+al_20}.  We don't know how this apparent action at a distance is achieved.

It is important that the ability to do physics experiments in living systems extends far beyond measuring protein concentrations and counting mRNA molecules.  Indeed,  studying systems in which signals are carried by changing concentrations of particular molecules is challenging in part because monitoring each different species of molecule could require a different probe, especially if we want to make real time measurements on live cells.  In contrast,  cells in the brain communicate by generating voltage differences across their membranes, and of course if we can record voltage in one cell we can in principle record from all cells with the same methods.  The currents that support transmembrane voltage changes are large enough that they result in voltage differences that are measurable in the salty water outside the cell; in particular many cells generate discrete pulses termed action potentials or spikes, and these are easier to identify.   Action potentials from single neurons were first measured nearly one hundred years ago \cite{adrian_28}.  The exploration of electrical activity in single neurons has been a source of productive interactions between physics and biology, exploiting very sensitive electronics and shaping theoretical ideas about  the microscopic mechanisms of this activity \cite{hh_52,aidley_98} and about the abstract structure of spike sequences as a code for sensory inputs and motor outputs \cite{spikes}.

\begin{figure}
\centerline{\includegraphics[width=\linewidth]{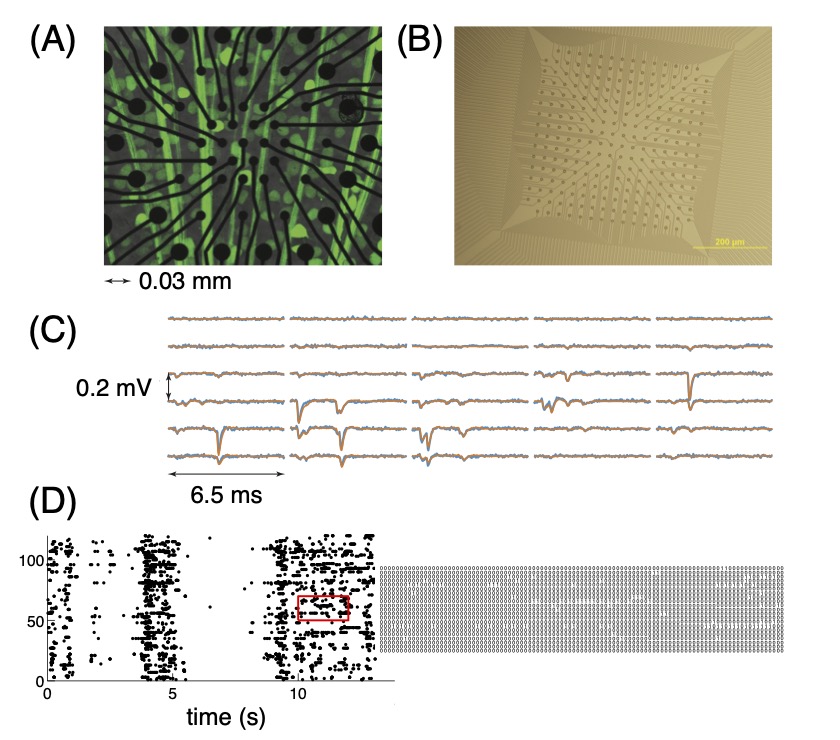}}
\caption{Recording action potentials from retinal ganglion cells. (A) Salamander retina on an array of electrodes. The electrodes, and the leads that carry signals away from the electrodes, are black features on a transparent slide; the ganglion cells of the retina have been filled
with a green dye. Round green objects are cell bodies, and long lines are bundles of axons that eventually converge to form the optic nerve. Note that the number of electrodes is comparable to the number of cells (B) The next generation of electrode arrays. (C) Voltage traces from a selection of these electrodes during an experiment on the salamander retina. Blue traces are the actual voltages, and orange traces are a reconstruction of the voltages as a superposition of stereotyped waveforms---action potentials (spikes) from individual neurons---learned from a different part of the data. (D) At left, a raster plot of the action potentials found in (C) for $100+$ cells, a dot showing the time of a spike from each neuron. At right, expanded version of the red box, with responses expressed as binary (spike/silence) words in $20\,{\rm ms}$ bins.   (A--C) reproduced from Ref \cite{bialek_12}, with permission; data from experiments by MJ Berry II and colleagues, including D Amodei, O Marre, JL Puchalla, and R Segev, with thanks \cite{segev+al_04,marre+al_12}. \label{neuro_array}}
\end{figure}

There has been dramatic progress in our ability to monitor the activity of many single neurons simultaneously, where ``many'' began with $\sim 10$ and now is $\sim 10^5$ and  soon $10^6$.   An early strategy was to focus on relatively flat pieces of brain tissue, such as the retina, which can be dissected out and placed onto an array of electrodes (Fig \ref{neuro_array}).  This led to experiments monitoring essentially all of the hundreds of cells that provide the brain with information about a small patch of the visual world as the retina is driven by complex visual inputs, including fully naturalistic movies \cite{segev+al_04,litke_04,marre+al_12}.  There are three--dimensional arrays of electrodes that can be inserted into thicker tissue \cite{campbell+al_91,jun+al_17}, and most recently there are flexible polymer based electrodes \cite{chung+al_19}.   An alternative to direct electrical measurements is to genetically engineer organisms so that neurons make proteins whose fluorescence is sensitive to electrical activity.  The ideal would be to have proteins that insert into the cell membrane and report directly on trans--membrane voltage, but these have developed slowly \cite{gong_15, yang+st-pierre_16,platisa+al_22,evans+al_23}.  Much better established are fluorescent proteins that respond to changes in calcium concentration inside the cell, which are (slower) corollaries of electrical activity \cite{tian+al_09,tian+al_12,zhang+al_23}.

\begin{figure}
\includegraphics[width=\linewidth]{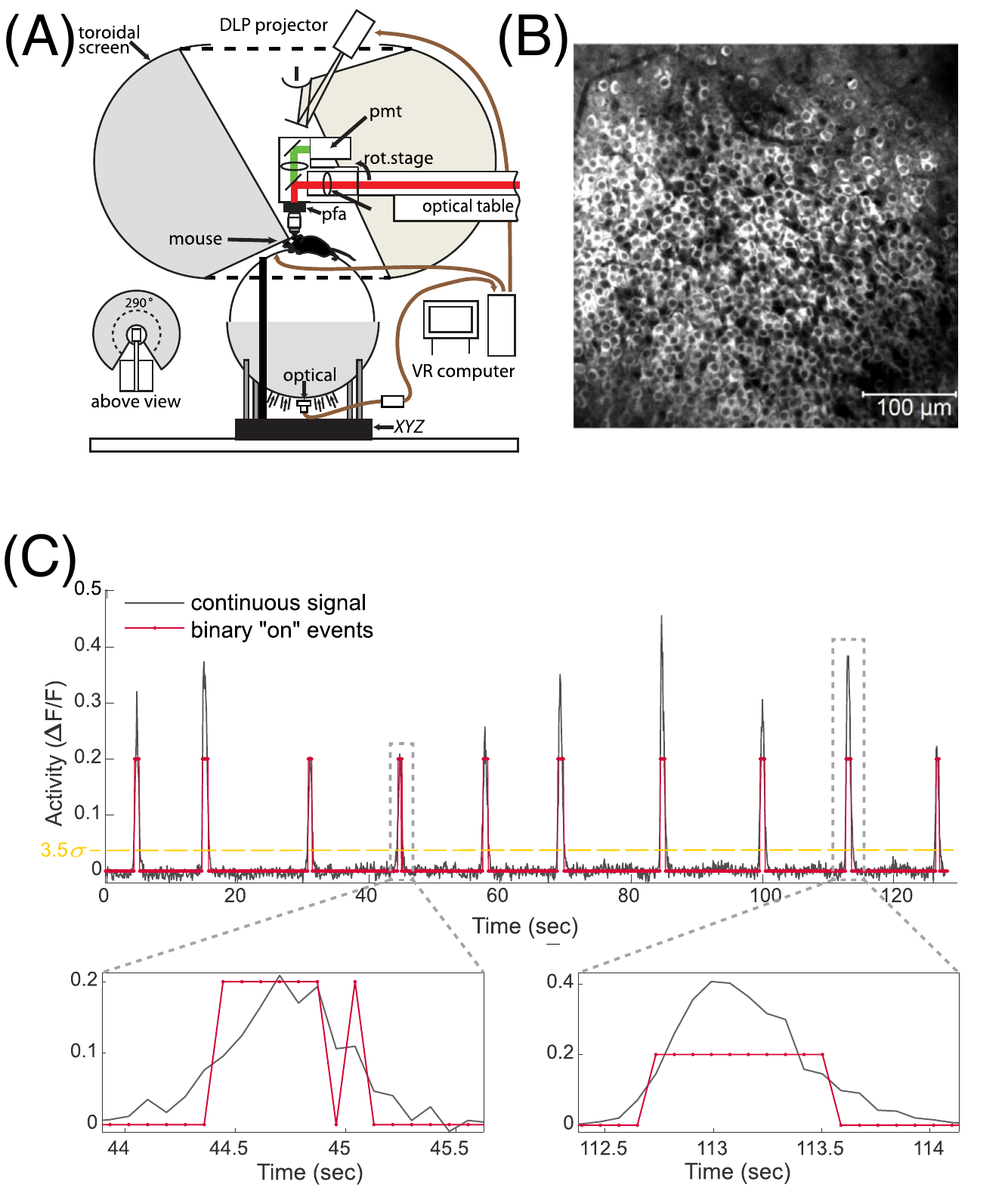}
\caption{Monitoring electrical activity in the brain by imaging of calcium sensitive fluorescent proteins as a mouse moves in virtual reality.  (A) Schematic of the experiment, with mouse (in black) running on a styrofoam ball. Motion of the ball advances the position of a virtual world projected on a surrounding toroidal screen, and a scanning two--photon microscope is focused on a single layer of cells in the hippocampus. (B) Fluorescence image of these neurons expressing calcium sensitive fluorescent protein. (C) Integrated fluorescence from a single cell vs time.  Note the very quiet baseline interrupted by discrete events.  These can be binarized, and (as shown in the insets) we understand enough about the dynamics of the indicator molecule to assign unusually slow transient decays as flickering off and on.  (A) and (B) reproduced from Ref \cite{meshulam+al_19}; (C) reproduced from Ref \cite{meshulam+al_17}, with permission.  With thanks to  L Mehsulam, JL Gauthier, CD Brody, and DW Tank.  \label{neuro_opt}}
\end{figure}

Fluorescent calcium indicators turn the problem of recording from large numbers of neurons into a problem of imaging  (Fig \ref{neuro_opt}).  To reach cellular resolution requires sophisticated microscopy methods, often built around scanning two--photon microscopy \cite{denk+al_90,dombeck+al_07}, with recent developments involving more advanced optics to allow better access to depth \cite{song+al_17,weisenburger+al_19,demas+al_21}.  To obtain high resolution images it is easiest if the brain is not moving, but many aspects of brain function are tightly coupled to behavior, and one solution is to construct virtual reality for experimental animals \cite{harvey+al_09}. The combination of genetic engineering, state of the art microscopy, and virtual reality in these experiments is impressive.  
In particular, note in Fig \ref{neuro_opt}C the very quiet baseline in individual cells, which demonstrates the generally low noise levels that can be achieved with these methods.  These tools are being used in a wide variety of organisms, from worms to mammals, and in many different regions of the brain devoted to different tasks.  

For smaller animals, such as the worm {\em C elegans} or the larval zebrafish, we are getting close to recording from every single neuron in the brain at reasonable time resolution \cite{ahrens+al_13,nguyen+al_16}.  The same techniques of genetic engineering that have led to calcium--sensitive fluorescent proteins have also led to proteins that act as light--sensitive ion channels \cite{nagel+al_03,han+boyden_07,zhang+al_08}.
Combining these tools means that one can both inject currents into neurons and record their responses, all using light and working at single cell resolution.  In {\em C elegans} this has led  to a nearly complete ``pump--probe'' experiment probing signal transfer among all $\sim 10^4$ pairs of neurons \cite{randi+al_23}.

It had long been possible to do physics experiments on isolated parts of living systems, culminating in the observation and manipulation of single molecules \cite{svoboda+al_94,abbondanzieri+al_05,ueno+al_05}, but what has emerged more recently is the ability to tame complexity and do physics experiments on an ever wider range of intact, functioning systems.  The genetic and neural networks discussed here are good examples, but one can also look to measurements on flocks of birds and swarms of insects \cite{cavagna+al_17,cavagna+al_18}, populations of bacteria \cite{qin+al_20,copenhagen+al_21}, pattern formation in groups of stem cells \cite{warmflash+al_14,shahbazi+al_19}, and much more.\footnote{Again, references are illustrative rather than exhaustive.  Broader coverage, including historical context, can be found in the first Decadal Survey of our field \cite{decadal}.}  While many systems remain to be tamed, the enormous variety of systems where it is possible to do physics experiments has removed a major obstacle to theorizing.  Importantly, as data improve we should expect more from our theories.

\subsection{Agenda}

This explosion of experimental developments obviously creates a need for new methods of data analysis. Indeed, as these approaches to high dimensional data collection have penetrated the mainstream of biology,  the biology community itself has emphasized the urgency of this need for mathematical analysis.  But for physicists theory is more than data analysis.  The most powerful analyses are grounded in theories, and there is a strong case that {\em all} analysis methods embody some theoretical prejudice.\footnote{In particle physics, for example,  the signals from thousands of detector elements are reduced to a plot of the rate for some class of events vs some energy variable.  An enormous amount of theoretical understanding is contained in the idea that this is the right thing to plot, even before we ask theory to predict what the plot shows.}   Something we do well in physics is to make these theoretical prejudices explicit.\footnote{At the risk of being pedantic, consider the simple  idea that high dimensional data---expression levels of hundreds of genes, electrical activity of thousands of neurons, and more---live in low dimensional spaces.  This seems  ``theory free,'' testable  by standard methods for linear \cite{shlens_14} or nonlinear \cite{roweis+saul_00} dimensionality reduction.  But low dimensionality is a theoretical claim:  what is the principle that limits the dimensionality of the dynamics?  More subtly, to measure how well a low dimensional description works, we need a metric, and this metric is a theoretical claim about which variations are most relevant, or perhaps which variations are measured most reliably.  } 
Returning to the fly embryo, as we discussed at the outset any attempt to give a realistic quantitative description immediately leads  into a dense forest of parameters.  Do we have clues about what principle(s) might cut through this complexity?    

\begin{figure}[t]
\includegraphics[width=\linewidth]{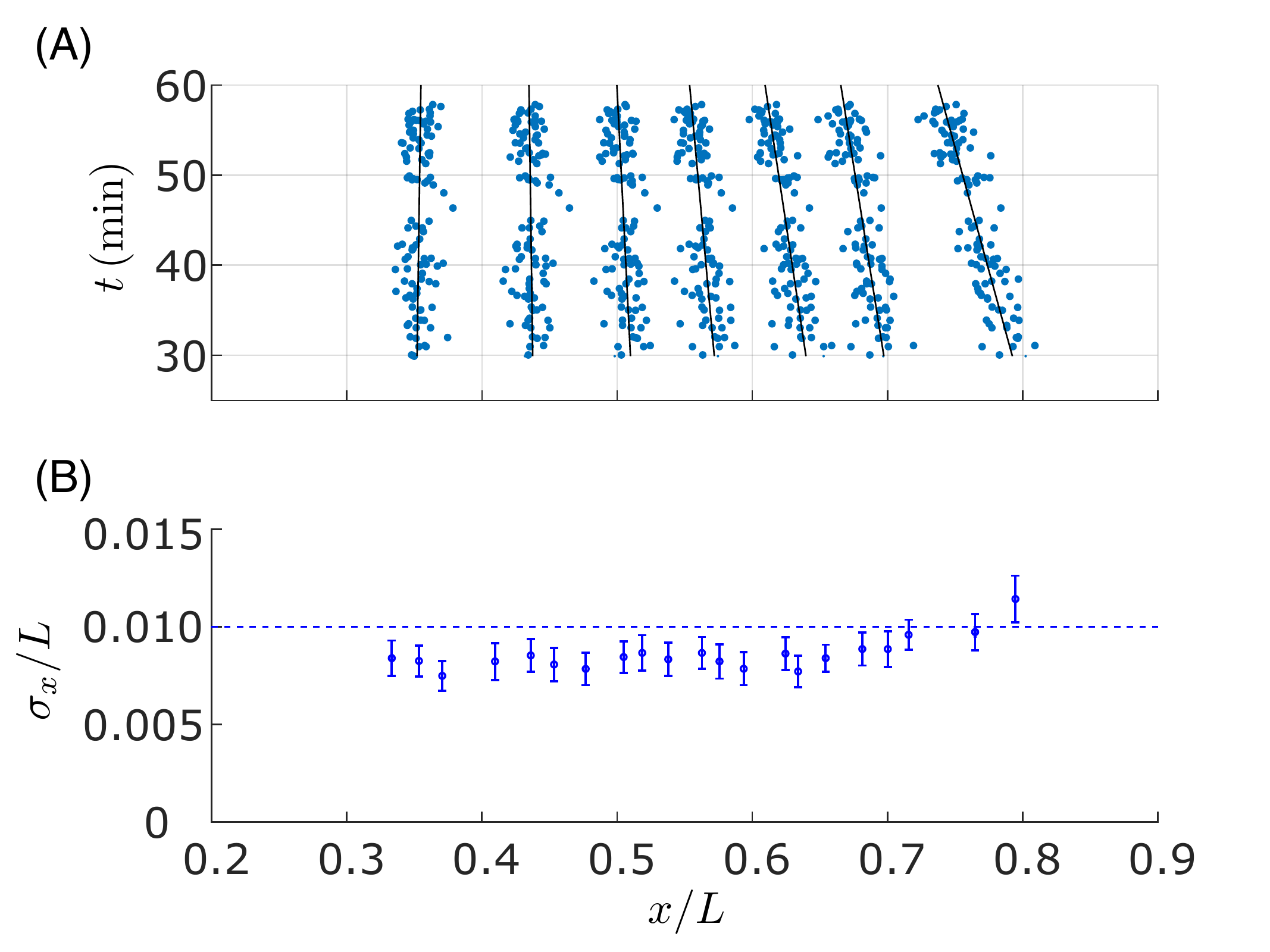}
\caption{Pair--rule stripes have positional noise $\sigma_x/L \sim 0.01$. (A) Positions of the seven pair--rule stripes for the gene shown at bottom in Fig \ref{iflow} drift with time during the fourteenth cycle of nuclear divisions, illustrated here for 100+ embryos.  (B) Once corrected for drift, the variance of stripe positions is very small.  Results for seven stripes from three different pair--rule genes (Fig \ref{maggot+stripes}, bottom); error bars are standard errors of the mean.  Dashed line is a positional noise equal to one percent of the embryo length.  Redrawn from Refs  \cite{dubuis+al_13b,mcgough+al_23}, with thanks to H Casademunt,  JO Dubuis, T Gregor, L McGough, M Nikoli\'c, MD Petkova, G Tka\v{c}ik, and EF Wieschaus. \label{fig:stripes}}
\end{figure}

{\em The system is extraordinarily precise.}  It has been thought for decades that just $\sim3\,{\rm hrs}$ after the egg is laid, every cell along the anterior--posterior axis knows its fate \cite{gergen+al_86}.  Apparently the genetic network schematized in Fig \ref{iflow} carries enough information to do this.  Since cell fates are tied to positions, and there are fewer than 100 rows of cells along the anterior--posterior axis, this means that the concentrations of just a handful of molecules specify position with $\sim 1\%$ accuracy.  This can be made explicit,  e.g. by measuring the reproducibility of the pair--rule stripe positions as in Fig \ref{fig:stripes} \cite{dubuis+al_13b,mcgough+al_23}.

{\em The concentrations of relevant molecules are low.} Almost all the molecules in the network of Fig \ref{iflow} are transcription factors---proteins that bind to specific sites along DNA and regulate the expression of other genes, in this case other genes in the network.  Again it has been known for decades that transcription factors function at concentrations measured in nanoMolar \cite{ptashne_92,pedone+al_96,winston_99}, and there is no reason to doubt that this is true of the relevant molecules in the embryo \cite{gregor+al_07b}.  Cell nuclei have dimensions measured in microns, and $1\,{\rm nM} = 0.6\,{\rm molecules}/\mu{\rm m}^3$, so even the absolute numbers of molecules can't be very large.

These two facts (here somewhat stylized) might be in conflict---at low concentrations things are noisy (because of physics not biology), and  it is hard for molecules to convey much information.  On the other hand there might be a principle that reconciles the conflict: parameters of the relevant networks have been selected to convey as much information as possible from these physically limited signals.  This principle  will be the focus of what follows.

Before getting started let me acknowledge that optimization principles engender strange reactions.  For some, optimization is obvious because evolution has had billions of years to get things right.  For others, optimization is nonsense because evolution is not about being best, but about being better than the competition.  Things get worse when we are optimizing abstract quantities such as information---why should the organism care about bits?  These discussions can devolve into debates about beliefs rather than evidence.  Nobody knows how to do a calculation that weighs the benefits of optimizing performance (e.g., counting single photons in vision) against the costs of the underlying mechanisms (energy dissipation in the biochemical amplification of single molecular events), and we certainly don't know enough to calculate how long it would take evolution to find the optimal tradeoff.  

Optimization comes along with an aesthetic that you might or might not find appealing. But what matters is that optimization principles make quantitative predictions that can be tested in modern experiments.\footnote{See also Chapter 3 in Ref \cite{bialek_12}.}  In many cases, as we will see, these predictions are essentially parameter free and accomplish the goal of circumventing highly parameterized models.  Importantly, the claim that information flow is  being optimized can be tested by measuring the information flow itself, in bits or as an effective noise level against which independently measured signals must be compared in order for the system to function.  
If we go back to the picture of information flow from maternal inputs through the gap genes to the pair--rule stripes, we can identify three distinct opportunities for optimization.

{\bf Optimal decoding.}  Information about position is encoded in the concentrations of the four gap gene products.  Given the (measured) noise levels in these concentrations, there is an optimal strategy for decoding this information, extracting an estimate of position that is as accurate as possible.  Does this optimal precision match the observed $\sim 1\%$ precision of the pair--rule stripes? Can we test whether the embryo really implements something functionally equivalent to the optimal decoding algorithm?

{\bf Matching distributions.}  The information that a system's output provides about its inputs depends not only on the internal dynamics and noise level in the system, but also on the distribution of its inputs.  In transmitting positional information, the embryo can't choose the distribution of positions, but it can adjust the representation of position by the maternal morphogens, which provide the direct input to the gap gene network.  How should these inputs be adjusted to optimize information transmission?  Can we see signatures of this optimization?

{\bf Network architecture.} Finally, we can go back to the 50+ parameters of the gap gene network and ask if these parameters have been set to optimize information transmission.  This is a hard problem, and we have tried to make progress by breaking off small pieces.  Recently there has been a major step forward, optimizing the whole gap gene network over a class of models that is almost (rich) enough to include the real network.  We are coming close to deriving the properties of this network from a  general physical principle, with no free parameters.  

We will take these problems in sequence, and in each case we'll see how the same principles are relevant to very different systems.  The first efforts to use these principles came in the context of neural information processing, and the idea that neural and genetic networks might be shaped by common principles is appealing in itself.
I take the liberty of including a few subsections that I didn't have time for in the lectures, and hopefully these provide more context.

\section{Optimal decoding}
\label{sec:decoding}

If we think that the genetic regulatory networks in the embryo have been selected to transmit as much positional information as possible while using a limited number of molecules, it would be very odd if the cells then used this information inefficiently.  So it is reasonable to ask how the embryo could use the concentrations of the gap proteins to draw the most reliable inferences about position, and whether we can find signatures of this optimization.  We'll need to build up some tools to answer these questions.

\subsection{A warmup exercise}
\label{sec:warmup}

The essential problem of inference is that we can only measure things that are related to what we care about, we can't get direct access.  This is familiar in the physics lab, where (for example) we measure currents in response to applied voltages to estimate the resistance of a material.  What we measure in this case is proportional to what we want to know, but also noisy.  So the simplest version is that we want to know $x$ but we measure $y$, and these are connected by
\begin{equation}
y = x + \eta .
\end{equation}
If, as often is the case, the noise is Gaussian with zero mean, then
\begin{equation}
P(y|x)  = {1\over\sqrt{2\pi\langle \eta^2\rangle}}\exp\left[ - {{(y-x)^2}\over{2\langle\eta^2 \rangle}}\right] .
\end{equation}
This predicts the values of $y$ that we will observe if we control $x$.  But the problem we face is that $x$ varies in ways outside our control, and we would like to infer these variations based on measurements of $y$.  Everything that we can say about this inference problem is contained in the conditional probability distribution $P(x|y)$.\footnote{I am a little embarrassed that what I give here as a warmup exercise also appeared in my Les Houches lectures 20+ years ago \cite{bialek_02}. It's still a good exercise, but I'll go more quickly, as I did in the 2023 lectures.}

We recall that what we need can be constructed from Bayes' rule:
\begin{eqnarray}
P(x,y) &=& P(y|x)P_X(x) = P(x|y) P_Y(y)\\
\Rightarrow P(x|y)  &=& {1\over{P_Y(y)}} P(y|x)P_X(x)  .
\end{eqnarray}
Just to be clear, $P(x,y)$ is the probability (density) for observing particular values of $x$ and $y$ together, while $P_X(x)$ and $P_Y(y)$ are the probabilities for observing each value independent of the value for the other.\footnote{I am being a bit pedantic to emphasize that the two marginals are different functions.  I think most of us would write $P(x)$ and $P(y)$ when calculating in private, and trust that we can keep things straight.} 
Bayes' rule tells us that to make inferences we combine the data, which sits in $P(y|x)$, with prior knowledge or expectations, encoded in $P_X(x)$.  Let's try assuming that $x$ is drawn from a Gaussian distribution with zero mean, so that
\begin{equation}
P(x|y) = {1\over{P_Y(y)}} {1\over\sqrt{2\pi\langle \eta^2\rangle}}\exp\left[ - {{(y-x)^2}\over{2\langle\eta^2 \rangle}}\right] {1\over\sqrt{2\pi\langle x^2\rangle}}\exp\left[ - {{ x^2}\over{2\langle x^2 \rangle}}\right] .
\end{equation}
We could work a little harder on the algebra, but you can see that this is a Gaussian function of $x$.  The mean is the same as the most likely value, which we can find from
\begin{eqnarray}
0 &=& {\partial\over{\partial x}} \left[  {{(y-x)^2}\over{2\langle\eta^2 \rangle}} + {{ x^2}\over{2\langle x^2 \rangle}} \right]_{x = x_*}\\
\Rightarrow x_* &=&{{\langle x^2\rangle}\over{\langle x^2\rangle + \langle \eta^2\rangle}} y.
\label{x*1}
\end{eqnarray}
This most likely value of $x$ given our observation of $y$ is one definition of the ``best estimate.''  Another definition is to find the estimator $x_{\rm est}(y)$ which makes the smallest mean--square error
\begin{equation}
\chi^2 = \int dx\int dy \,P(x,y) \left[ x_{\rm est}(y) - x\right]^2 .
\end{equation}
The best estimate in this sense, that is the solution to $\delta \chi^2 /\delta x_{\rm est}(y) = 0$, can be found for arbitrary distributions, and is equal to the conditional mean,
\begin{equation}
x_{\rm est}^{\rm opt}(y) = \int dx\, x P(x| y).
\end{equation}
In the case where both the signal $x$ and the noise $\eta$ are Gaussian, as above, then these two different definitions of the optimal estimate agree.  More generally if different but plausible definitions of ``best'' lead to significantly different estimators, it probably is a sign that $P(x|y)$ has a complicated structure, such as multiple peaks, so that inference is not just noisy but genuinely ambiguous.

Notice that with Gaussian signals $x$ and Gaussian noise $\eta$, and a linear input/output relation for the $x \rightarrow y$ transformation, the optimal estimate of $x$ from $y$ also is linear.  This doesn't generalize.  Suppose that 
\begin{equation}
P_X(x) = {a\over 2} e^{-a|x|} .
\end{equation}
Then 
\begin{equation}
P(y|x) \propto \exp\left[ - a|x| - {{(y-x)^2}\over{2\langle\eta^2 \rangle}}\right] ,
\end{equation}
and one can see that the most likely $x$ is a thresholded function of $y$,  
\begin{eqnarray}
x_*(y) &=& 0 \,\,\,\,\,\,\,\,\,\, |y| < a\langle\eta^2 \rangle \label{x*2a}\\
x_*(y) &=& y - a\langle\eta^2 \rangle {\rm sgn} (y) \,\,\,\,\,\,\,\,\,\, |y| > a\langle\eta^2 \rangle .
\label{x*2b}
\end{eqnarray}
If we compute the conditional mean then the threshold is softened but the optimal estimator still is nonlinear.  This emphasizes that the statistical structure of the inputs can shape the {\em qualitative} structure of the estimator.

One other point  is that the optimal estimator is not a perfect estimator.  There is noise, which leads to random errors, but also there are systematic errors.  If you work through, you can see that (for example) Eq (\ref{x*1}) describes an estimator that systematically underestimates the magnitude of $x$, and this also is obvious in Eqs (\ref{x*2a}, \ref{x*2b}).  There is a tradeoff between systematic and random errors, and there is a best tradeoff, but  no way to escape from both.

With these remarks in mind, let's do a real problem.

\subsection{Counting photons and estimating motion}
\label{sec:photoncount}

The ability of visual systems to count single photons remains, for me, one of the most striking facts about the physics of life.  There is an obvious sense in which this provides an example of near optimal performance, close to the limits of what the laws of physics allow.  To reach this level of performance one has to think about physics at many scales \cite{bialek_12}:  
\begin{itemize}
\item the dynamics of the rhodopsin molecules, where photon--driven transitions are so fast that they compete with the loss of quantum coherence, and spontaneous transitions are so slow that individual molecules are stable for a thousand years; 
\item the biochemical mechanisms of amplification, which allow the photoreceptor cell to ``smell'' one rhodopsin molecule out of one billion that has changed structure in response to photon absorption, and generate a macroscopic response; 
\item the circuitry of the retina, which preserves and processes the single photon responses of individual receptor cells amid a sea of noise from other cells; 
\item and neural computations that combine single photon signals with prior expectations, for example to compensate for long delays in the retinal response.
\end{itemize}
There is much to discuss here, much of it now classical but still some questions remain open.  I want to focus on one part of the retinal circuitry problem.

Let's consider the limit where each receptor cell $\rm i$ receives either zero or one photon.  Each photon triggers a  rather reproducible pulse of current, and we can choose units in which this pulse has unit amplitude.  Then the signal from each cell becomes
\begin{equation}
y_{\rm i} = n_{\rm i} + \eta_{\rm i} ,
\label{photon1}
\end{equation}
where $n_{\rm i} =0$ or $1$ is the number of photons and $\eta_{\rm i}$ is a Gaussian noise source with variance $\sigma^2$.  While $\sigma \ll 1$, there is a problem in combining signals from many cells.  As an example, the initial experiments showing that the statistics of human responses to dim light flashes are consistent with photon counting involved flashes that delivered $\sim 5$ photons distributed over $\sim 500$ receptor cells.  In some species we know that integration over such a large area happens as receptor cell signals cross the first synapse, converging on the bipolar cell.  If the bipolar cell just adds up the signals, then the fluctuations in the sum are $> 20\sigma$, and now single photon signals will be lost.

Although $y_{\rm i}$ is on average equal to $n_{\rm i}$ the best estimate of photon count is not the receptor output itself.  Following the argument above, we have
\begin{eqnarray}
P(n_{\rm i}|y_{\rm i}) &=& {1\over{P(y_{\rm i})}} P(y_{\rm i}|n_{\rm i}) P(n_{\rm i})\\
P(y_{\rm i}) &=& \sum_{n_{\rm i}} P(y_{\rm i}|n_{\rm i}) P(n_{\rm i}).
\end{eqnarray}
From Eq (\ref{photon1}) we have
\begin{equation}
P(y_{\rm i}|n_{\rm i}) = {1\over\sqrt{2\pi\sigma^2}} e^{-(y_{\rm i} - n_{\rm i})^2/2\sigma^2} .
\end{equation}
Further, since the lights are low the only nonzero values for $P(n_{\rm i})$ are $P(0)$ and $P(1)$, so that
\begin{eqnarray}
P(n = 1| y) &=& {{P(1) \exp\left[- (y-1)^2/2\sigma^2\right]}\over
{P(1) \exp\left[- (y-1)^2/2\sigma^2\right] + P(0) \exp\left[- (y)^2/2\sigma^2\right]}}\\
&=&{1\over{1 + e^{-\beta (y-\theta)}}} ,
\label{photon_est1}
\end{eqnarray}
which is a sigmoidal function.  In this regime the conditional mean of $n_{\rm i}$, which we recall is the best estimate in the least squares sense, is just $P(n_{\rm i} = 1| y_{\rm i})$.  So the best estimate of the photon count is the sigmoidal function in Eq (\ref{photon_est1}), with ``threshold'' 
\begin{equation}
\theta = \sigma^2 \ln\left[ {{P(0)}\over{P(1)}}\right] + {1\over 2}
\end{equation}
and sensitivity
\begin{equation}
\beta = {1\over{\sigma^2}} .
\end{equation}
After passing through this near threshold nonlinearity, we can sum the signals from many cells with much less sensitivity to the background noise.

\begin{figure}[t]
\centerline{\includegraphics[width = \linewidth]{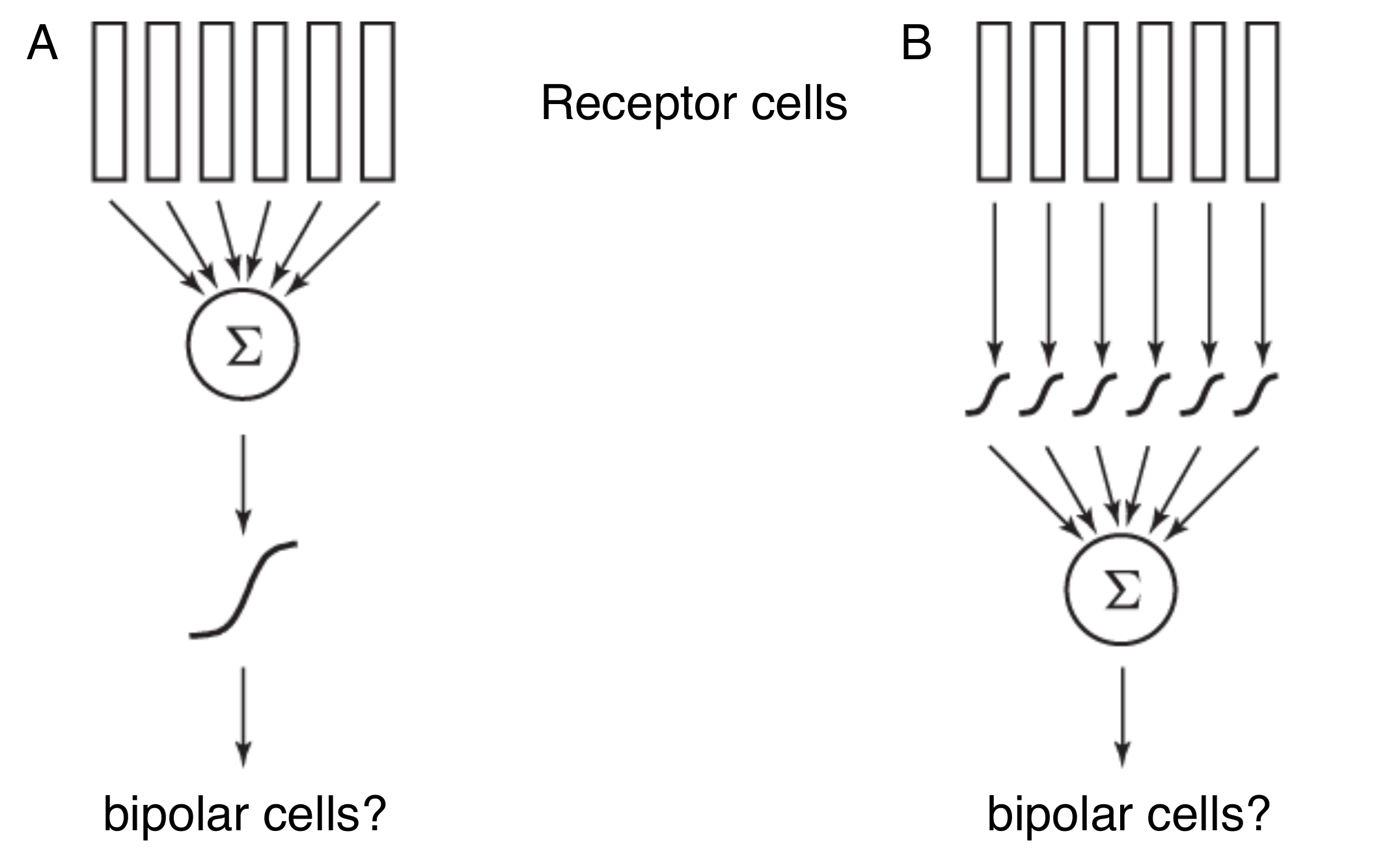}}
\caption{Two models of integrating single photon signals at the first synapse in the retina \cite{bialek_12}. (A) Receptor cell signals are summed, and the target (bipolar) cell applies a nonlinearity, as in conventional neural network models. (B) Each receptor cell signal is passed through a nonlinearity, as suggested by Eq (\ref{photon_est1}), and the resulting estimates of photon count are summed.
\label{bipolar}}
\end{figure}

We can think of the arguments as contrasting two models, shown in Fig \ref{bipolar}. In (A) we see something like the conventional neural network model:  inputs to a single neuron sum linearly, and a sigmoidal nonlinearity is applied after the summation \cite{block_62,lecun+al_15}. In (B) we have the opposite, where nonlinearities are applied to each input signal as it crosses the synapse, and these processed signals are summed.  Despite expectations, optimal estimation of photon counts requires something more like (B).  Happily, these synaptic nonlinearities have been detected directly in the mouse retina, in at least semi--quantitative agreement with theory \cite{field+rieke_02}; importantly these nonlinearities do not appear at synapses to bipolar cells which are not involved in processing single photon signals. This discussion has emphasized instantaneous nonlinearities, but temporal filtering can also help to separate single photon signals from noise at this first synapse \cite{rieke+al_91}.

In addition to detecting light and estimating its intensity, the brain of course draws many more sophisticated inferences from its visual inputs.  An interesting example, especially appealing for physicists, is the inference of movement.  The fact that we have the {\em appearance} of motion from discrete flashing lights is also among the foundational observations of gestalt psychology.  We can study visual motion perception in humans, and the behavioral responses of insects to visual motion, and there are striking similarities across this enormous evolutionary distance.  In particular, humans and insects make similar systematic errors in estimating movement velocities, especially in scenes with low contrast.

One might think that estimating movement velocity is not so hard.  Let's work in one dimension (which admittedly hides some important issues), so the image that falls on our retina is $I(\phi , t)$ where $\phi$ is the azimuthal angle.  If a small patch of the visual world is moving relative to us at velocity $v$, then we should have
\begin{equation}
I(\phi , t) = I_0(\phi-v t) .
\end{equation}
This suggests that
\begin{equation}
v_{\rm est} =  v_{\rm deriv} = - {{\partial I/\partial t}\over{\partial I/\partial \phi}} ,
\label{vest_grad}
\end{equation}
provides a direct estimate of velocity as the ratio of temporal and spatial derivatives.  This is overly optimistic, because we have neglected noise; more subtly we have neglected any dynamics in the image that cannot be ascribed directly to movement. In the presence of noise it is dangerous to differentiate,  because noise typically extends to higher frequencies than the signal, and it is dangerous to divide, because the denominator might fluctuate to zero.  In Eq (\ref{vest_grad}) we commit both these sins.

Equation (\ref{vest_grad}) says that, in the presence of motion, the spatial and temporal derivatives should be proportional to one another.  A gentler statement is that these derivatives are correlated, and the strength of this correlation should be related to the movement velocity.  This suggests that we might be able to estimate
\begin{equation}
v_{\rm est} = v_{\rm corr} \propto  {{\partial I}\over{\partial t}} \times {{\partial I}\over{\partial \phi}} .
\label{vest_corr}
\end{equation}
Notice that if we double the variations in light intensity, then this estimate will increase by a factor of four, unless we do some sort of normalization.  But this confounding of contrast and velocity is one of the systematic errors made by humans and insects alike, at least at low contrast.  The idea that brains estimate motion based on spatiotemporal correlation goes back to classical experiments on insect behavior \cite{hassenstein+reichardt_56,reichardt+poggio_76} and reappears decades later as a model of ``motion energy'' in human and non--human primate vision \cite{adelson+bergen_85,santen+sperling_85}.

It is especially attractive to study visual motion estimation in flies because there is a very accessible and beautifully laid out population of neurons that encode these estimates and ultimately drive behaviors such as flight control \cite{strausfeld_76,hausen_82,stavenga+hardie_89}.  In this system one can also give a detailed characterization of signals and noise in the photoreceptors and second order neurons, including evidence that photon shot noise is dominant at counting rates up to $\sim 10^6\,{\rm s}^{-1}$ \cite{ruyter+laughlin_96a,ruyter+laughlin_96b}.  By the early 1990s, a theory/experiment collaboration with Rob de Ruyter van Steveninck had shown that sequences of action potentials from the motion sensitive neurons in flies encoded motion estimates with a reliability close to the physical limits set by noise in the photoreceptors and diffraction blur in the compound eye \cite{bialek+al_91,ruyter+bialek_95}.

Motivated by the observation of near--optimal performance, we developed a theory of optimal motion estimation following the lines sketched above \cite{potters+bialek_94}.  We were excited that, as a function of signal strength or noise level the optimal estimator interpolated between something like the ratio of derivatives in Eq (\ref{vest_grad}) and the correlator in Eq (\ref{vest_corr}).  This suggested that some of the systematic errors of real visual motion estimation might actually be features of the optimal estimator.  The problem is that the detailed form of the optimal estimator depends, as expected from \S\ref{sec:warmup}, on the statistical structure of the visual inputs, and in the early work we just had to guess at this.  In particular, depending on this structure, the crossover between correlator and ratio estimators might or might not occur in a regime that is relevant for real brains under natural conditions.

\begin{figure}[t]
\centerline{\includegraphics[width = \linewidth]{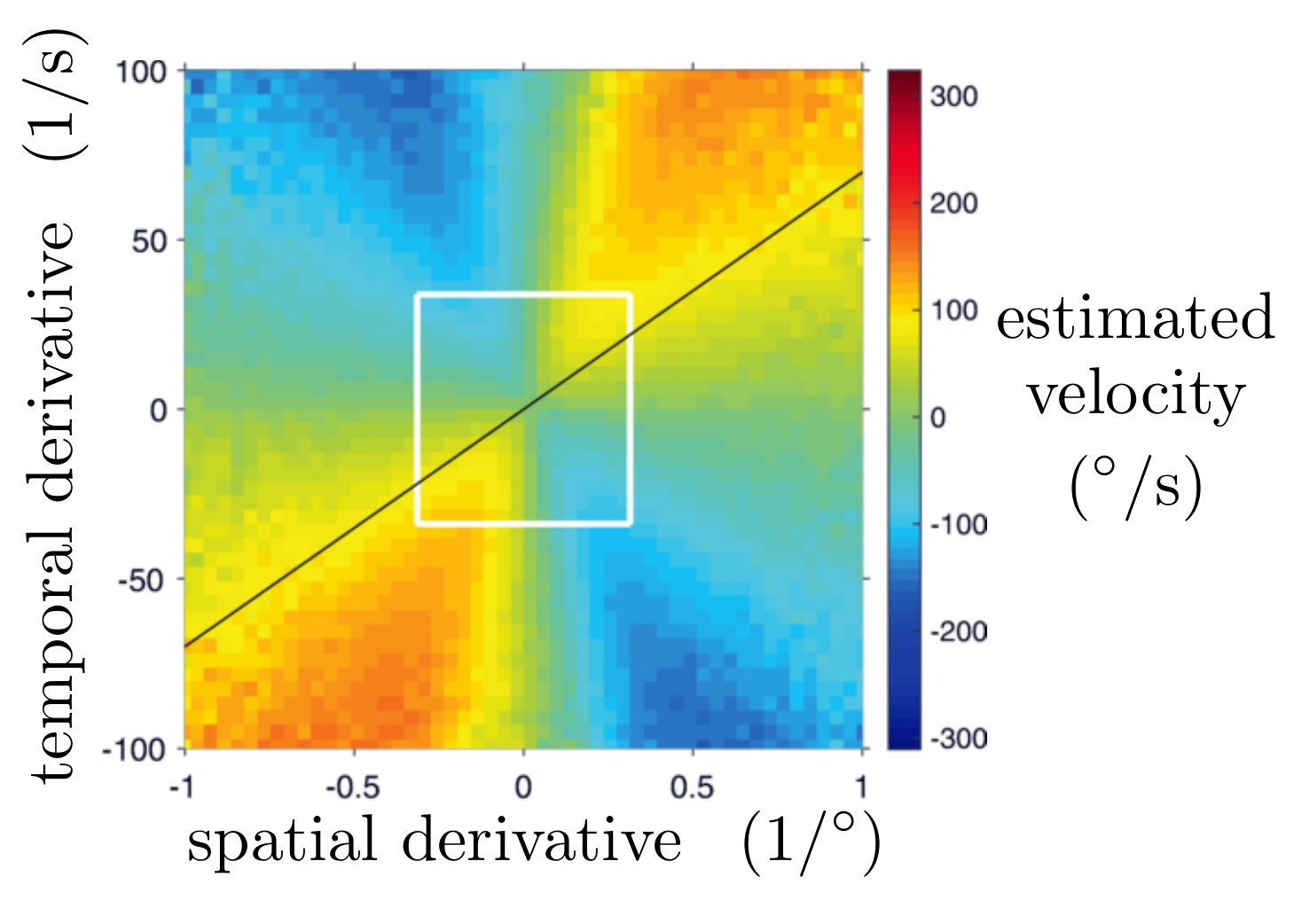}}
\caption{Estimation of visual motion from local derivatives. The optimal estimator in Eq (\ref{vest_final}) is computed from data collected by a special purpose camera during a walk in the woods. Black line indicates the predictions of Eq (\ref{vest_grad}) for $v_{\rm deriv} = 70\,^\circ/{\rm s}$.  White square encloses $90\%$ of the data.  Redrawn from Ref \cite{sinha+al_21}, with thanks to  S Sinha and RR de Ruyter van Steveninck.
\label{fig:vest}}
\end{figure}

To make progress let's focus on motion estimates derived from very small patches of the visual world; in the fly this could just mean a handful of neighboring receptors on the lattice of the compound eye.  Rob and his colleagues built a special purpose camera that samples the visual world at high frame rates and with optics that matches those of the compound eye \cite{sinha+al_21}.  From a long walk in the woods, one can derive many samples of the local image derivatives and the angular velocity of the camera at the same moment in time.  To remove the (largely irrelevant) absolute light level we can take the log intensity as the raw data, so we can think of this experiment as providing an estimate of the joint distribution $P(\partial_t \ln I, \partial_\phi \ln I, v)$.  Then we can form the conditional distribution $P(v|\partial_t \ln I, \partial_\phi \ln I)$, and finally the optimal estimator 
\begin{equation}
v_{\rm est}(\partial_t \ln I, \partial_\phi \ln I ) = \int dv\, v P(v|\partial_t \ln I, \partial_\phi \ln I) .
\label{vest_final}
\end{equation}
Results are shown in Fig \ref{fig:vest}.

If the optimal estimator were the ratio of derivatives, then contours of constant $v_{\rm est}$ would be straight lines in Fig \ref{fig:vest}, as with the black line at $70\,^\circ/{\rm s}$.  We see that this is a decent description of the estimator at large spatial and temporal derivatives.  On the other hand, the correlator model predicts that contours of constant $v_{\rm est}$  are hyperbolic, and this curvature is what we see at small values of the derivatives; we can make this clearer by taking different slices through the data \cite{sinha+al_21}.  If we test the system with a rigidly moving spatial pattern then in the small derivative regime the optimal estimator will be systematically wrong, with these errors arising as a by product of insulation against random errors.  What we have called ``small'' and ``large'' can be read from Fig \ref{fig:vest}, but importantly real derivatives are mostly small: the white box shows a range of spatial and temporal derivatives that contains $90\%$ of the samples collected on an hour long walk through the woods.    These results provide direct evidence that motion estimation in a naturalistic context really is in the regime where correlation is optimal. 

There is much more to be done.  We have not yet added back the effects of photon shot noise\footnote{The camera is built to have a much larger collecting area than the fly's receptor cells, so the signals analyzed here are essentially noiseless.} and other sources of noise in the receptor cells; these will widen the dynamic range over which the correlator model is optimal. Asymmetries in the underlying distributions should lead to asymmetries in the optimal estimator, which are barely visible in Fig \ref{fig:vest} and should be connected to the separate processing of on and off signals \cite{fitzgerald+al_11,behnia+al_14}. If the walk takes us through regions of very different statistical structure we may be able to divide the data accordingly and predict adaptation of the optimal estimator to the input statistics, perhaps connecting to adaptation seen in the responses of motion--sensitive neurons.  It also will be interesting to understand the rules for optimal combination of these local estimators into wide field motion signals. 

After many decades we have gotten used to the idea that the visual system can count single photons, and perhaps we forget that this provides evidence for optimal performance---functional behavior near the limits of what is allowed by the laws of physics.  It is tempting to think that such physical limits are irrelevant to vision on a bright sunny day, but Fig \ref{fig:vest} shows us that this is not true.  From data collected literally at noon, we see that the physical structure of visual input is such that to make maximally precise estimates the brain must do unexpected things, including making systematic errors of the same form made by humans and insects.  These errors are driven by physics, not by biological limitations.

\subsection{Concentration measurements, revisited}
\label{lec2-embryo}

How do these ideas play out in the fly embryo?  Roughly three hours after the egg is laid, individual cells have access to the concentrations (expression levels) of the four proteins encoded by the gap genes.  In order to do the right thing, cells need to know where they are in the embryo.  So it is natural to ask how a cell could infer its position from the gap  gene expression levels.  This idea that cells extract positional information from the concentration of specific morphogen molecules is very old \cite{wolpert_69}.  The fact that in the fly we can identify {\em all} the relevant molecules gives us a chance to turn these words into mathematics.

\begin{figure}
\centerline{\includegraphics[width = \linewidth]{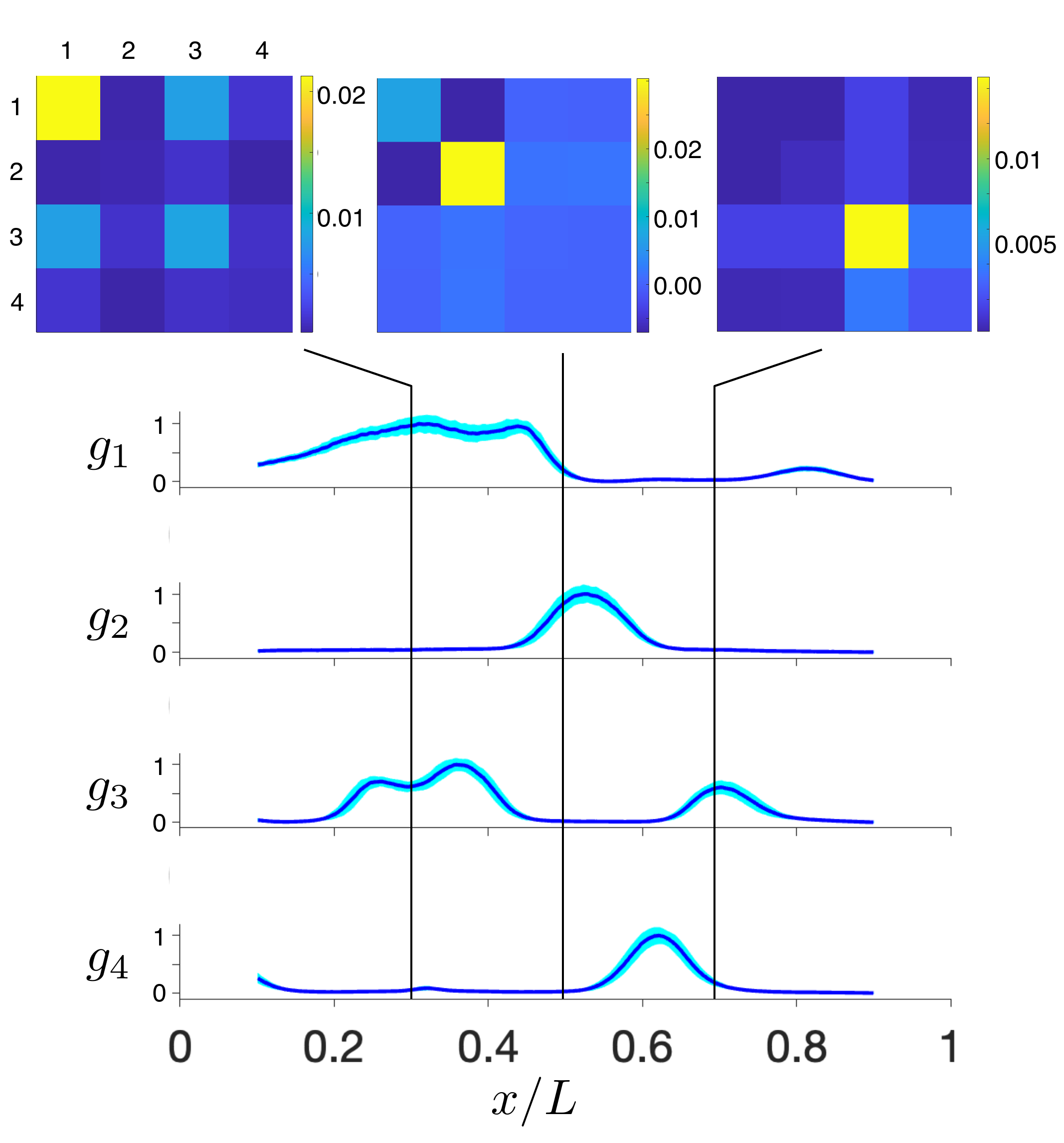}}
\caption{Expression levels of the four gap genes as a function of position along the long (anterior--posterior) axis of the embryo.  Bottom panels show the means $\langle g_{\rm i}(x)\rangle$ (blue lines) and standard deviations $\sqrt{\langle[\delta g_{\rm i}(x)]^2\rangle}$ (cyan shading).  Top panels show examples of the covariance matrix $C_{\rm ij} (x)  = \langle \delta g_{\rm i}(x) \delta g_{\rm j}(x)\rangle$ at three positions, $x/L = 0.3, 0.5,$ and $0.7$.  Data from Ref  \cite{petkova+al_19}, with thanks to T Gregor, MD Petkova, G Tka\v{c}ik, and EF Wieschaus.
\label{gapgenes}}
\end{figure}

Figure \ref{gapgenes} shows the spatial profiles of the gap gene expression levels along the long (anterior--posterior) axis of the embryo.\footnote{I tried to give this course without mentioning the names of these (and other) molecules, because I don't think it matters.  But,  to connect with the literature, the names are (1) Hunchback, (2) Kr\"uppel, (3) Giant, and (4) Knirps.} These data are extracted from the fluorescent antibody staining experiments discussed in \S\ref{lec1-expts}.  We will refer to the concentration of each molecule at position $x$ as $g_{\rm i}(x)$, with the index ${\rm i} = 1,\, 2,\, 3,\, 4$.  Solid lines show the mean concentrations $\langle g_{\rm i}(x)\rangle$, with cyan shading to indicate the standard deviation of fluctuations around this mean.\footnote{Recall from Fig \ref{KrAb} that we choose units such that $\langle g_{\rm i}(x/L) \rangle$ runs between 0 and 1 for each gene.}  An important qualitative observation is that these fluctuations in fact are quite small.  These data sets are large enough that we can estimate, reliably, the $4\times4$ covariance matrix of fluctuations at each point, 
\begin{equation}
\left[ \hat C (x)\right]_{\rm ij} = C_{\rm ij} (x) = \langle \delta g_{\rm i} (x) \delta g_{\rm j}(x) \rangle,
\end{equation}
where as usual 
\begin{equation}
\delta g_{\rm i} (x) = g_{\rm i} (x) -\langle g_{\rm i}(x)\rangle.
\end{equation}  
If we can make the approximation that fluctuations $\delta g$ are Gaussian, then armed with these measurements we can write the probability distribution
\begin{eqnarray}
P\left( \{g_{\rm i}\} | x\right) &=& {1\over Z(x)} \exp\left[ -{1\over 2} \chi^2 (\{g_{\rm i}\}; x)\right] \label{condP1}\\
\chi^2 (\{g_{\rm i}\}; x) &=& \sum_{\rm ij}\delta g_{\rm i} (x) \left[\hat C^{-1} (x)\right]_{\rm ij} \delta g_{\rm j} (x) \label{condP2}\\
Z(x) &=& \sqrt{(2\pi)^4 \det \hat C (x)} ,\label{condP3}
\end{eqnarray}
where $\hat C^{-1}$ is the inverse of the matrix $\hat C$ and $\det \hat C$ is its determinant.

As is in the examples above, the problem facing the embryo is inverse to the problem we face in characterizing the patterns of gene expression.  A cell (again, more precisely, a nucleus) has access to the concentrations $\{g_{\rm i}\}$ and must infer its position $x$.  Everything that can be known about $x$ by observing $\{g_{\rm i}\}$ is contained in the conditional probability distribution
\begin{equation}
P\left( x | \{g_{\rm i}\} \right) = { {P\left( \{g_{\rm i}\} | x\right)  P_X(x)} \over 
{P\left( \{g_{\rm i}\} \right)}} .
\end{equation}
In the embryo all positions are equally likely a priori, so $P_X(x) = 1/L$, and
\begin{equation}
P\left( \{g_{\rm i}\} \right) = \int dx\, P\left( \{g_{\rm i}\} | x\right)  P_X(x) .
\end{equation}
From the distribution $P\left( x | \{g_{\rm i}\} \right)$ we can compute many things.  

In particular it is tempting to think about constructing a single estimator $x_{\rm est}\left(\{g_{\rm i}\}\right)$.  As above, this could be the optimal estimator in the least--square sense,  the conditional mean
\begin{equation}
x_{\rm est}^{(1)} \left(\{g_{\rm i}\} \right) = \int dx\, x P\left( x | \{g_{\rm i}\} \right),
\end{equation}
or it could be the maximum likelihood estimator
\begin{equation}
x_{\rm est}^{(2)} \left(\{g_{\rm i}\} \right) = \arg\max_x P\left( x | \{g_{\rm i}\} \right).
\end{equation}
If the distribution $P\left( x | \{g_{\rm i}\} \right)$ has a single sharp peak in the neighborhood of 
$x_{\rm est}^{(2)} \left(\{g_{\rm i}\} \right)$, then these two estimators will be very close to one another, and to any other reasonable estimator, e.g. the one that minimizes the mean absolute error (the $L_1$ estimator).  On the other hand, if there is genuine ambiguity, so that $P\left( x | \{g_{\rm i}\} \right)$ has more than one peak, or if the estimation is very uncertain, so that $P\left( x | \{g_{\rm i}\} \right)$ is extremely broad, then no single estimator really captures what a cell ``knows'' based on the expression levels $\{g_{\rm i}\}$.  At the start, it is not obvious that cells won't be in one of these ambiguous or uncertain situations, so we would like to keep all the available information.  This requires us to visualize $P\left( x | \{g_{\rm i}\} \right)$ more directly.

\begin{figure}
\includegraphics[width = \linewidth]{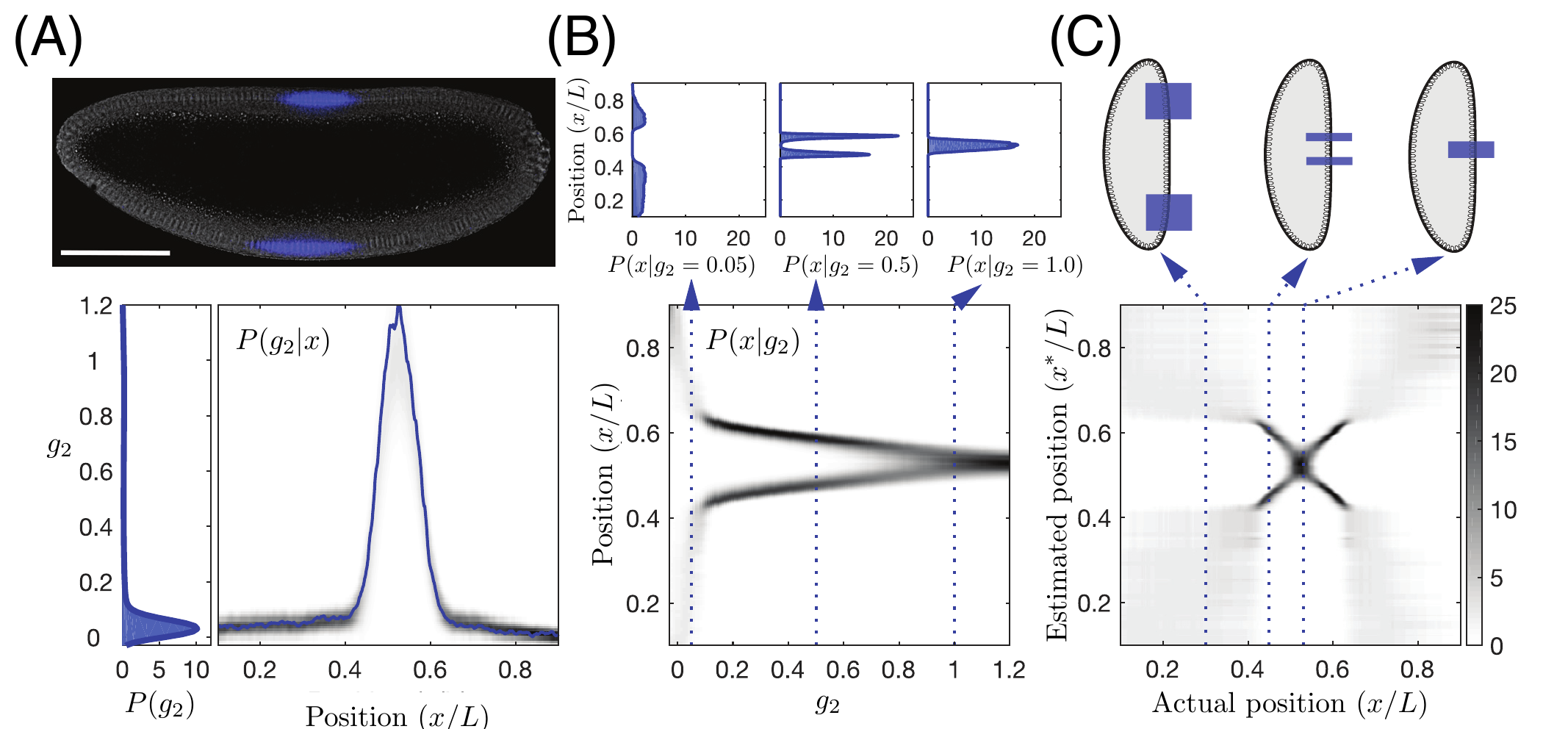}
\caption{Decoding position from a single gene. (A) Top panel is an optical section through the midsagittal plane of a {\em Drosophila} embryo with immunofluorescence labeling for the protein encoded by a single gap gene, corresponding to $g_2$ in Fig \ref{gapgenes} (scale bar $100\,\mu{\rm m}$). Bottom panel shows the normalized $g_2(x)$ for this particular embryo (blue) and a gray--level visualization of $P(g_2 | x)$; $P(g_2)$ is at left.
(B) The distribution $P(x|g_2)$, shown in gray levels (bottom) and slices at fixed $g_2$ (top). (C) The average decoding map $P(x^*|x)$ from Eq (\ref{mapdef}).  Top panel schematizes the combination of ambiguity and uncertainty. Reproduced from Ref \cite{petkova+al_19}, with permission. 
\label{KrMap}}
\end{figure}

Consider a cell at (actual) position $x$ along the anterior--posterior axis.  In one particular embryo $\alpha$, this cell has expression levels $\{g_{\rm i}^\alpha (x)\}$ at this position.  If we ask what this cell knows about its possible or estimated position $x^*$, it is chosen from
\begin{equation}
P^\alpha (x^* | x) = P\left( x^* | \{g_{\rm i}\} \right){\bigg |}_{\{g_{\rm i} = g_{\rm i}^\alpha (x)\}} .
\end{equation}
For simplicity it is useful to look at the average of these ``decoding maps'' across all $N_{\rm em}$  embryos in an experimental ensemble,
\begin{equation}
P (x^* | x) = 
{1\over {N_{\rm em}} }
\sum_{\alpha =1}^{N_{\rm em}} P\left( x^* | \{g_{\rm i}\} \right){\bigg |}_{\{g_{\rm i} = g_{\rm i}^\alpha (x)\}} 
\rightarrow \int \left( \prod_{\rm i} dg_{\rm i}\right) P\left( x^* | \{g_{\rm i}\} \right) P\left( \{g_{\rm i}\} | x\right).\label{mapdef}
\end{equation}
To understand how this works, let's start not with all four gap genes but with one, as shown in Fig \ref{KrMap}, which focuses on the information contained in $g_2$.

The concentration $g_2$ peaks roughly in the middle of the embryo, and falls to be very low in both the front quarter and the back quarter.  Thus cells in these regions would be very uncertain about their position if they had access to only this one gene.  In contrast, cells in the middle of the embryo experience near maximal concentrations and this ``points'' to a relatively narrow region along the anterior--posterior axis.  This peak rises to an amplitude $P(x^* /L | x/L) \sim 25$ which means that the width of the distribution must be $\sigma_x/L \sim 1/25 \sim 4\%$.  Because the mean profile $\langle g_2 (x) \rangle$ is non--monotonic, rising and falling almost symmetrically around the peak, cells that are on these flanks have a two--fold ambiguity in the inference of $x^*$ from $g_2$.  This combination of uncertainty at the ends, precision in the middle, and ambiguity on the flanks gives rise to the X--like pattern that we see when representing $P(x^*|x)$ in gray levels in Fig \ref{KrMap}C.  If $g_2$ were the only signal available, embryos literally would not know their head from their tail, and no single cell could reach the precision of $\sigma_x/L \sim 1\%$ that is seen all along the anterior--posterior axis (\S\ref{lec3-embryo}).

\begin{figure}[b]
\includegraphics[width = \linewidth]{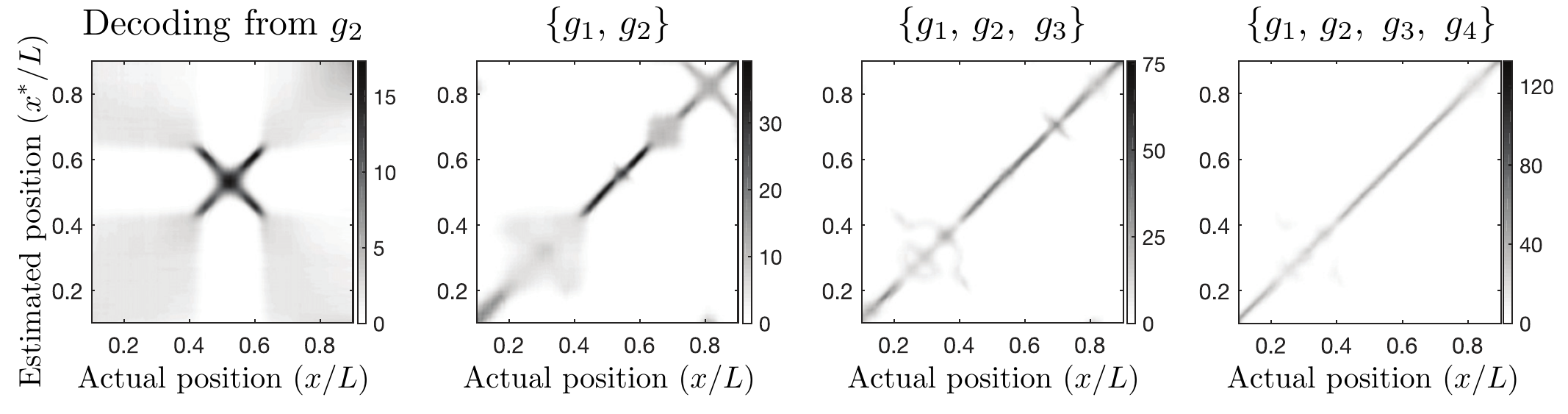}
\caption{Decoding position from an increasing number of gap genes reduces uncertainty and eliminates ambiguity. Redrawn from Ref \cite{petkova+al_19}, with permission.
\label{4genes}}
\end{figure}

Fortunately cells have access to multiple gap genes.  We see in Fig \ref{4genes} that as we use more of these genes to infer position, ambiguities are removed and uncertainty is reduced.  Finally when we use all four genes the distribution $P(x^* | x)$ narrows essentially to a single band around the diagonal $x^* = x$.  The peak height is $\sim 100$, which means that the width is $\sim 0.01$.  Thus, the gap genes provide enough information to specify position with $1\%$ accuracy, but only if cells read this information optimally.

We recall that the pair--rule stripes are positioned with $\sim 1\%$ accuracy, and similarly the location of the cephalic furrow is reproducible from embryo to embryo with $\sim 1\%$ precision.  A possible conclusion is that embryos ``read out'' the information carried by local gap gene concentrations and use this to guide subsequent events.  This read out is optimal, and  sets the precision of the body plan.  Certainly what we see is {\em consistent} with this conclusion, but the evidence is not unambiguous.

As an example, we could imagine that cells make dramatically sub--optimal use of the local concentration information, but compensate at the next stage through interactions among many cells. More subtly, driven by the nature of experiments we have emphasized reading the signals from gap genes at a single moment in time, while any realistic mechanism will involve some integration over time, perhaps providing another opportunity to reduce noise.  There are reasons to think that these options are more limited than they seem:  noise in the expression levels of the gap genes is correlated over long distances \cite{krotov+al_14,mcgough+al_23}, and momentary expression levels are the output of a network whose dynamics imposes temporal correlations on these noise levels; both these effects limit the possibilities for noise reduction by averaging.  Still, one would like a more positive and convincing test of the idea that cells are performing the optimal readout of the positional information encoded in the gap genes.

\begin{figure}
\centerline{\includegraphics[width = 0.9\linewidth]{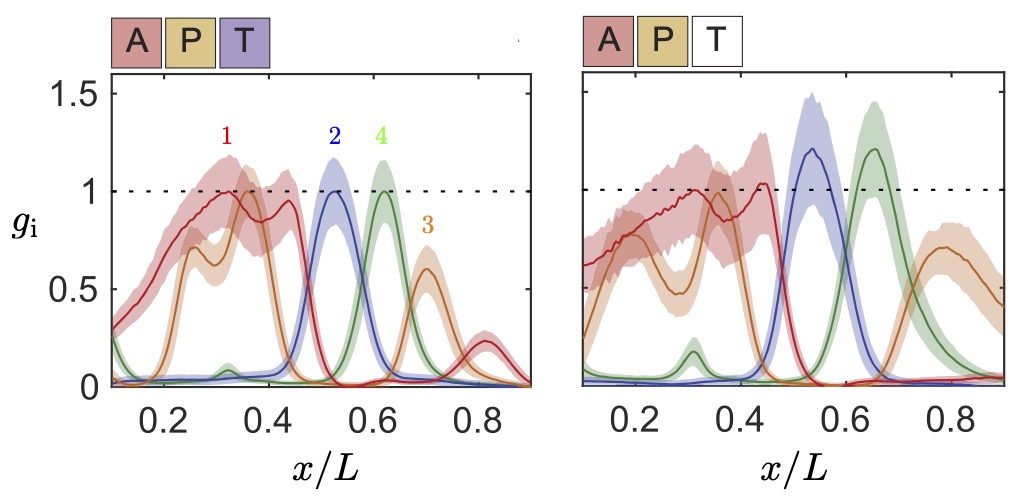}}
\caption{Gap gene concentration profiles change in mutant flies missing one of the maternal inputs.  At left are the profiles in wild type flies,  as in Fig \ref{gapgenes} but with the four genes represented by different colors.  At right are the results of the same measurements done in flies that are missing the terminal inputs T.  Reproduced from Ref \cite{petkova+al_19}, with permission.  It is essential that although concentrations are normalized (maximum mean concentration is one in the wild type), the normalization is the same in both panels so that concentrations can be compared meaningfully.
\label{knockout_T}}
\end{figure}

If the embryo performs an optimal readout, then if the spatial patterns of gap gene expression are perturbed cells will get the wrong answer for their estimates of position, and this effect should be both systematic and predictable.  We can perturb the gap gene patterns by knocking out one or two of the three maternal inputs to the gap gene network.  We recall that one of the maternal inputs has high concentration at the anterior (A) of the embryo, one has high concentration at the posterior (P), and one has high concentration at both ends (the ``terminal'' inputs T).  Figure \ref{knockout_T} shows what happens in mutant flies that are missing the T inputs.  As expected there is very little change to the gap gene expression levels in the middle of the embryo, with larger perturbations at both the anterior and posterior extremes, but in general it seems fair to say that these perturbations are fairly gentle.

\begin{figure}[t]
\centerline{\includegraphics[width = 0.9\linewidth]{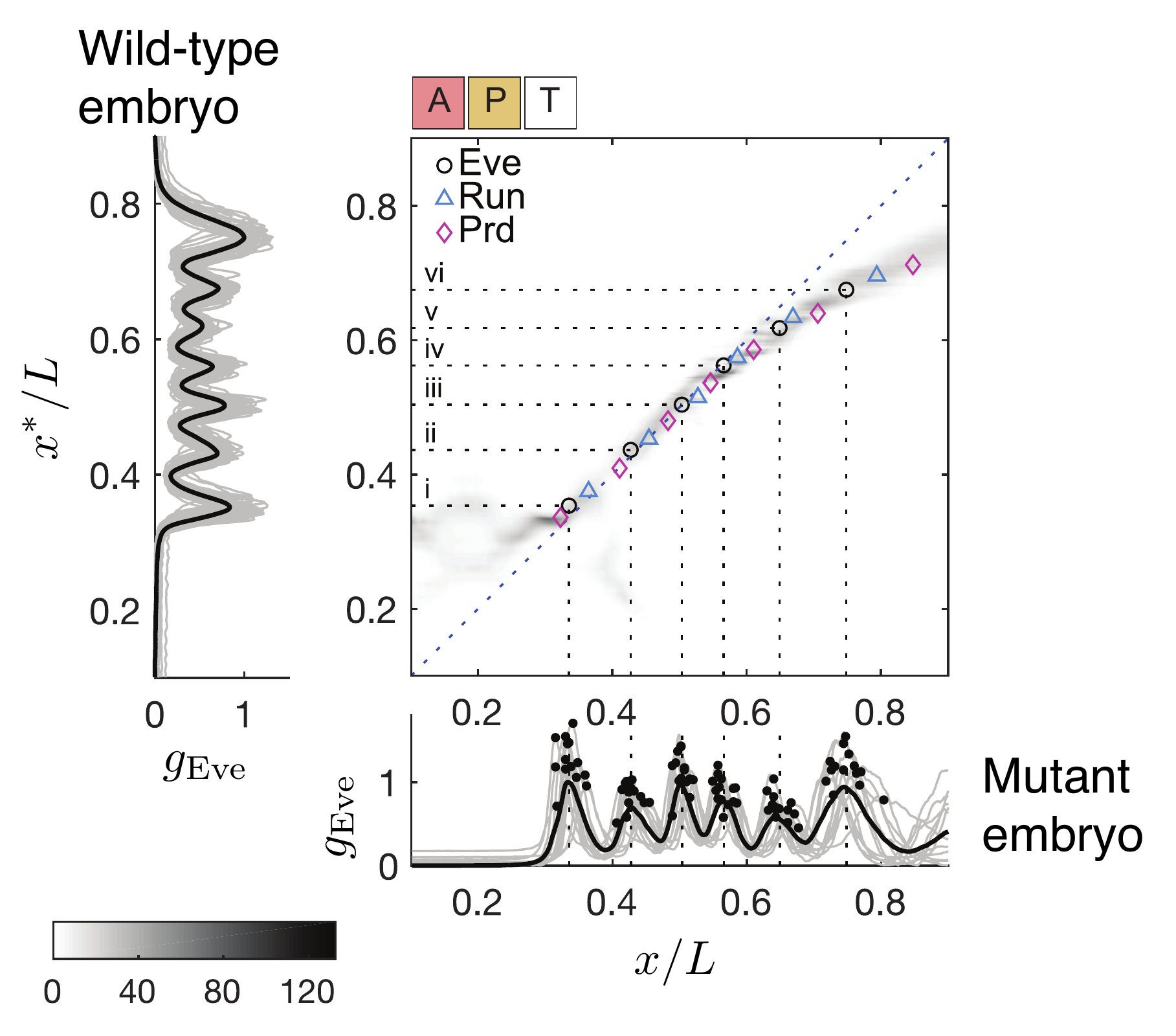}}
\caption{Optimal decoding predicts pair--rule stipe shifts in mutant embryos.  We use the distribution $P(x^*|\{g_{\rm i}\})$ constructed from data on wild type flies to interpret the gap gene expression signals in flies missing terminal inputs T (Fig \ref{knockout_T}).  Results are shown as the average decoding map $P_{\rm mut}(x^* | x)$ from Eq (\ref{mutmap}); density in grey scale (bottom left).  At left is the expression of Eve, one of the pair--rule genes; individual embryos in grey, mean in black, illustrating the seven stripes at positions $x_{\rm i}$.  When $P_{\rm mut}(x^* = x_{\rm i}|x)$ is large we expect a stripe at (real) position $x$   in the mutant, as confirmed at the bottom.  Markers  summarize these results, and those for two other pair--rule genes. Reproduced from Ref \cite{petkova+al_19}, with permission. 
\label{minusTmap}}
\end{figure}

Figure \ref{minusTmap} shows what happens when we try to decode the patterns of gap gene expression found in the mutants of Fig \ref{knockout_T}.  The idea is that we take the measured $\{g_{\rm i}\}$ in the mutant and pass it through the function $P(x^*|\{g_{\rm i}\})$, which was constructed from data taken in the normal (wild--type) embryos, and then we average as before to get a new map $P_{\rm mut}(x^* | x)$.  By analogy with Eq (\ref{mapdef}) we can write
\begin{equation}
P_{\rm mut}(x^* | x) = \int \left( \prod_{\rm i} dg_{\rm i}\right) P\left( x^* | \{g_{\rm i}\} \right) P_{\rm mut} \left( \{g_{\rm i}\} | x\right) .
\label{mutmap}
\end{equation}
As expected, the map is only slightly perturbed in the central region of the embryo.  Signals at small $x/L$ are noisy and ambiguous, while at large $x/L$ the ridge of maximum probability is systematically at $x^* < x$, along a gently curving trajectory.

We can test the predicted $P_{\rm mut}(x^* | x)$ with a simple idea.  If the only information that cells have about position is their estimate $x^*$, then pair--rule stripes should be at locations $x^* = x_{\rm s}$ not $x = x_{\rm s}$ as in the wild--type embryo, for each stripe $\rm s$; this is the construction shown by dashed lines in Fig \ref{minusTmap} for  the gene {\em eve}.\footnote{This is one place where I'll use names.  Eve is short for ``even--skipped,'' which gives a sense for what goes wrong in the structure of the mutant embryo.  The convention is that genes are named in lower case italics, while the protein products are described with capitalized Roman text.  Also, beware that biochemists name molecules after what they do, while geneticists name molecules after what goes wrong when they are mutated.}  We can test all of the stripes, for each of several pair--rule genes, and we see that the results track along the ridge of $P_{\rm mut}(x^* | x)$, with high precision.

We can redo the analysis of Fig \ref{minusTmap} six times, deleting the three maternal inputs singly and then in pairs.  A sanity check is that if we delete all three inputs there is no positional information, so any nonzero $g_{\rm i}$ is uniform in $x$.  Detailed descriptions of the results can be found in the original paper \cite{petkova+al_19}, so let me draw attention here to a few points:
\begin{itemize}
\item In most cases when we delete maternal inputs the density in $P(x^*|x)$ fails to intersect $x^* = x_{\rm s}$ for many values of the stripe index $\rm s$.  This predicts that certain stripes should be missing, and these  (many) predictions are correct.  
\item We can analyze  maps $P^\alpha_{\rm mut}(x^*|x)$ constructed from data on individual mutant embryos (before averaging), and in some cases the density is sufficiently variable at $x^* = x_s$ that we predict stripes to be present in some but not all of the embryos.  This variability never happens in wild--type embryos but it happens in the mutants, where we predict it.\footnote{A limitation of the experiments is that they measure all the gap genes or the pair--rule genes, but not in the same embryo.  Thus we know that pair--rule stripes are variable where we predict them to be variable based on measurements of the gap genes in many mutant embryos, but we don't know if the presence or absence of stripes is connected deterministically to the gap gene profiles in single embryos, as we predict.}
\item In mutants where the anterior input is deleted, the most likely $x^*$ is a non--monotonic function of $x$.  Most of the {\em eve} stripes are predicted to be missing, but the seventh stripe is predicted to be duplicated, and this happens at the predicted location.  Details of the underlying molecular biology show that this really is a duplicate of stripe seven and not a shifted version of a more anterior stripe.
\item In mutants where both the anterior and posterior inputs are deleted, we predict that there should be only two {\em eve} stripes located symmetrically along the anterior--posterior axis, and this is confirmed.
\end{itemize}
In total we have 70 of these examples, and almost all predictions are confirmed within experimental error.  More subtly, the predicted noise in stripe position, from the width of $P_{\rm mut}(x^*|x)$, agrees with the measured variability.  The few errors are in places where the map $P_{\rm mut}(x^*|x)$ has discontinuities, so that a little bit of spatial averaging (which we neglect) would have large effects, or where expression levels in the mutant are very deep in the tails of   $P(\{g_{\rm i}\})$.

We have constructed the optimal decoder $P(x^*|\{g_{\rm i}\})$ from measurements in a small window of time during the fourteenth nuclear cycle.  This window is chosen to surround the point at which positional information is maximized, and is as narrow as possible while still leaving a reasonable number of samples.  But gap gene expression profiles vary slowly throughout cycle fourteen.  If the embryo implements the optimal decoding, tuned to the time of maximal positional information, then pair--rule stripes are predicted to evolve with time as well.  This effect has been known for a long time, and subject to multiple interpretations.  We were surprised to find that these details---shifts of $\Delta x(t)/L$ corresponding to just a few percent over half an hour---are predicted correctly as well.

Optimization principles provide a compact formulation for much of physics.  As applied to living systems we typically use such principles to select the behavior of the particular systems that we see in nature out of the universe of possibilities available.   In the past, this sort of argument has led to a single number, or a scaling relation between different numbers.  What is new, I think, in this analysis of positional information in the fly embryo is that a single optimization principle leads to a rich set of subtle parameter free predictions, essentially all of which agree quantitatively with experiment.  It is the same physical principle that leads to predictions about visual motion estimation in Fig \ref{fig:vest}.

\section{Matching distributions}
\label{sec:matching}

In the fly embryo information flows from maternal inputs to the gap gene network to the striped patterns of pair--rule gene expression.  The previous lecture was about optimization at the output of the gap gene network.  Can we also optimize the inputs?  This is part of a more general question:  given some signal processing system with fixed signal and noise characteristics, how can we choose the inputs to optimize information flow?  

\begin{figure}[t]
\centerline{\includegraphics[width = 0.9\linewidth]{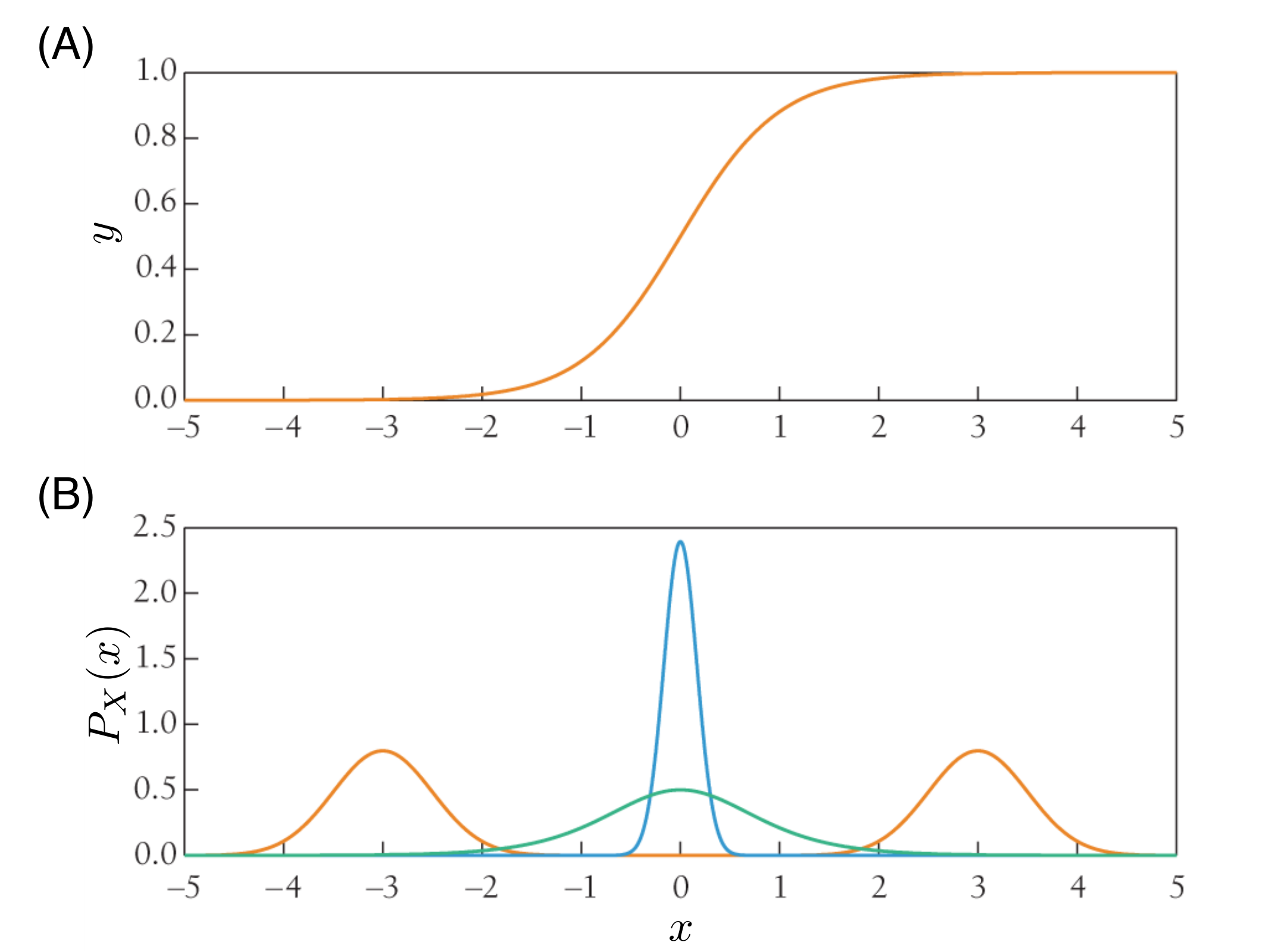}}
\caption{Matching distributions to the input/output relation \cite{bialek_12}. (A) Example of an input/output relation. (B) Different possible probability distributions for the inputs. With either of the orange distributions, the output is either fully on or fully off and provides no information about the input signal.  With the blue distribution the modulations of the output are very small, and likely to be obscured by noise (which is missing from the figure).   The green distribution seems to be a better match, moving the output through its full dynamic range.
\label{match_dist}}
\end{figure}

To be more explicit, a random cell along the length of the embryo encounters concentrations of maternal input morphogens that are drawn out of a distribution.  We would like to formulate an optimization principle for this distribution.  Some intuition for this is given in Fig \ref{match_dist}, where we compare the distribution of inputs to the structure of a hypothetical input/output relation.   In this example, if the inputs $x$ are concentrated at high or low values, then the output $y$ is ``stuck'' in a fully on or fully off state, and not sensitive to any variations in the input.  If the inputs are concentrated near the point of maximum sensitivity (here, zero input) then the responses will be larger, but if the distribution is too narrow than the variations in the output still will be small and can easily be masked by noise.  It seems sensible that the best we can do is to have inputs that are centered on the point of maximum sensitivity and have a distribution wide enough to drive the outputs through their full dynamic range.

The intuitions of Fig \ref{match_dist} can be formalized in the language of information theory.  Information theory is a beautiful subject that has deep connections to statistical mechanics, but physicists vary in their level of comfort and familiarity with the ideas.  I gave the lecture in Les Houches as if people knew the basics, but there was a strong desire for a more pedagogical introduction.  So we followed the regular lecture with a tutorial \cite{thome+bialek_24}, which we plan to have as an Appendix to the Summer School proceedings.\footnote{My sincere thanks to Tarek Tohme, who will co-author this Appendix, having recorded the tutorial and helped turn it into coherent prose.  He even captured many of the excellent questions from the students.} For completeness let me also note that (in contrast to many other subjects) one really can learn much of what you need to know about information theory by reading Shannon's original paper \cite{shannon_48}.  The  standard textbook is by Cover and Thomas \cite{cover+thomas_91},  a version aimed  at the physics community is by M\'ezard and Montanari \cite{mezard+montanari_09}, and I have tried to give a fuller account of these ideas in the context of biophysics \cite{bialek_12}.  Each of these texts provide  more than you need to make sense out of these lectures, but perhaps you will want to explore more deeply.

\subsection{One input, one output}
\label{one-to-one}

The simplest version of information transmission is a system in which one input $x$ drives one output $y$.  The mutual information $I(x;y)$ between these variables can be thought of either as the information that the output provides about the input or vice versa, as a measure of control power.  This measure is unique, and can be written as
\begin{equation}
I(x;y) = \int dx\int dy \,P(x,y) \ln\left[{{P(x,y)}\over{P_X(x) P_Y(y)}}\right],
\end{equation}
where, as usual,  $P(x,y)$  is the joint distribution while $P_X(x)$ and $P_Y(y)$ are the two marginals.

Mutual information is a measure of correlation between $x$ and $y$.  If the variables were independent, then the entropy of the joint distribution would be equal to the sum of the entropies of the two marginal distributions, but in fact it is smaller.  The mutual information is exactly this decrease in entropy:
\begin{equation}
I(x;y) = S[P_X(x)] + S[P_Y(y)] - S[P(x,y)],
\end{equation}
where $S[P] $ is the entropy of the distribution $P$,
\begin{equation}
S[P(z)] = -\int dz\, P(z) \log P(z) .
\end{equation}
We also can think of the mutual information as a functional of the two distributions,
\begin{eqnarray}
I(x;y) &=& I[P_X(x) , P(y|x)]  = \int dx\,P_X(x) \int dy\, P(y|x) \log\left[{{P(y|x)}\over{P_Y(y)}}\right]\\
P_Y(y) &=& \int dx\, P(y|x) P_X(x) .
\end{eqnarray}
The convexity of the entropy implies that $I(x;y)$ has a maximum with respect to variations in $P_X(x)$ and a minimum with respect to variations in $P(y|x)$.  Thus if we fix the way in which $y$ responds to $x$, as encoded in $P(y|x)$, we can maximize information transmission by adjusting the distribution of inputs.

To be concrete, let's assume that
\begin{equation}
P(y|x) = {1\over\sqrt{2\pi\sigma_y^2(x)}}\exp\left[ - {{\left(y - \bar y(x)\right)^2}\over{2\sigma_y^2(x)}}\right] .
\label{Py_given_x}
\end{equation}
Notice  that 
\begin{eqnarray}
I(x;y) &=& \int dx\,P_X(x) \int dy\, P(y|x) \log\left[{{P(y|x)}\over{P_Y(y)}}\right]
\nonumber\\
&=& - \int dy\, P_Y(y) \log P_Y(y) -\int dx\, P_X(x) \left[ \int dy\, P(y|x) \log P(y|x)\right]\\
&=& S[P_Y(y)] - \langle S[P(y|x)]\rangle^{(x)} ,
\label{Iequalsdiffentropy}
\end{eqnarray}
where $\langle \cdots \rangle^{(x)}$ denotes an average over the distribution $P_X(x)$.  We can compute the conditional entropy from Eq (\ref{Py_given_x}), now using natural logs:
\begin{eqnarray}
S[P(y|x)] &\equiv& -\int dy\, P(y|x) \ln P(y|x)\\
&=& -\int dy\, {1\over\sqrt{2\pi\sigma_y^2(x)}}\exp\left[ - {{\left(y - \bar y(x)\right)^2}\over{2\sigma_y^2(x)}}\right] \left[ -{1\over 2}\ln[2\pi\sigma_y^2(x)] - 
{{\left(y - \bar y(x)\right)^2}\over{2\sigma_y^2(x)}}\right]\nonumber\\
&&\\
&=& {1\over 2}\ln[2\pi\sigma_y^2(x)] + {1\over 2}\\
&=& {1\over 2}\ln[2\pi e \sigma_y^2(x)] .
\end{eqnarray}
This result for the entropy of Gaussian distributions is very useful.  

\begin{figure}[t]
\includegraphics[width=\linewidth]{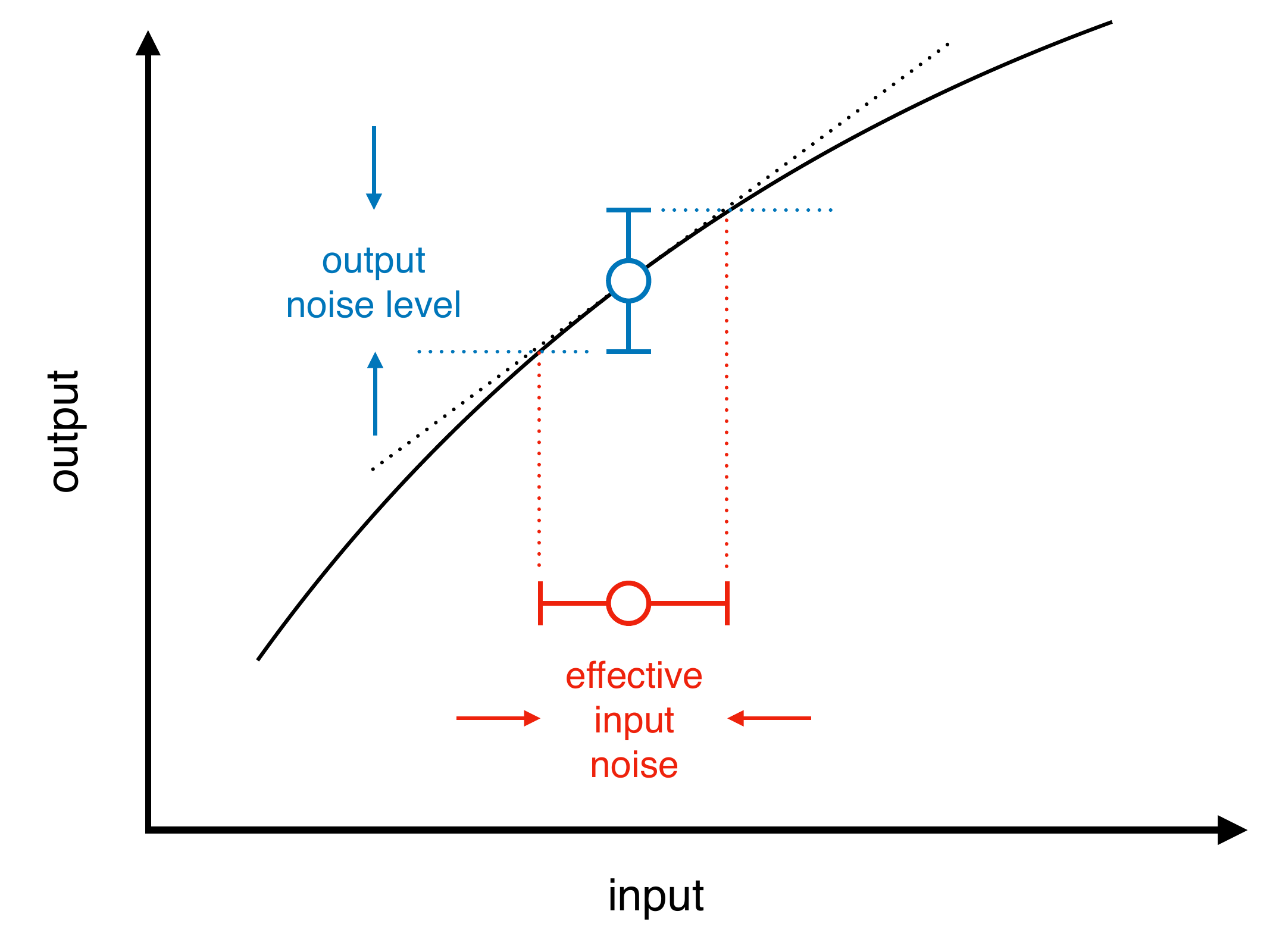}
\caption{Error propagation, from Eq (\ref{errorprop_eq}).  The  noise level at the output---which often can be estimated directly from experiment---is translated to an effective noise level at the input, using the local slope of the input/output relation. \label{errorprop_fig}}
\end{figure}

Substituting into Eq(\ref{Iequalsdiffentropy}), we have
\begin{eqnarray}
I(x;y) &=&S[P_Y(y)] - \langle S[P(y|x)]\rangle^{(x)}\nonumber\\
&=& -\int dy\, P_Y(y) \ln P_Y(y) - \int dx\, P_X(x) {1\over 2}\ln[2\pi e \sigma_y^2(x)] .
\end{eqnarray}
Now if the function $\bar y (x)$ is monotonic, and the noise is small, we can approximate
\begin{equation}
P_Y(y) dy \approx P_X(x) dx ,
\label{pydy=pxdx}
\end{equation}
so that 
\begin{eqnarray}
I(x;y) 
&\approx& -\int dx\, P_X(x) \ln \left[ P_X(x) {\bigg |} {{dy}\over {dx}}{\bigg |}^{-1}\right] - \int dx\, P_X(x) {1\over 2}\ln[2\pi e \sigma_y^2(x)]\\
&=& -\int dx\, P_X(x) \ln \left[ P_X(x) 
\sqrt{2\pi e}\sigma_x^{\rm eff}(x) \right] ,
\label{Ixy4}
\end{eqnarray}
where we identify
\begin{equation}
{1\over {\sigma_x^{\rm eff}(x)}} =  {1\over {\sigma_y(x)}}{{d\bar y(x)}\over{dx}}  .
\label{errorprop_eq}
\end{equation}
We can understand this as the propagation of the noise $\sigma_y$ back into an estimate of $x$ with effective noise level $\sigma_x^{\rm eff}(x)$, illustrated in Fig \ref{errorprop_fig}.  This approximation is self--consistent if $\sigma_x^{\rm eff}(x)$ is small on the scale over which $\bar y(x)$ and $P_X(x)$ vary.

Starting from Eq (\ref{Ixy4}), we can vary the input distribution $P_X(x)$ to maximize the information $I(x;y)$, introducing a Lagrange multiplier to enforce normalization:
\begin{eqnarray}
0 &=& {\delta\over{\delta P_X(x)}} \left[ I(x;y) - \lambda \left( \int dx\, P_X (x) - 1\right)\right] \\
&=& {\delta\over{\delta P_X(x)}} \left[-\int dx\, P_X(x) \ln \left[ P_X(x) 
\sqrt{2\pi e}\sigma_x^{\rm eff}(x) \right] - \lambda \left( \int dx\, P_X (x) - 1\right)\right] \\
&=& -\ln \left[ P_X(x)  \sqrt{2\pi e}\sigma_x^{\rm eff} (x)\right]  - 1 -\lambda \\
\Rightarrow P_X(x) &=& {1\over Z} {1\over{\sigma_x^{\rm eff}(x)}},\label{match1}
\end{eqnarray}
where we collect all the normalization constants into 
\begin{equation}
Z = \int {{dx}\over{\sigma_x^{\rm eff}(x)}} .
\label{Zdef}
\end{equation}
Notice that with this result  $\ln \left[ P_X(x) 
\sqrt{2\pi e}\sigma_x^{\rm eff} (x)\right] = \ln[Z/\sqrt{2\pi e}]$, so that
\begin{equation}
I_{\rm max} (x;y) = \ln\left( {Z\over{\sqrt{2\pi e}}}\right)  .
\label{Imax}
\end{equation}
The crucial result here is Eq (\ref{match1}):  to transmit the maximum  information (at low noise levels) we should use inputs in  inverse proportion to their effective noise levels.  In this sense the optimal input distribution matches the input/output relation and the associated noise levels.

The result that  inputs should be used in inverse proportion to their noise level is a precise version of familiar ideas.  When writing we avoid words that we don't know how to spell, and when speaking a foreign language we avoid constructions which we suspect we might get wrong.  Different species of frogs call to one another in different frequency bands, and these match the bands where they hear best.  In the low noise limit we can think of the matching condition as applying equally well to the input or the output,
\begin{equation}
P_X(x) \propto {1\over{\sigma_x^{\rm eff}(x)}} \,\,\,\,\,\,\,\,\,\, \Rightarrow \,\,\,\,\,\,\,\,\,\, P_Y(y) \propto {1\over{\sigma_y(x)}}{\bigg |}_{x = \tilde x(y)} ,
\label{match2}
\end{equation}
where $\tilde x (y)$ is the inverse of the function $\bar y(x)$.

\subsection{Neural input/output relations}
\label{lec3-neurons}

An important special case of these arguments is where the noise at the output is constant, $\sigma_y(x) = \sigma_y$.  Then Eq (\ref{match2}) tells us that optimal information transmission corresponds to $P_Y(y)$ also being constant. Outputs always have limited dynamic range, so this means that the system transmits the maximum information by using this dynamic range uniformly, maximizing the entropy of the outputs.\footnote{Maximizing information transmission generally is {\em not} the same as maximizing the entropy of the outputs.  The difference arises both because the noise can have structure [$\sigma_y(x) \neq \sigma_y$] and because we can depart from the low noise limit where Eqs (\ref{match1}, \ref{match2}) were derived.  This will be important below.} A familiar example is where the outputs are quantized, as with a digital image.  Then the dominant source of (effective) noise often is the discretization itself, and thus is constant---we distinguish $1$ vs $2$ as reliably as we distinguish $254$ vs $255$.  Then optimal information transmission occurs when all the output values are used equally often, and this is called ``histogram equalization'' or adaptive binning.  

We can go one step further and choose units such that $0 < y < 1$, so the optimized uniform distribution of outputs is $P_Y(y)=1$. In the low noise limit we have Eq (\ref{pydy=pxdx}), so that
\begin{eqnarray}
P_Y(y) dy &=& P_X(x) dx \nonumber\\
dy &=& P_X(x) dx \\
{{dy}\over{dx}} &=& P_X(x).
\end{eqnarray}
Again because this is the low noise limit the $y$ which appears here can be taken as $\bar y(x)$, and so we have 
\begin{equation}
{{d\bar y (x)}\over{dx}} = P_X(x).
\label{laughlin_match}
\end{equation}
Integrating, the optimal input/output relation becomes the cumulative distribution of inputs.  

Two interesting things just happened.  First, we started by optimizing the distribution of inputs and ended up by expressing the result as an optimal input/output relation.  Second, we have a prediction that the input/output relation should match the distribution of inputs, quantitatively and with no free parameters.  If the distribution $P_X(x)$ has a single peak,  $\bar y(x)$ should be roughly sigmoidal, as one finds for the input/output behaviors of many biological systems.

\begin{figure}[t]
\centerline{\includegraphics[width=0.75\linewidth]{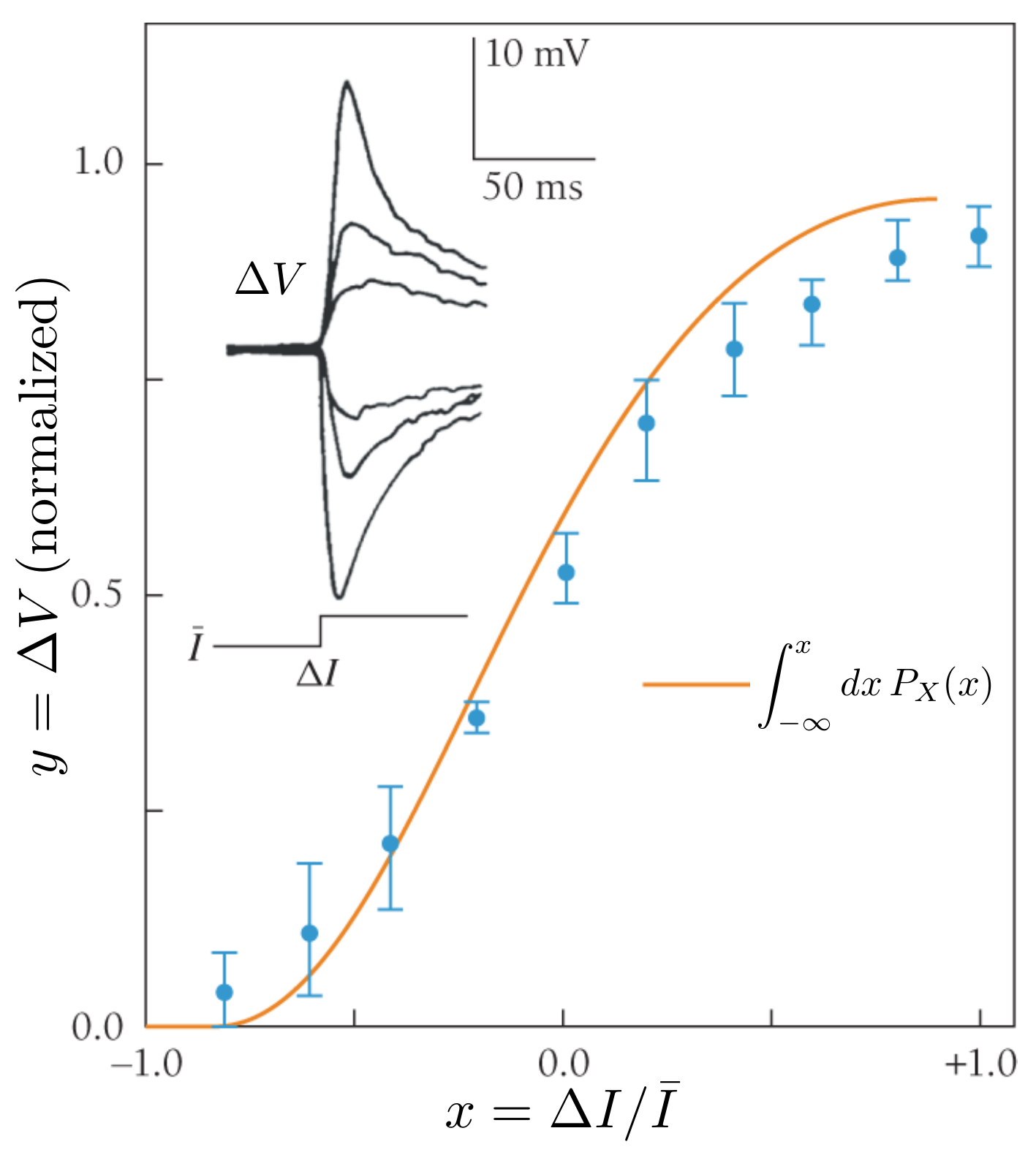}}
\caption{Input/output relations match the distribution of inputs. Brief changes in light intensity $\Delta I$ above or below background $\bar I$ produce transient voltage changes in the large monopolar cells (inset), and the peaks of these responses are taken as the cell’s output $\Delta V$ . Normalized responses are compared to the cumulative probability distribution of light intensities, testing the predictions of Eq (\ref{laughlin_match}). Data redrawn from Ref  \cite{laughlin_81}, under the Creative Commons Attribution NC-ND 3.0 License.\label{fig:laughlin}}
\end{figure}

In an inspiring paper, now 40+ years ago, Laughlin took these theoretical ideas seriously and applied them to the responses of neurons in the fly retina, the ``large monopolar'' cells (LMCs) that take inputs directly from the photoreceptors \cite{laughlin_81}.\footnote{These cells are in the same position in the fly's retina as the bipolar cells  in the vertebrate retina (\S\ref{sec:photoncount}).}. These cells produce a graded voltage in response to changes in light intensity around a background.  Laughlin built a photodetector to match the optics of a single receptor cell in the fly's eye and measured the distribution of light intensities found by scanning natural scenes.  He then measured the input/output relations of the LMCs, with the comparison shown in Fig \ref{fig:laughlin}.  Note that there are no free parameters.

To be fair, the input/output relation is not a very complicated function, so saying that we predict its form with no free parameters may be an overstatement.  What we predict is that, in the normalized units of Fig \ref{fig:laughlin},  the maximum slope should be at a location near $x=0$ and the width of the response should span $x \in [-0.5, 0.5]$; these predictions come from the shape of the distribution of light intensities.  

The light intensity that we see is the product of a source intensity and the reflectivity of the surface we are looking at.  As the overall brightness changes, e.g. from dawn through noon to dusk, reflectivities are constant.  This suggests that the distribution of log intensity should keep roughly the same shape, just shifting along the intensity axis.  The matching condition Eq (\ref{laughlin_match}) then predicts that the neural responses to log intensity should keep the same shape, just shifting their midpoints, and this is a good zeroth order theory of what happens during light and dark adaptation.

There are many over--simplifications here, but the idea is powerful.  When we ask why the input/output relation of a neuron looks the way it does, the standard answer is to start explaining all the molecular, cellular, and synaptic mechanisms that lead to the final phenomenology.  Such explanations, as emphasized in the introductory lecture, lead us into a forest of  parameters.  Worse yet, these parameters are adjustable, for example if the cell in question had expressed different ion channels, different neurotransmitter receptors, etc.  The idea here is different:  input/output relations have the form that they do because they are matched to their inputs, optimizing information transmission.  The fact that the predictions from such a potentially general, parameter free theory are even approximately right is very encouraging.

If we think that input/output relations are matched to the distribution of inputs, it is natural to ask on what time scale this matching occurs.  More critically, one might worry that ``the'' distribution is such a dynamic object that it is not well defined.  But maybe this is a good thing, since we know that input/output relations in neurons are themselves dynamic objects.

The fact that input/output relations of neurons change in response to background conditions is called adaptation.\footnote{Surely ``adaptation''  is one of the most over used words in the description of biological systems.} In the original descriptions of adaptation in sensory neurons, the focus was on the fading response to constant stimuli, in effect subtracting a constant from the output, but this is far from the whole story.  The time scales of response often change with the background, e.g. as visual responses become slower in the dark.  More subtly there are changes in the gain of the response, e.g. as one extra photon produces a smaller response when added to a brighter background of light.  All of these effects, however, can be seen as driven by the mean background signal.  Maximizing information transmission predicts that neurons should match their input/output relations to the whole {\em distribution} of input signals.

Maybe the simplest question is whether neurons adapt to the variance of inputs as well as to the mean.  As an example, if we look at the activity of the neurons that carry the output of the retina to the brain (retinal ganglion cells), then stimulate the retina with time varying light intensity, there is a clear and rather immediate response to changing the mean intensity and that gradually relaxes as the retina adapts.  The same thing happens if we suddenly change the variance of the light intensity, or its spatial correlations \cite{smirnakis+al_97}, which certainly is consistent with the idea that input/output relations are changing to match the input distribution.

We would like to map the input/output relations in the steady states that are reached after the system has adapted to different distributions of inputs.    How to do this is a subject in itself.  Briefly, we are interested here in neurons that respond to inputs with a sequence of discrete, identical electrical pulses called action potentials or spikes. We will assume for simplicity that the input is a single function of time $s(t)$.  One characterization of the input/output relation is then to give the probability per unit time that the neuron will generate a spike near time $t$ given the history of inputs up to this moment, $s(t-\tau)$ for $\tau > 0$.  If the inputs were weak one could try a linear approximation,
\begin{equation}
P_{\rm spike}(t) = \bar r \left[ 1 + \int_0^\infty d\tau\, g(\tau) s(t-\tau) + \cdots \right],
\end{equation}
where $\bar r$ is the mean probability per unit time or rate of generating spikes, and $g(\tau )$ is the linear response function.  This is too restrictive, and easily runs up against the constraint that $P_{\rm spike} \geq 0$.  A natural extension is to say that only the filtered inputs are important, but these could be processed nonlinearly,
\begin{equation}
P_{\rm spike}(t) = \bar r F \left[ \int_0^\infty d\tau\, g(\tau) s(t-\tau)  \right] ,
\label{lin-nonlin1}
\end{equation}
where $F$ is arbitrary but a natural choice might be something sigmoidal as in Fig \ref{fig:laughlin}.

The ``linear--nonlinear'' model in Eq (\ref{lin-nonlin1}) is quite popular \cite{rieke+al_97,dayan+abbott_01}.  It is interesting in part because it is tractable.  If the input signals $s(t)$ are Gaussian with zero mean,\footnote{I think in the computer science and applied mathematics literature one would say ``Gaussian stochastic process.'' Physicists use ``Gaussian'' more vaguely to describe anything from a single random variable to a free field theory.} then we can separate the filter from the nonlinearity by computing a correlation function between the spike sequence and the input, or equivalently an average of the input triggered on spike times:
\begin{equation}
\langle s(t_{\rm spike} - t)\rangle \propto \int d\tau g(\tau ) \langle s(t) s(\tau ) \rangle .
\label{revcor}
\end{equation}
In the simplest case where the inputs are both Gaussian and white, so that $\langle s(t) s(\tau ) \rangle \sim \delta (t-\tau)$, we have just $\langle s(t_{\rm spike} - t)\rangle \propto g(t)$.   This strategy for describing neural responses came from work on the neurons that first encode sound in the inner ear, as people tried to separate the mechanical filtering of sound from the nonlinearity of spiking \cite{deboer+kuyper_68}.

Another way of expressing Eq (\ref{lin-nonlin1}) is to say that the input is a high dimensional vector---the function $s(t < t_{\rm spike})$, perhaps sampled at discrete times---and that the neural response depends on only one projection of this vector.  Then there is a natural generalization 
\begin{eqnarray}
P_{\rm spike}(t) &=& \bar r F \left[ s_1(t),\, s_2 (t) ,\, \cdots ,\, s_K(t)  \right] 
\label{lin-nonlin2}\\
s_\mu (t) &=& \int_0^\infty d\tau\, g_\mu (\tau) s(t-\tau) .
\end{eqnarray}
This description emerged from analysis of the distribution of inputs conditional on the occurrence of a spike \cite{ruyter+bialek_88} and would be used as a model of responses only later \cite{brenner+al_00,bialek+ruyter_05,rust+al_05}.  Again the key is that triggered averages are connected to the underlying structure.  If we compute the covariance of inputs in the neighborhood of a spike, and compare with the total (not triggered) covariance, then Eq (\ref{lin-nonlin2}) implies that
\begin{equation}
\Delta C(t,t') = \left[ \langle s(t_{\rm spike} - t)s(t_{\rm spike} - t')\rangle - \langle s(t_{\rm spike} - t)\rangle\langle s(t_{\rm spike} - t')\rangle\right] - \langle s( t)s(t')\rangle 
\end{equation}
is an operator of rank $K$.  Further, the eigenfunctions of this operator span the same space as the filters $\{g_\mu (\tau )\}$, blurred by the input correlation function.  If the rank $K$ is reasonably small this gives us a path to identify the relevant projections of the input and map the input/output relation $F[\{s_\mu\}]$.  We can then change the distribution of inputs $P[s]$ and ask if this relation changes in ways that we expect for the optimization of information transmission.

To be fair, we don't really know how to solve the relevant optimization problem for spiking neurons.  Obviously the assumption of constant noise at the output, as above, doesn't really make sense when the output is the probability of generating a spike.  But if the noise levels are small, then in any reasonable scenario the scales of the optimal input/output relation $F[\{s_\mu\}]$ will be set by the scale of the probability distribution $P[s(t)]$ itself.  Concretely, if we rescale the input signals we expect a compensatory rescaling of the response function,
\begin{equation}
s\rightarrow \lambda s \,\,\,\,\,\,\,\,\,\Rightarrow\,\,\,\,\,\,\,\,\,\, F[\{s_\mu\}] \rightarrow F[\{s_\mu/\lambda\}] .
\label{rescale}
\end{equation}
Maybe a clearer way to say this is that the input/output relations should be different in response to different distributions of input, but these should collapse if we plot the response not vs the projected signals $s_\mu$ but rather vs the normalized projections $s_\mu/ s_\mu^{\rm rms}$.

\begin{figure}[t]
\includegraphics[width=\linewidth]{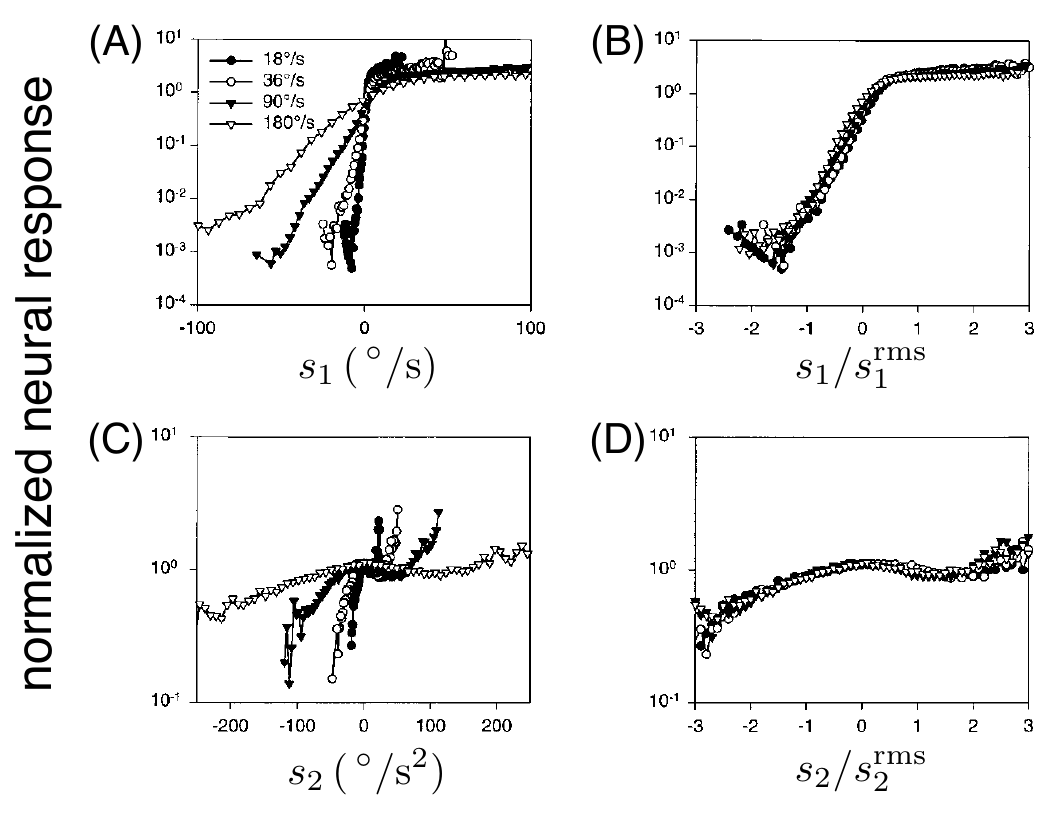}
\caption{Adaptive rescaling in the responses of a motion--sensitive neuron in the fly visual system.  Normalized neural response $F(s_1, s_2)$ from Eq (\ref{lin-nonlin2}) projected onto stimulus features $s_1$ in  (A, B) and  onto $s_2$ in (C, D).  (A, C) Input/output functions are different when inputs are chosen from different distributions and plotted in physical units; inset in (A) indicates the standard deviation  of the input velocities in the four distributions.  (B, D) Input/output relations collapse in normalized units, confirming the prediction of Eq (\ref{rescale}).  Reproduced from Ref  \cite{brenner+al_00}, with permission, and thanks to N Brenner and RR de Ruyter van Steveninck. \label{fig:naama}}
\end{figure}

These predictions were tested in experiments on motion--sensitive neurons in the fly visual system \cite{brenner+al_00}.  The fly watches a movie of a random spatial pattern that moves with velocity $s(t)$ chosen from a Gaussian distribution with a short ($2\,{\rm ms}$) correlation time, and the standard deviation of this distribution could be changed by rescaling as above.    Mean rates of spiking were $\bar r \sim 70\,{\rm spikes}/{\rm s}$, and the spike--triggered covariance $\Delta C(t,t)$ had only two significantly nonzero eigenvalues.  The filter $f_1(\tau )$ smooths the velocity over a $\sim 50\,{\rm ms}$ window, while $f_2(\tau )$ is almost exactly the derivative of $f_1(\tau )$.  It then is natural to express $s_1$ in the same physical units as the input angular velocity ($^\circ/{\rm s}$), and  $s_2$ as an angular acceleration ($^\circ/{\rm s}^2$).  Figure \ref{fig:naama} shows the projections of $F(s_1, s_2)$ onto the two axes as measured under four different conditions where the standard deviation of the input velocity varies by an order of magnitude.

Figures \ref{fig:naama}A and C show the input/output relation in physical units.  We see that when the same velocity is chosen from a different distribution, the response of the neuron can differ by orders of magnitude.  But if we rescale the inputs by their standard deviations, these enormous variations collapse onto a single function, as in Figs \ref{fig:naama}B and D.  As an aside, we note that this approach maps the input/output relations over a dynamic range of $\sim 10^3$, and the rescaling is essentially perfect across this full range.  Although harder to visualize, we can see the same rescaling in the $(s_1, s_2)$ plane.  This is exactly what we expect from Eq (\ref{rescale}).

One might worry that the changing input/output relations are an artifact of sampling a fixed function of many variables in different parts of the space as we change distributions.  We are reassured by the fact that we see the same behavior along two dimensions, separately and together, and that $\Delta C$ is very accurately of low rank.  We can also check that the information that single spikes convey about the pair $(s_1, s_2)$ is equal to the information that is conveyed about the stimulus as a whole, within experimental error.   The same effects can be recapitulated in response to inputs that have orders of magnitude slower time scales.

These results show that the system can implement not just a single input/output relation $F(s_1,s_2)$ but rather a whole family of relations $F(bs_1 , bs_2)$, where the scale factor is inversely proportional to the standard deviation of the stimulus, $b = \lambda/s_{\rm rms}$.  The real system is characterized by a particular value of $\lambda$, which depends on exactly how we normalize the filters $f_\mu (\tau)$; as we did things the real $\lambda = 1$.  But we could imagine systems that achieve adaptive rescaling, stretching their input/output relation as the stimulus variance is increased, but the value of $\lambda$ could be different.  If we make a model of this behavior, we find that information transmission is maximized at the observed $\lambda = 1$:  not only does the system rescale as predicted by our optimization principle, but the precise rescaling factor is optimal \cite{brenner+al_00}.  If we make a sudden shift in the input variance, then we can ``catch'' the neuron in an intermediate state, before adaptation, and in this state each spike really does transmit less information about the input \cite{fairhall+al_01}.  The recovery of the information is very fast, perhaps as fast as the system can reliably infer that the input distribution has changed, although longer time scale dynamics help to encode the input variance, resolving ambiguities.

The same adaptive rescaling phenomena have been seen in the bird ``field L,'' which is an analog of our auditory cortex \cite{nagel+doupe_06}; in the rat ``barrel cortex,'' which responds to whisker movements \cite{maravall+al_07};  and many other systems. 
Related effects are  seen in midbrain  regions of mammalian auditory processing, and there is direct evidence that these adaptations improve information transmission by the population of neurons as a whole \cite{dean+al_05}.  The dynamics of these adaptation processes, as in the example of the fly \cite{fairhall+al_01}, span many time scales, which is a separately fascinating subject \cite{wark+al_07}.  In the retina one can make progress toward identifying the molecular mechanisms responsible for adaptation to the distribution \cite{kim+rieke_01,kim+rieke_03}.

\subsection{Positional error in the embryo}
\label{lec3-embryo}

The gap gene expression levels $\{ g_{\rm i}\}$ encode information about position $x$ along the anterior--posterior axis.  In \S\ref{lec2-embryo} we have discussed how to decode this information, but we haven't talked about how much information is present.  In general this is hard to estimate, but a useful observation is that noise levels are small.  Let's explore \cite{dubuis+al_13b,tkacik+al_15}.

Formally we have Eqs (\ref{condP1}--\ref{condP3}),
\begin{eqnarray}
P\left( \{g_{\rm i}\} | x\right) &=& {1\over Z(x)} \exp\left[ -{1\over 2} \chi^2 (\{g_{\rm i}\}; x)\right]  \\
\chi^2 (\{g_{\rm i}\}; x) &=& \sum_{\rm ij}\delta g_{\rm i} (x) \left[\hat C^{-1} (x)\right]_{\rm ij} \delta g_{\rm j} (x)  \\
Z(x) &=& \sqrt{(2\pi)^4 \det \hat C (x)} , 
\end{eqnarray}
where as usual $\delta g(x)$ is the fluctuation around the mean value at $x$,
\begin{equation}
\delta g_{\rm i} (x) = g_{\rm i} (x) -\langle g_{\rm i}(x)\rangle ,
\end{equation}
and $\hat C(x)$ is the covariance matrix of these fluctuations 
\begin{equation}
\left[ \hat C (x)\right]_{\rm ij} = C_{\rm ij} (x) = \langle \delta g_{\rm i} (x) \delta g_{\rm j}(x) \rangle .
\end{equation}
Suppose that the concentrations $\{g_{\rm i}\}$ are those found in a cell at position $x= x_{\rm true}$.  Then it is convenient to expand $\chi^2 (\{g_{\rm i}\}; x)$, and we'll assume that the covariance varies more slowly than the mean. So we start with
\begin{equation}
\langle g_{\rm i} (x)\rangle = \langle g_{\rm i} (x_{\rm true})\rangle +\left[ {{d\langle g_{\rm i} (x)\rangle}\over{dx}}\right]_{x = x_{\rm true}} (x - x_{\rm true}) + \cdots ,
\end{equation}
which then implies
\begin{eqnarray}
 \chi^2 (\{g_{\rm i}\}; x) &=& \chi^2 (\{g_{\rm i}\}; x_{\rm true} ) - 2 A (x-x_{\rm true})  + B (x-x_{\rm true} )^2 + \cdots\\
A &=& \left[ \sum_{\rm ij}  { {d\langle g_{\rm i} (x)\rangle} \over{dx} }  \left[\hat C^{-1} (x)\right]_{\rm ij} (g_{\rm j} - \langle g_{\rm j} (x)\rangle) \right]_{x = x_{\rm true}} \\
B&=& \left[ \sum_{\rm ij} {{d\langle g_{\rm i} (x)\rangle}\over{dx}} \left[\hat C^{-1} (x)\right]_{\rm ij} {{d\langle g_{\rm j} (x)\rangle}\over{dx}} \right]_{x = x_{\rm true}} 
\end{eqnarray}
If the distribution of positions $P(x)$ is smooth, then the conditional distribution $P(x|\{g_{\rm i}\})$ becomes a Gaussian with a mean that is slightly shifted from $x_{\rm true}$ by terms related to the noise $\delta g$ and a variance $\sigma_x^2 = 1/B$.  All of these approximations are self--consistent if the values of $\sigma_x$ that we compute in this way comes out to be small enough.

It is useful to think of these results as a generalization of error propagation, as in the discussion surrounding Fig \ref{errorprop_fig}.  If the position $x$ is encoded by a single variable $g$ that has variance $\sigma_g^2$, then the effective variance in $x$ is determined by
\begin{equation}
{1\over{\sigma_x^2}} = {1\over {\sigma_g^2}} \left[ {{d\langle g \rangle}\over{dx}}\right]^2 .
\end{equation}
Measurements are like springs that hold our estimate of the underlying variable close to its true value.  As with the thermal fluctuations of a particle hanging from a spring, the variance is inversely proportional to the spring constant.  If we make multiple independent measurements the spring constants should add, reducing the total variance,
\begin{equation}
{1\over{\sigma_x^2}} = \sum_a {1\over {\sigma_{g_a}^2}} \left[ {{d\langle g_a \rangle}\over{dx}}\right]^2 .
\label{indmeas}
\end{equation}
In fact we have
\begin{equation}
{1\over{\sigma_x^2}} = \sum_{\rm ij} {{d\langle g_{\rm i} (x)\rangle}\over{dx}} \left[\hat C^{-1} (x)\right]_{\rm ij} {{d\langle g_{\rm j} (x)\rangle}\over{dx}} ,
\label{sigmax-final}
\end{equation}
but this reduces to Eq (\ref{indmeas}) if we rotate into the eigenbasis of the covariance matrix $\hat C (x)$.   Notice that in this small noise approximation, the effective positional noise $\sigma_x$ depends on the true position but not on the particular concentrations $\{g_{\rm i}\}$ that we happen to observe.  Also, in contrast to the discussion of decoding in \S\ref{lec2-embryo}, this analysis is local, and does not address ambiguities; again, it makes sense in a small noise limit, where ambiguities are resolved.

\begin{figure}
\includegraphics[width=\linewidth]{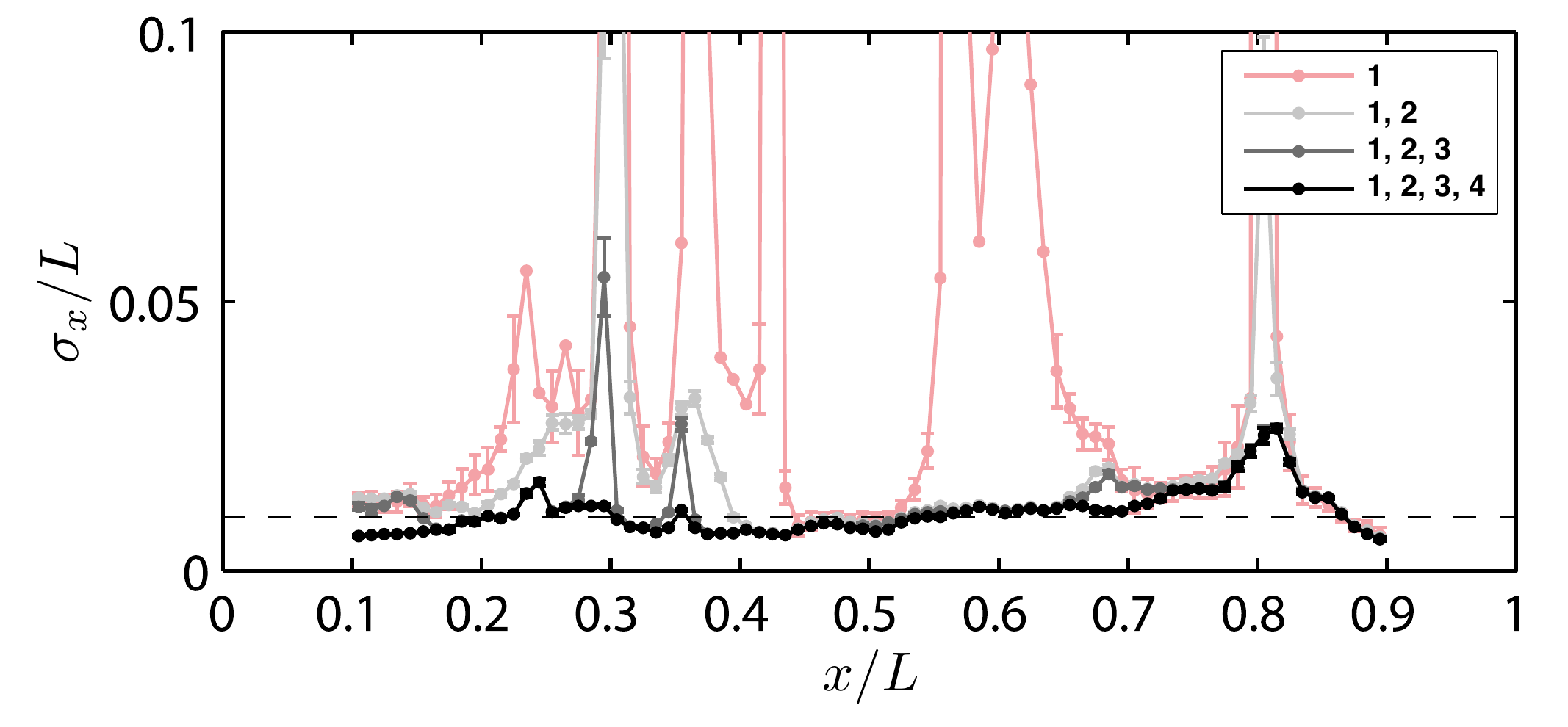}
\caption{Effective positional noise levels from gap gene expression levels.  Calculations of $\sigma_x$ are done in the small noise limit, from Eq (\ref{sigmax-final}), using the data illustrated in Fig \ref{gapgenes}.  As indicated in the inset, different curves correspond to including just $g_1$, the combinations $\{g_1,\, g_2\}$ and $\{g_1,\, g_2,\, g_3\}$ and finally all four gap genes $\{g_1,\, g_2,\, g_3,\, g_4\}$.  With all the genes we find that the positional error is small along the entire anterior--posterior axis, and close to $\sigma_x/L = 0.01$ (dashed line).  Reproduced from Ref \cite{dubuis+al_13b}, with thanks to JO Dubuis, G Tka\v{c}ik, EF Wieschaus, and T Gregor. \label{fig:sigma_x}}
\end{figure}

Everything that we need to compute the effective positional error is contained in the data of Fig \ref{gapgenes}, where we see the mean expression levels $\langle g_{\rm i}(x)\rangle$ and covariance matrices $\hat C(x)$ as functions of position. Although Eq (\ref{sigmax-final}) includes multiple genes, we can start with just one, corresponding to $g_1$ in Fig \ref{gapgenes}.  We see that with this one gene the positional noise drops to $\sigma_x/L \sim 0.01$ in a window near the middle of the embryo, but is much larger outside this window.  As we add more of the gap genes, the very largest values of $\sigma_x$ are reduced dramatically, and we end up with a nearly uniform $\sigma_x/L \sim 0.01$ across the full length of the embryo.  Among other things, this is small enough that our approximations above are self--consistent.  One should also see this as quantifying the progressive improvement of the decoding maps as we add more genes in Fig \ref{4genes}.\footnote{Although the analysis of positional errors came before the full decoding maps, by roughly five years \cite{petkova+al_19,dubuis+al_13b}.}

There are at least two distinct points of interest in Fig \ref{fig:sigma_x}.  First, the scale of the positional errors, $\sigma_x/L\sim 0.01$.  This is the same as the scale of positional errors in pair--rule stripe positions, and also in the placement of the ``cephalic furrow'' \cite{liu+al_13}, which is the first macroscopic structural change that happens after the embryo is separated into discrete cells.  The conventional perspective on this system has been that information flow from maternal inputs to gap genes to pair--rule genes entails a gradual refinement, with noisy inputs ultimately generating  a precise and reproducible pattern \cite{arias+hayward_06,lacalli_22}.  We see here that the precision visible at the pair--rule genes already is present in the gap genes, so information must be transformed and preserved rather than refined. 

Second, the positional error is nearly uniform along the anterior--posterior axis.  This is surprising because individual genes have complicated patterns of expression and noise, and provide precise information only in limited regions.  Indeed, as we saw in \S\ref{lec2-embryo}, information carried by single gap genes is not only inhomogeneous but also ambiguous.  Somehow the signal and noise (and even the covariances) of all four gap genes conspire to generate nearly constant precision.

To understand why the constancy of precision is so interesting, let's finish our project of estimating the information, in bits carried by $\{g_{\rm i}\}$.  We can write
\begin{equation}
I(\{g_{\rm i}\} ; x) = S[P_X(x)] - \langle S[P(x|\{g_{\rm i}\})] \rangle ,
\label{I60}
\end{equation}
where as before $S[P]$ is the entropy of the distribution $P$.  But we have seen that $P(x|\{g_{\rm i}\})$ is nearly Gaussian, which means immediately that
\begin{equation}
S[P(x|\{g_{\rm i}\})] = {1\over 2}\ln\left[ 2\pi e \sigma_x^2 \right] .
\end{equation}
Further, we have seen that $\sigma_x$ doesn't depend on the precise expression levels, but only on the position of the cell in which these expression levels are observed.  Thus the average in Eq (\ref{I60}) can be written as an average over positions, so that
\begin{eqnarray}
I(\{g_{\rm i}\} ; x) &=& -\int dx\,P_X(x)\ln P_X(x)  - {1\over 2} \int dx\,P_X(x) \ln\left[ 2\pi e \sigma_x^2 (x) \right] \\
&=& -\int dx\, P_X(x) \ln \left[ P_X(x)  \sqrt{2\pi e}\sigma_x (x) \right] 
\label{I63}
\end{eqnarray}
which brings us back exactly to Eq (\ref{Ixy4}).  Although there are many dimensions to the output of this system, the fact that there is only one dimension at the input means that, at least in the low noise limit, we can get back to a picture very much like the one input, one output system that we started with in \S\ref{one-to-one}.

In particular, if we imagine that the embryo could adjust the distribution of cell positions, then the optimal information transmission would occur when
\begin{equation}
P_X(x) \propto {1\over{\sigma_x (x)}} .
\end{equation}
Since the real $P_X(x)$ is essentially uniform, optimal information transmission predicts that $\sigma_x(x)$ should be uniform, and this is what we see.  Again, it requires considerable conspiracy among the four gap genes to achieve uniform positional error, so this seems like a significant success for the theory of optimal information transmission.  There are two problems.

The first problem is that $\sigma_x(x)$ is not exactly uniform, but theory actually gives us a way to measure the significance of these deviations.  With uniform $P_X(x) = 1/L$ the actual information transmitted is, substituting into Eq (\ref{I63}),
\begin{equation}
I = {1\over L}\int_0^L dx\, \ln \left[ {L\over{\sqrt{2\pi e} \sigma_x (x)}}\right] .
\label{Ifromsigma}
\end{equation}
On the other hand, if we are free to optimize $P_X(x)$ over the range $0 < x < L$ then from Eq (\ref{Imax}) the maximum information possible given the measured $\sigma_x(x)$ becomes
\begin{equation}
I_{\rm max}  =  \ln \left[ {1\over{\sqrt{2\pi e}}}\int_0^L {{dx}\over{\sigma_x (x)}} \right] .
\end{equation}
If we plug in the results from our analysis of the experimental data, we find\footnote{Estimating information theoretic quantities from real data can be challenging, and there are interesting theoretical questions about how best to do this, especially if you need control over the error bars (as in this case). See, for example, Appendix A.8 of Ref \cite{bialek_12}.}
\begin{equation}
{I\over{I_{\rm max}}} = 0.984\pm 0.003 .
\end{equation}
Thus optimization of information transmission predicts a match between the positional noise levels and the distribution of cell positions, and this match is sufficiently good that it brings the embryo within $2\%$ of the optimum \cite{dubuis+al_13b}.

The second problem is that it seems a bit weird to talk about redistributing cells along the embryo's axis.  We don't really need to do this.  The gap gene network output can be thought of not as providing information about position but rather about the concentrations of the maternal input molecules.  The mapping between position and input concentration is something that could be different with different parameters of the relevant dynamics, different anchoring of the mRNA molecules, etc..  Thus we {\em can} imagine a family of embryos with different possible distributions of inputs to the gap gene network.   But if the mapping between position and input concentration is deterministic and invertible, then information about concentration is the same as information about position, and all of the arguments here about matching go through.

To get a rough estimate of the total positional information conveyed by the gap genes we can use Eq (\ref{Ifromsigma}) but with $\ln \rightarrow \log_2$ so the units are bits, and the approximate $\sigma_x/L \sim 0.01$, which gives $I \sim \log_2(100/\sqrt{2\pi e}) \sim 4.6\,{\rm bits}$.  We can't see all of this because we are measuring only in the central $80\%$ of the anterior--posterior axis ($0.1< x/L < 0.9$); outside this region imaging becomes prone to systematic errors from the curvature of the embryo.  But if we correct for this then the integral gives a result very close to the rough answer, and also very close to a more brute force integration over the four dimensional space $\{g_{\rm i}\}$, which does not require any small noise or Gaussian approximations \cite{dubuis+al_13b,tkacik+al_15}.

The actual number of bits is not that much more than four, and there are four gap genes ... maybe it's just one bit per gene after all?  The way to test this is to imagine that the readout mechanisms can only resolve two levels of expression, high and low; formally this means transforming
\begin{equation}
g_{\rm i} \rightarrow \sigma_{\rm i} = H(g_{\rm i} -\theta_{\rm i}),
\end{equation}
where $H$ is the step function and $\theta$ is a threshold that divides the two levels.  If we know the thresholds then we can compute $I_{\mathbf \theta }(\{\sigma_{\rm i}\}; x)$, where the notation reminds us that the answer depends on the thresholds ${\mathbf \theta } = \{\theta_{\rm i}\}$.  To be generous, we can optimize the thresholds, maximizing how much information this on/off description of gene expression can capture. The answer is that $I (\{\sigma_{\rm i}\}; x) < 3\,{\rm bits}$ \cite{dubuis+al_13b}.  Further, this information is distributed very inhomogenously, so that some ``binary words'' $\{\sigma_{\rm i}\}$ point to $x$ values that span $\sim 10\%$ of the anterior--posterior axis.  Thus, to extract $\sim 4\,{\rm bits}$ of positional information from four genes requires mechanisms that have much more than this capacity to read out the expression levels.

Recently we returned to the issue of readout precision \cite{bauer+al_21,bauer+bialek_23}.  We know, as just explained, that reading each concentration with one bit of precision is not enough to extract all the available information.   On the other hand, our discussion of decoding positional information from the gap genes (\S\ref{sec:decoding}) assumed that cells had access to the true measured concentrations of each molecule, which surely is unrealistic.
How can the embryo best trade bits of precision in the readout against the bits of relevant positional information that are preserved? 

Formally we can imagine that the expression levels of the gap genes are mapped into some intermediate variable, such as the occupancy of binding sites along the DNA; let's call this intermediate variable $C$.  Inevitably this mapping is noisy, and this means that the information which $C$ can carry about 
$\{g_{\rm i}\}$, $I(C; \{g_{\rm i}\})$, is limited.  Given this limitation, what mapping $\{g_{\rm i}\}
\rightarrow C$ will maximize the information about position $I(C; x)$?  This defines a new optimization problem
\begin{equation}
\max_{P(C|\{g_{\rm i}\})} \left[ I(C; x) - T I(C; \{g_{\rm i}\})\right],
\end{equation}
where $T$ is a Lagrange multiplier and we make explicit that the mapping $\{g_{\rm i}\}
\rightarrow C$ is probabilistic; this is an example of the information bottleneck problem \cite{tishby+al_99}. To extract all the available information, that is to have $ I(C; x)\approx  I(\{g_{\rm i}\}; x)$, requires readout mechanisms with a capacity $I(C; \{g_{\rm i}\})$ of at least $8\,{\rm bits}$ \cite{bauer+al_21}.  Further, the most efficient mechanisms involve $C$ being sensitive to combinations of the different expression levels; if we have separate sensors $g_{\rm i} \rightarrow C_{\rm i}$ they need to have vastly more capacity.  Finally, the embryo is in a regime where the trading of $I(C; x) $ vs $I(C; \{g_{\rm i}\})$ has a universal form \cite{bauer+bialek_23}.
If we take seriously the idea that $C$ is something like the occupancy of binding sites, or the collective states of the ``enhancers'' into which these sites are grouped, then these arguments about trading bits provide a path to predict the molecular mechanisms that instantiate the optimal decoding strategies of \S\ref{lec2-embryo}.

Before leaving this topic I want to emphasize that Figs \ref{fig:laughlin}, \ref{fig:naama}, and \ref{fig:sigma_x} all are testing the {\em same}  optimization principle.  Thus we are making predictions about neurons that generate graded voltages, neurons that generate spikes, and networks of genes, all with the same theoretical idea and all connecting to experiment with no free parameters.

\section{Network architecture}
\label{sec:network}

In the flow of positional information from the maternal inputs to the gap genes to the pair--rule genes, we have seen evidence for optimization in the distribution of inputs (\S\ref{lec3-embryo}) and in the processing of the outputs (\S\ref{lec2-embryo}).  We now have to ask if we can apply optimization principles to the network itself.  Doing this involves returning to the challenge laid out in the first lecture:  realistic descriptions of the gap gene network require 50+ parameters, and are in some obvious sense very complicated.  To approach optimization of such a complex network it will be useful to break off smaller pieces of the problem and gain intuition.  

The idea that we can derive the functional behavior of a network from the optimization of information transmission is quite old, having its origins in the context of sensory information processing by the brain.  Just a decade after Shannon's original work on information theory the National Physical Laboratory in the UK hosted a remarkable {\em Symposium on the Mechanization of Thought Processes}.  Among other presentations, Horace Barlow spoke about ``Sensory mechanisms, the reduction of redundancy, and intelligence'' \cite{barlow_59}.  This was, I think, the first place where optimization principles for neural information processing were articulated.\footnote{Barlow revisited these ideas at another conference a few years later, and this is the more widely cited version of his ideas about ``efficient coding''  \cite{barlow_61}.  Thinking about our modern publication habits, it seems worth noting that this work had impact on generations of scientists even though there is no regular journal article to cite.}  It also seems worth emphasizing the ambition that one sees in these titles, both of Barlow's paper and for the symposium as a whole.  

\subsection{Linear filtering in neural networks}
\label{lec4:linearnets}

\begin{figure}[t]
\includegraphics[width=\linewidth]{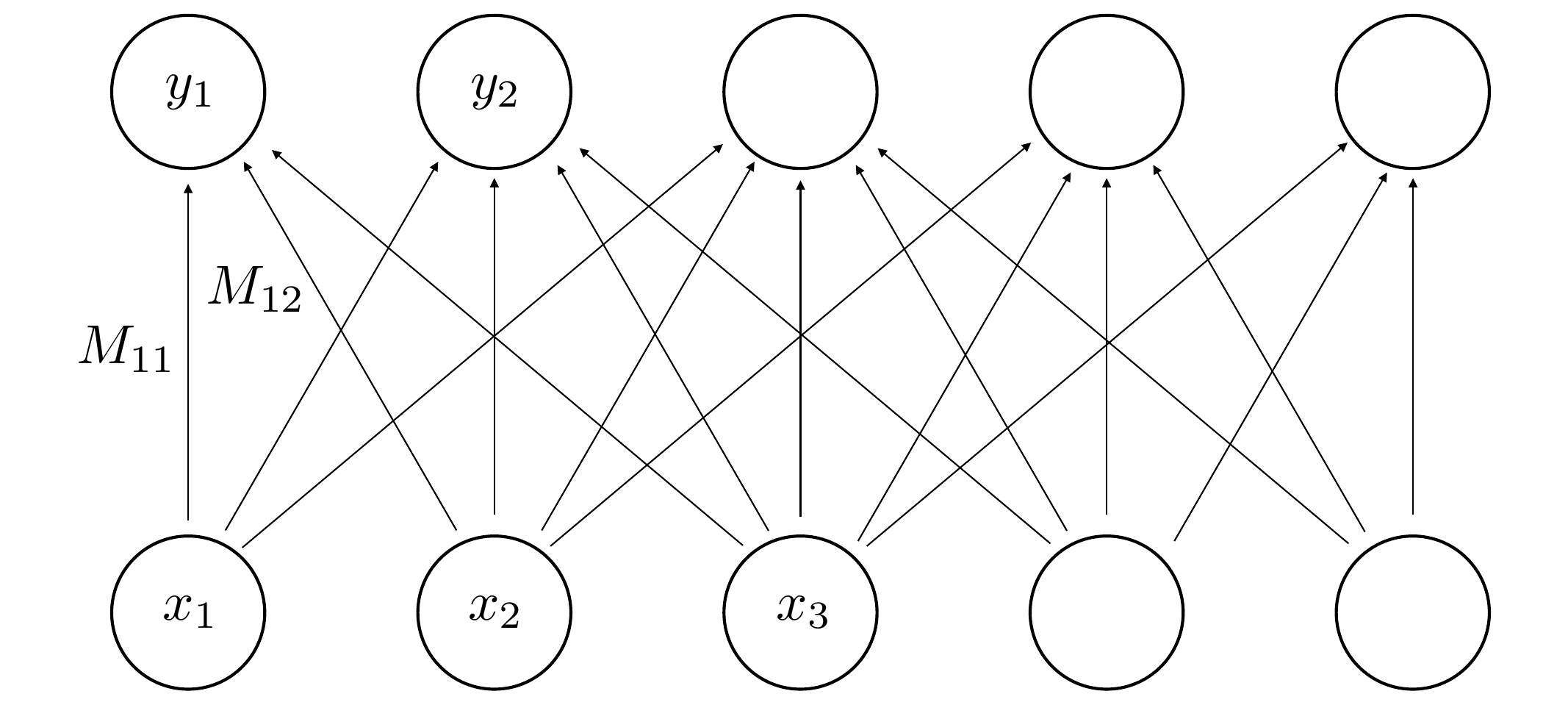}
\caption{Transformation from inputs $\{x_{\rm i}\}$ to outputs $\{y_{\rm i}\}$, as in Eq (\ref{linmodel1}).  Only a fraction of possible connections $M_{\rm ij}$ are illustrated; fewer still are labelled.  Not shown are the independent noise sources added to each output element. \label{fig:Mij}}
\end{figure}

Let's dive in and think about a network in which a layer of input variables $\{x_{\rm i}\}$ drives a layer of output variables $\{y_{\rm i}\}$, as in  Fig \ref{fig:Mij}.  The simplest possibility is that the transformation is a noisy linear mapping, so that
\begin{equation}
y_{\rm i} = \sum_{\rm j} M_{\rm ij} x_{\rm j} + \zeta_{\rm i} ;
\label{linmodel1}
\end{equation}
again to keep things simple we imagine the number of inputs and outputs are the same so ${\rm i} = 1,\, 2,\, \cdots ,\, N$ and similarly for $\rm j$.   Let the noise $\zeta_{\rm i}$ be Gaussian and independent at every site $\rm i$.  In this case optimizing the network means choosing the transformation matrix $\hat M$ to maximize the mutual information between  inputs and outputs.  If the variance of the noise is fixed, then one can always increase the mutual information by increasing the magnitude of $\hat M$, which seems like cheating.  To have a well defined problem we should bound the magnitudes, so that the scale of the noise has meaning in relation to the inputs; a conventional choice is to constrain the summed variances of the outputs.  This leaves us with  the optimization problem
\begin{equation}
\max_{\hat M} \left[ I\left( \{y_{\rm i}\} ; \{x_{\rm i}\}\right) -\mu \sum_{{\rm i}=1}^N \langle y_{\rm i}^2\rangle \right] .
\label{optlin1}
\end{equation}
The structure of this problem depends on the distribution $P(\{x_{\rm i}\})$ from which inputs are drawn.

The general version of Eq (\ref{optlin1}) is challenging.  It is much easier if we can approximate $P(\{x_{\rm i}\})$ as Gaussian.  Then we have
\begin{eqnarray}
P(\{x_{\rm i}\}) &=& {1\over\sqrt{(2\pi )^N}} \exp\left[ - {1\over 2} \ln\det \hat C_x - {1\over 2} \sum_{{\rm i},{\rm j}} x_{\rm i} \left(\hat C_x^{-1}\right)_{\rm ij} x_{\rm j}\right] \\
P(\{y_{\rm i}\}| \{x_{\rm i}\}) &=& {1\over\sqrt{(2\pi )^N}} \exp\left[ - {N\over 2} \ln \langle\zeta^2\rangle  - {1\over {2\langle\zeta^2\rangle}} \sum_{{\rm i}}  \left(y_{\rm i} - \sum_{\rm j}M_{\rm ij} x_{\rm j} \right)^2   \right] \\
P(\{y_{\rm i}\} ) &=& {1\over\sqrt{(2\pi )^N}} \exp\left[ - {1\over 2} \ln\det \hat C_y - {1\over 2} \sum_{{\rm i},{\rm j}} y_{\rm i} \left(\hat C_y^{-1}\right)_{\rm ij} y_{\rm j}\right] ;
\end{eqnarray}
 the covariance matrices are
 \begin{eqnarray}
 \left(\hat C_x \right)_{\rm ij} &=& \langle x_{\rm i}x_{\rm j}\rangle \\
 \left(\hat C_y \right)_{\rm ij} &=& \langle y_{\rm i}y_{\rm j}\rangle = M_{\rm ik}  \left(\hat C_x \right)_{\rm kl} M_{\rm jl} + \langle\zeta^2\rangle \delta_{\rm ij}\\
\Rightarrow \hat C_y  &=& \hat M \hat C \hat M^T + \langle\zeta^2\rangle \mathbb{1} .
\end{eqnarray}
We can substitute into the mutual information
\begin{eqnarray}
I\left( \{y_{\rm i}\} ; \{x_{\rm i}\}\right) &\equiv& \int d^Nx \int d^Ny\, P(\{y_{\rm i}\}| \{x_{\rm i}\})P(\{x_{\rm i}\})\log\left[ {{P(\{y_{\rm i}\}| \{x_{\rm i}\})}\over{P(\{y_{\rm i}\} ) }}\right]\\
&=& {\bigg\langle} \log\left[ {{P(\{y_{\rm i}\}| \{x_{\rm i}\})}\over{P(\{y_{\rm i}\} ) }}\right]{\bigg\rangle}\\
&=& - {N\over 2} \ln \langle\zeta^2\rangle  - {1\over {2\langle\zeta^2\rangle}} \sum_{{\rm i}} {\bigg\langle} \left(y_{\rm i} - \sum_{\rm j}M_{\rm ij} x_{\rm j} \right)^2 {\bigg\rangle} \nonumber\\
&&\,\,\,\,\,\,\,\,\,\,\,\,\,\,\,\,\,\,+ {1\over 2} \ln\det \hat C_y + {1\over 2} \sum_{{\rm i},{\rm j}} \langle y_{\rm i}  y_{\rm j}\rangle \left(\hat C_y^{-1}\right)_{\rm ij}  .
\end{eqnarray}
Notice that, from Eq (\ref{linmodel1}), we have
\begin{equation}
y_{\rm i} - \sum_{\rm j}M_{\rm ij} x_{\rm j} = \zeta_{\rm i} \Rightarrow {1\over{2\langle\zeta^2\rangle}}\sum_{\rm i} {\bigg\langle} \left(y_{\rm i} - \sum_{\rm j}M_{\rm ij} x_{\rm j} \right)^2 {\bigg\rangle}  = {1\over{2\langle\zeta^2\rangle}}\sum_{\rm i}\langle \zeta^2\rangle  = {N\over 2} .
\end{equation}
Similarly 
\begin{equation}
\langle y_{\rm i}y_{\rm j} \rangle = \left(\hat C_y\right)_{\rm ij} 
\Rightarrow 
{1\over 2} \sum_{{\rm i},{\rm j}} \langle y_{\rm i}  y_{\rm j}\rangle \left(\hat C_y^{-1}\right)_{\rm ij} = {1\over 2}\sum_{{\rm i},{\rm j}}\left(\hat C_y\right)_{\rm ij}\left(\hat C_y^{-1}\right)_{\rm ij}  = {N\over 2} .
\end{equation}
Thus the complicated looking summations cancel and we have
\begin{equation}
I\left( \{y_{\rm i}\} ; \{x_{\rm i}\}\right)  = - {N\over 2} \ln \langle\zeta^2\rangle +  {1\over 2} \ln\det \hat C_y =  {1\over 2}{\rm Tr}\ln\left[ \mathbb{1} + {1\over {\langle\zeta^2\rangle}} \hat M \hat C \hat M^T\right]
\end{equation}
If the eigenvalues of $\hat M \hat C \hat M^T$ are $\lambda_{\rm n}$, then the optimization problem in Eq (\ref{optlin1}) becomes
\begin{equation}
\max\left[ {1\over 2}\sum_{\rm n}\ln\left( 1 + \lambda_{\rm n}/\langle\zeta^2\rangle\right) - \mu \sum_{\rm n}\lambda_{\rm n} \right] .
\end{equation}
The optimum is reached where all the $\lambda_{\rm n}$ are equal.

The correlation structure of the inputs usually is non--trivial, so the eigenvalues of the covariance matrix $\hat C_x$ are spread over some spectrum.  Optimizing information transmission in the class of problems defined by Eq (\ref{linmodel1}) means rearranging these inputs to ``whiten'' this spectrum.\footnote{The name comes from the fact that in truly white light all components of the spectrum have equal weight.} Notice that if all eigenvalues of $\hat M \hat C_x \hat M^T$ are equal then $\hat C_y$ is proportional to the unit matrix and hence different signals $y_{\rm i}$ and $y_{{\rm j}\neq{\rm i}}$ are independent of one another.  Thus the optimal $\hat M$ diagonalizes the covariance matrix of the input signals, transforming to principal components, and then rescales these so that they have equal variances.

A simple but important example of these ideas is color vision.  Recall that in daylight our vision is based on three kinds of cones, each tuned to a different range of photon energies.  We can think of these as sensitive approximately to red, green, and blue, though you should never say this in front of someone who actually studies color vision---``red'' is a percept, not the label of a cell type; the convention is to describe the different cones as sensitive to long, medium, and short  wavelengths.  It is an old idea that the three cone signals are processed in well defined combinations corresponding roughly to the percepts of luminance, red vs green, and blue vs yellow.  The history of these ideas is complicated, with all sorts of heroic figures shouting at one another.  Things were confusing in part because the ``opponent process'' theory seemed to involve four color axes (red, green, blue, and yellow) while the alternative ``trichromatic'' theory involved only three (which we now attach to the three types of cones).  It seems to have been Schr\"odinger who sorted this out in 1925 \cite{schrodinger_25}.  We now know that the three cone signals indeed are grouped together in the three combinations already in the retina, and it is these combined signals that are transmitted to the brain \cite{gegenfurter+sharpe_01}.

In 1983, Buchsbaum and Gottschalk suggested that the transformation into luminance, red vs green, and blue vs yellow might be the solution to the optimization problem we are discussing here \cite{buchsbaum+gottschalk_83}.  In order to see if this makes sense, we need to know the covariance matrix of the three cone signals.  This depends strongly on the absorption spectra of the cone pigments, with large positive correlations arising simply because these spectra overlap.   But the cone signal statistics also depend on the spectral composition of the images that the eye is looking at.  To get at these, Ruderman, Cronin, and Chiao used a hyperspectral camera to take pictures in natural environments---woodlands, forests and rainforests, and a mangrove swamp  \cite{ruderman+al_98}.  In effect this camera does spectroscopy in each pixel of a digital image, and then these spectrally resolved images can be projected onto the known sensitivities of the individual cone pigments to estimate the photon capture in each of the three cones looking at the same point.  In keeping with the discussion of adaptation above (\S\ref{lec3-neurons}) the cone signals $x_{\rm i}$ were taken as the log of the number of photon counts.\footnote{We use our cones in bright light, so there is no worry about zero counts.} 

The results obtained by Ruderman et al are remarkably simple and crisp.  In our notation, the optimal matrix $\hat M$ takes the form
\begin{equation}
\hat M = 
\left[
\begin{array}{ccc}
{1\over{\sqrt{3}\sigma_\ell}}  &  0 & 0  \\
 0 &  {1\over{\sqrt{6}\sigma_{by}}} &  0  \\
0  &  0 &   {1\over{\sqrt{2}\sigma_{rg}}} 
\end{array}
\right]
\left[
\begin{array}{rrr}
1.004  & 1.005   &  0.991   \\
 1.014   & 0.968   &  -2.009   \\
0.993    & -1.007    &   0.016  
\end{array}
\right] .
\end{equation}
We see that $y_1$ is the sum of the signals in the three cones, and the weights are equal in the second decimal place or better.  Similarly $y_3$ is the difference between the signals in long and medium length cones, with almost no contribution from the short wavelength cones; this is the ``red minus green'' opponent channel.  Finally $y_2$ combines the long and medium cone signals and subtracts the short wavelength signal, corresponding to ``yellow minus blue.''  As with the luminance channel $y_1$, the combinations of cone signals in the two opponent channels are in integer ratios, with one percent accuracy.  The signals along the different channels have standard deviations in the ratio $\sigma_\ell : \sigma_{by} : \sigma_{rg} \sim 1:0.2:0.02$.  While we have a great appreciation for the subtleties of color, most of the variance in the images that we see is in the luminance channel.\footnote{In fact this might be a cautionary tale about dimensionality reduction. We can capture $\sim 95\%$ of the variance in what we see with just one dimension, corresponding to a completely greyscale world.} Embarrassingly, I don't know if the combination of cone signals into neural signals has been measured precisely enough to test these predictions of integer ratios.

Staying with the visual system, we can think of the $\{x_{\rm i}\}$ as signals from cones at different positions (now neglecting color).  If the cones form a regular lattice, and the input visual signals are translationally invariant, then everything will be diagonal after a Fourier transform.  Let's call $\mathbf k$ the spatial Fourier variable, and then
\begin{equation}
\tilde y({\mathbf k}) = \tilde M ({\mathbf k}) \tilde x({\mathbf k}) + \tilde \zeta ({\mathbf k}) .
\end{equation}
Further, the transformation $\tilde M ({\mathbf k})$ that maximizes information transmission will obey
\begin{equation}
| \tilde M ({\mathbf k})| \propto {1\over{\sqrt{S_x({\mathbf k})}}} ,
\label{white2}
\end{equation}
where $S_x({\mathbf k})$ is the power spectrum of the signals $\{x_{\rm i}\}$ \cite{atick+redlich_90,hateren_92}.  We know that the spatial power spectrum of natural images is scale invariant  \cite{ruderman+bialek_94},
\begin{equation}
S({\mathbf k}) = {A\over{|{\mathbf k}|^{2-\eta}}} ,
\end{equation}
so the prediction is $| \tilde M ({\mathbf k})| \propto |{\mathbf k}|^{1-\eta/2}$.  The growth with $|{\mathbf k}|$ is cut off by the effects of noise in the inputs $x_{\rm i}$, e.g. the random arrival of photons.

Given the statistical structure of natural images, optimization of information flow predicts that there will be zero gain for zero spatial frequency, $| \tilde M ({\mathbf k} = 0)| =0$.  The transformation $M_{\rm ij}$ thus serves roughly as a spatial differentiator, so that the output $y_{\rm i}$ is large when the pattern $\{x_{\rm j}\}$ includes a sharp edge, and is small when the $\{x_{\rm j}\}$ are nearly uniform.  Qualitatively these predictions are correct for neurons in the retina.  The output cells, which carry information from eye to brain, have long been known to have ``center--surround'' receptive fields \cite{barlow_53,kuffler_53}: the response of these neurons is driven by the difference between the average light intensity in a small central region and the average over a larger surrounding region, and in many cases the two regions have equal weight so that the response is differentiating.  Barlow understood that this sort of spatial differencing,  also called ``lateral inhibition''  \cite{hartline_69}, would remove the redundancy between signals in neighboring photoreceptors, enhancing the transmission of information by limited numbers of action potentials along limited numbers of neurons \cite{barlow_59,barlow_61}; the analysis here translates this qualitative observation into equations.

These arguments about whitening apply also in the time domain. In the simplest example 
\begin{equation}
y(t) = \int d\tau M(\tau ) x(t-\tau) + \eta (t) .
\end{equation}
If $\eta$ is white noise, $\langle \eta (t)\eta(t') \rangle = {\cal N}_0 \delta (t-t')$, then the rate at which information about $x(t)$ is conveyed by $y(t)$ is \cite{bialek_12,cover+thomas_91,shannon_49}
\begin{equation}
R_{\rm info} = {1\over 2}\int {{d\omega}\over{2\pi}} \ln\left[ 1 + {1\over {{\cal N}_0}} |\tilde M (\omega )|^2 S_x(\omega )\right] ,
\end{equation}
where $S_x(\omega)$ is the power spectrum of the input signal
\begin{equation}
\langle x(t)x(t') \rangle = \int {{d\omega}\over{2\pi}} e^{-i\omega(t-t')} S_x(\omega ).
\end{equation}
If we optimize $R_{\rm info}$ while holding fixed the output dynamic range $\langle y^2 \rangle$, then  we find an analog of  Eq (\ref{white2}),
\begin{equation}
|\tilde M (\omega )|\propto {1\over\sqrt {S_x(\omega )}} .
\end{equation}
If  input signals have scale invariant dynamics, $S_x(\omega ) \sim 1/|\omega|$, then we predict   $|\tilde M (\omega )| \sim\sqrt{ |\omega |}$.  This is weird, since it corresponds to the filter $M$ taking half of a derivative.

\begin{figure}
\includegraphics[width=\linewidth]{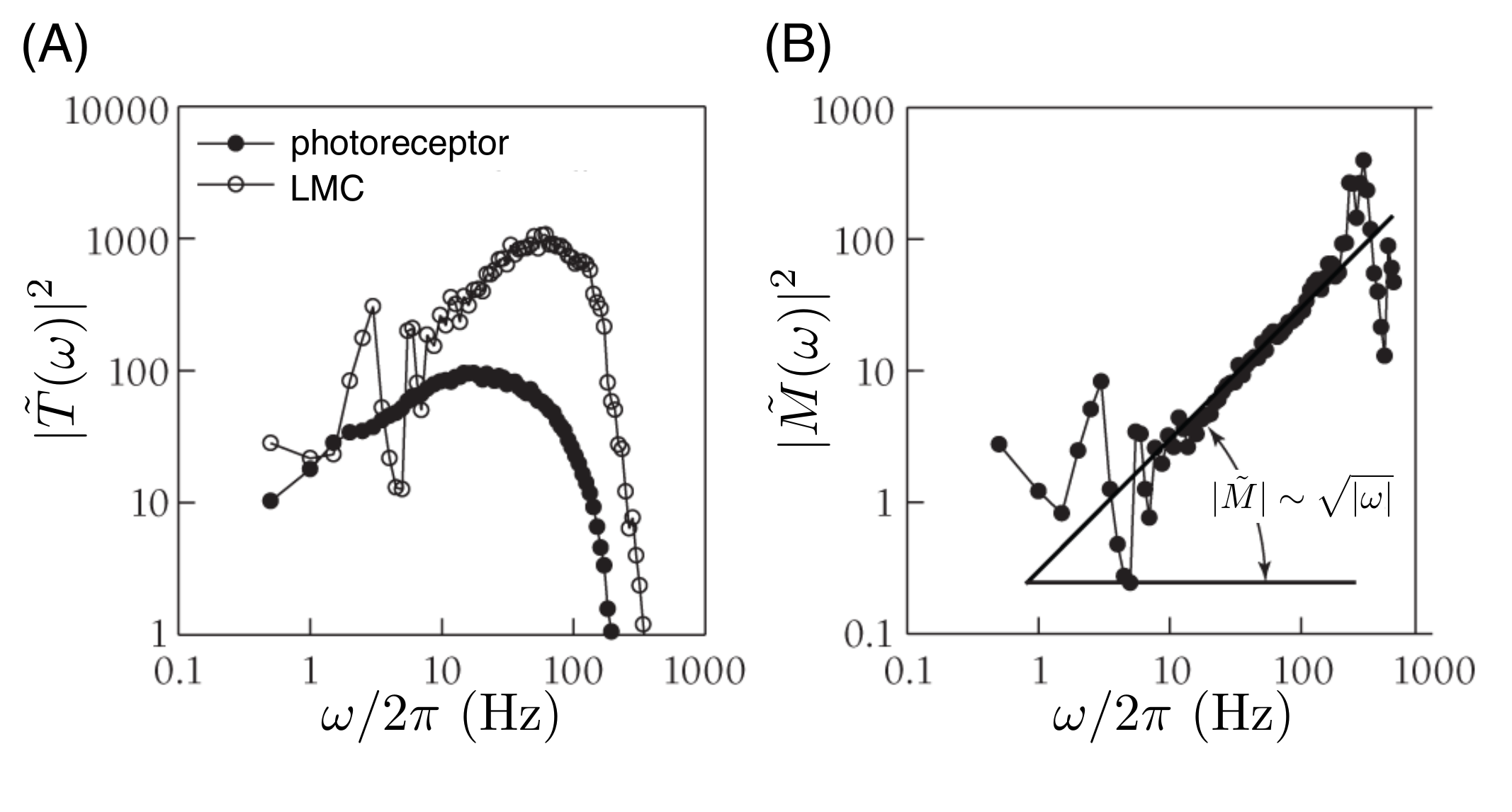}
\caption{Filtering at the first synapse in fly vision. (A) Linear response of voltage to temporal variations in light intensity, as in Eq (\ref{Tdef}), for photoreceptors and large monopolar cells (LMC). (B) Effective transfer function between photoreceptors and LMCs, from Eq (\ref{MfromT}).  My thanks to SB Laughlin and RR de Ruyter van Steveninck for sharing these data, from experiments described in Ref  \cite{laughlin+ruyter_96}. \label{fig:LMCfilter}}
\end{figure}

Direct measurements of the transformation $M$ between neurons are not so easy.  Our linear models make more sense if the cells have graded voltage responses, as in the first stages of vision.  In the fly retina, for example, both the photoreceptor cells and the next cells in line (large monopolar cells, LMCs from \S\ref{lec3-neurons}) have a large regime in which the average voltage $V(t)$ responds linearly to time variations in light intensity $I(t)$ around a background $I_0$,
\begin{equation}
\langle V(t)\rangle  - V_0 = {1\over{I_0}} \int d\tau \,T(\tau ) [I(t-\tau) - I_0]
\label{Tdef}
\end{equation}
defining a transfer function $T(\tau)$; it is more convenient to think about the Fourier transform $\tilde T (\omega)$.  These transfer functions can be measured in both the receptors and LMCs, with results in Fig \ref{fig:LMCfilter}A.  Then we can take the ratio to estimate the filter across the synapse,
\begin{equation}
\tilde M (\omega ) = {{\tilde T_{\rm LMC}(\omega )}\over{T_{\rm receptor} (\omega )}} ,
\label{MfromT}
\end{equation}
with results in Fig \ref{fig:LMCfilter}B.  Strikingly, we really do see $|\tilde M (\omega )| \sim\sqrt{ |\omega |}$ across a wide range of frequencies. Such fractional differentiation is much more widespread, perhaps serving to optimize the transmission of information about scale invariant signals throughout the brain \cite{lundtsrom+al_08,lundtsrom+al_10}.  These ideas have distant but fascinating precursors \cite{thorson_74}.

\subsection{Ingredients of a genetic network}

The simplest example of a genetic network was shown schematically in Fig \ref{regulation}.  To be more consistent with the notation below, let's describe this by saying that a single transcription factor (TF) at concentration $c$ controls the expression level $g$ of a target gene.  We'll assume for simplicity that the system is in steady state.

The information that $g$ provides about the input concentration $c$ is 
\begin{equation}
I(g;c) = \int dg\int dc P(g|c) P_{\rm in}(c) \log\left[ {{P(g|c)}\over{P_{\rm out}(g)}}\right],
\end{equation}
where the distribution of outputs 
\begin{equation}
P_{\rm out}(g) = \int dc\, P(g|c)P_{\rm in}(c).
\end{equation}
Consistent with the analysis of decoding in \S\ref{lec2-embryo}, we'll assume that the noise in the response of $g$ to the input $c$ is Gaussian, so that
\begin{equation}
P(g|c) = {1\over\sqrt{2\pi\sigma_g^2(c)}}\exp\left[ - {{(g - \bar g (c))^2}\over{2\sigma_g^2(c)}}\right] .
\end{equation}

We expect that the mean expression level $\bar g (c)$ has a sigmoidal dependence on the TF concentration, as in Fig \ref{regulation}, which we can write as
\begin{equation}
\bar g(c) = {{c^h}\over{K^h + c^h}};
\end{equation}
$h>0$ means that the TF activates expression and $K$ is the concentration at which this effect is half--maximal.  More quantitatively $h$ is a measure of sensitivity.  We can rationalize this form by imagining the $h$ molecules of the TF bind cooperatively to sites along the DNA, and this binding influences transcription.  To complete our description, and the calculation of the information transmission,  we need to know the noise level $\sigma_g(c)$.

The molecules whose concentration is measured by $g$ ultimately are made one at a time, and these are random events.  So we expect $\sqrt{N}$ fluctuations in making $N$ molecules.  We have to be a bit careful, first because we have chosen units in which the maximum $\bar g$ is unity, and second because many proteins can be translated from a single mRNA molecule before it is degraded.  Thus we can write the contribution of this counting noise as
\begin{equation}
\sigma_{g,{\rm count}}^2 (c) = {1\over {N_{\rm max}}} \bar g (c),
\label{sigmag-count}
\end{equation}
where $N_{\rm max}$ is the maximum number of {\em independent} molecules being made.  Probably this is the number of mRNAs, but we can be flexible.

In order to regulate gene expression, the TF molecules have to arrive at a small target along the DNA, where they bind to specific sequences.  We can thus think of the regulatory mechanism as a small sensor of transcription factor concentration, with gene expression as the readout.  The physical limits to sensing concentrations were first discussed by Berg and Purcell in the context of bacterial chemotaxis  \cite{berg+purcell_77}.  This remains one of the foundational papers of our field, so it is worth taking a detour to review their arguments, and some of the subsequent developments, leading up to Eq (\ref{sigmac_final}).

\begin{figure}[t]
\centerline{\includegraphics[width = \linewidth]{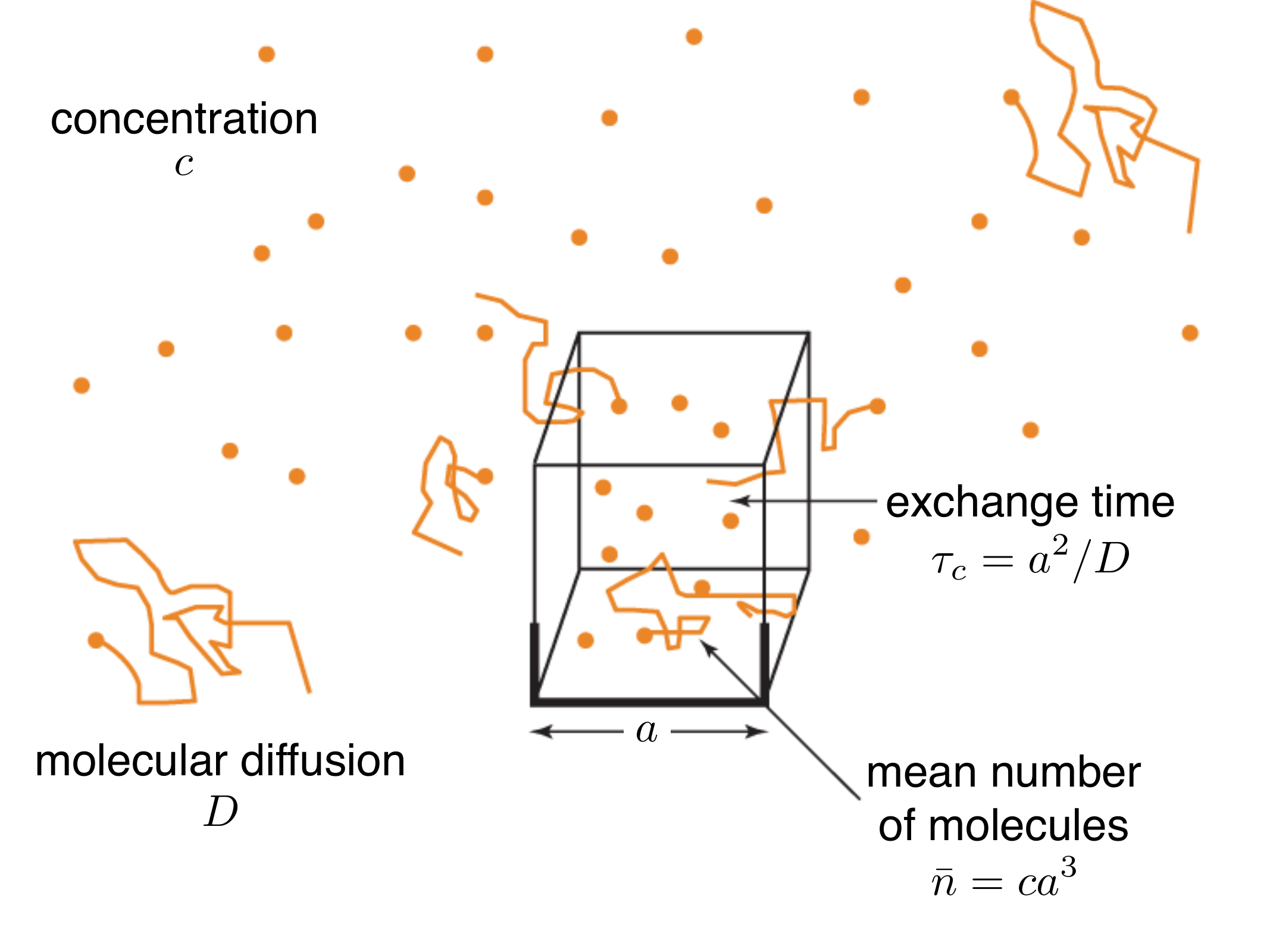}}
\caption{Understanding the physical limits to concentration measurements \cite{berg+purcell_77,bialek_12}. A receptor of linear size $a$ samples a volume $\sim a^3$ and thus counts a mean number of molecules $\bar n \sim ca^3$, where $c$ is the concentration. Molecules move randomly  in and out of the sensitive volume with a diffusion constant $D$, setting the  correlation time $\tau_c\sim a^2 /D$.  Figure adapted from Ref \cite{bialek_12}, with permission. \label{BPfig}}
\end{figure}

A sensor of linear dimension $a$ sitting in a solution of molecules with concentration $c$ will count $\bar n \sim ca^3$ molecules on average, as shown in Fig \ref{BPfig}.  But as molecules diffuse in and out of the sensitive volume, this molecule count will fluctuate by $\delta n \sim \sqrt{\bar n}$.  Since concentration is proportional to molecule number, this means that estimates of concentration will have a fractional fluctuation
\begin{equation}
{{\delta c}\over c}{\bigg |}_1 \sim {{\delta n}\over{\bar n}} \sim {1\over{\sqrt{ca^3}}} ,
\end{equation}
where the subscript reminds us that this is based on one snapshot of molecule counts.  We should be able to reduce the noise by averaging over time, but this works only if we make multiple {\em independent} measurements; just counting the same molecules again doesn't give a better estimate of the surrounding concentration.  It takes a time $\tau_c\sim a^2 /D$ for molecules in the sensitive volume to exchange with the bulk solution via diffusion, so if we average over time $\tau_{\rm avg}$ we can make $\tau_{\rm avg}/\tau_c$ independent measurements, and thus reduce the noise
\begin{equation}
{{\delta c}\over c} \rightarrow {{\delta c}\over c}{\bigg |}_1 \cdot {1\over \sqrt{\tau_{\rm avg}/\tau_c}} \sim {1\over\sqrt{D a c\tau_{\rm avg}}} .
\label{BP}
\end{equation}
We see that the different parameters of the problem combine in a very simple way.  This is the ``Berg--Purcell limit'' to concentration sensing.

It's always good to check that the units work out:
\begin{equation}
D a c\tau_{\rm avg} = \left[\ell^2 / t\right] \left[\ell\right] \left[ 1/\ell^3\right] \left[ t\right] = \left[\right],
\end{equation}
so this is a dimensionless combination.  Note that the $3$ in $[1/\ell^3]$ is from three dimensions, so the answer must be different if we are trying to sense the concentration of molecules diffusing in a membrane, or if there is a significant contribution from proteins sliding along the length of the (one--dimensional) DNA molecule \cite{tkacik+bialek_09}.

It is useful to note that $\sim D a c$ is the mean rate at which molecules arrive at a (three--dimensional) target of size $a$ via diffusion.  For example, if we have the chemical reaction  $A + B \rightarrow AB$, we write the dynamics of the concentration as
\begin{equation}
{{d[AB]}\over {dt}} = k_2 [A][B] ,
\end{equation}
where $k_2$ is a ``second order rate constant.''  This rate constant is bounded by the rate at which $A$ and $B$ can find each other, and this is $k_2 \sim Da$, where $a$ is the size of the molecules.  One can make this precise by solving the diffusion equation with appropriate boundary conditions,\footnote{See Problem 54 in Ref \cite{bialek_12}.} and develop more precise models in which the molecules have to approach one another at the correct orientation.  Thus one can think of $Dac$ as the rate at which molecules arrive at the sensor, and $Dac\tau_{\rm avg}$ as the mean number of molecules that our sensor counts. The Berg--Purcell limit then is the statement that there are $\sqrt{n}$ fluctuations in counting molecules, just as with counting photons from a conventional light source.

In this discussion the linear dimensions of the sensor $a$ is a rough concept.  In their original discussion,  Berg and Purcell were thinking about a bacterium sensing the concentration of attractive or repulsive molecules in its environment, and the initial guess is that $a$ is just the size of the bacterium itself.  But this is weird, since what really happens is that molecules bind to particular receptors on the cell surface.  These receptors are themselves of molecular dimensions---nanometers rather than the micron size of the whole cell---and receptors for any particular molecules cover only a tiny fraction of the cell's surface.  

If there is just one receptor of linear dimension $a_{\rm r}$ then the Berg--Purcell argument should still work.  If there are $N_{\rm r}$ of these receptors then it is plausible that noise is reduced by a factor $1/\sqrt{N_{\rm r}}$, which is equivalent to saying they act together as a single receptor with effective size $a_{\rm eff} \sim N_{\rm r}a_{\rm r}$.  But if the number of receptors becomes large enough that the area occupied by the receptors becomes comparable to the area of the cell surface, $N_{\rm r}a_{\rm r}^2 \sim a_{\rm cell}^2$, then it seems like the whole cell should act as one big receptor and $a_{\rm eff} \rightarrow a_{\rm cell}$.  How does this work?

In a fabulous bit of hand waving plus analogies, Berg and Purcell argued that the tendency of random walk trajectories to bounce along a surface leads to correlations in the encounters of individual molecules with nearby receptors.  With $N_{\rm r}$ receptors scattered over the surface of the cell, this leads to an effective size
\begin{equation}
a_{\rm eff} \sim a_{\rm cell} {{N_r a_{\rm r}}\over{N_r a_{\rm r} + a_{\rm cell}}}.
\label{aeff}
\end{equation}
Notice that this saturates when $N_r a_{\rm r} \sim  a_{\rm cell}$, at which the fractional coverage of the surface is $N_r a_{\rm r}^2   /a_{\rm cell}^2 \sim a_{\rm r}   /a_{\rm cell} \sim 10^{-3}$.  This is a dramatic effect, and matters enormously in the life of the cell, which can act as one big sensor even though only a small fraction of its surface is covered by any single receptor type.

The goal of Berg and Purcell's work was (in part) to compare the physical limits to concentration sensing with the performance of real bacteria as they navigate through chemical gradients.  The result was that it would be physically impossible for cells to make decisions with the observed reliability if they were making spatial measurements, comparing concentrations at head vs.~tail to see if they are moving in the right direction.  Instead they must measure changing concentrations in time along the path taken as they swim.  They need to average these time derivatives, effectively comparing the recent past with a longer term average, but the duration of this averaging is limited by the rotational Brownian motion of the cell itself.  The result is a semi-quantitative theory of what bacteria must do in order to achieve their observed chemotactic performance, and all of these predictions proved to be correct.  

One question is whether the Berg--Purcell arguments could be sharpened to give a more quantitative theory of chemotactic strategies, and this continues to be an interesting direction.  More relevant to our discussion is whether these limits to concentration sensing are sufficiently general that they can be applied, for example, to the problem of transcriptional control.  This matters because of a simple order--of--magnitude argument.  Transcription factors function at concentrations of tens of nanoMolar,
\begin{equation}
c\sim 10\,{\rm nM} = 10\times 10^{-9}\times (6\times 10^{23}) {1\over {10^3\, ({\rm cm})^3}} \times \left( {{10^{-4}\,{\rm cm}}\over{\mu{\rm m} }}\right)^3 \sim 6\, (\mu{\rm m})^{-3} .
\end{equation}
Diffusion constants for proteins in the cytoplasm are $D\sim 1\,\mu{\rm m}^2/{\rm s}$, and the target to which these molecules bind has dimensions $a\sim 3\,{\rm nm}$.  The result is that
\begin{equation}
Dac \sim 2\times 10^{-2}\, s^{-1} \Rightarrow 
{{\delta c}\over c} \sim {1\over{\sqrt{Dac\tau_{\rm avg}}}}
 \sim \left( {{1\,\rm {min}}\over{\tau_{\rm avg}}}\right)^{1/2} .
 \label{TF-BP}
\end{equation}
This suggests that reliable responses to ten percent differences in transcription factor concentration will require more than an hour of temporal integration.  Neighboring cells in the fly embryo experience maternal inputs that differ by $\sim 10\%$ in concentration, and they reliably express different combinations of gap genes; this certainly takes less than one hour \cite{gregor+al_07b}.

The conclusion from Eq (\ref{TF-BP}) is not that transcriptional regulation definitely reaches the physical limits to concentration sensing, but rather that it operates in a regime where these limits are relevant.  Even this more modest conclusion  hinges on applying the Berg--Purcell ideas far from their origins; this is made more uncertain by the beautifully intuitive but non--rigorous nature of the original arguments.\footnote{I never had the chance to discuss this with Purcell, but Berg was clear that their arguments were rough and that it would be nice to have something more rigorous.} Some years ago my colleagues and I started to worry about this \cite{bialek+setayeshgar_05,bialek+setayeshgar_08,tkacik+bialek_09}, and by now the literature has grown substantially \cite{vanzon+al_06,endres+wingreen_09,mora+wingreen_10,kaizu+al_14,mora_15,carballo-pacheco+al_19,mora+nemenman_19,wave+al_20}.  Here is what I think we know:
\begin{itemize}
\item One way to make the Berg--Purcell argument rigorous is to analyze the fluctuations in occupancy of a receptor binding site as it comes to equilibrium with molecules diffusing in the surrounding solution.  These fluctuations are a form of thermal noise, subject to the fluctuation--dissipation theorem.
\item Using the fluctuation--dissipation theorem we can see explicitly how diffusion among nearby receptors generates correlated noise and results with the flavor of Eq (\ref{aeff}).
\item Cooperativity among multiple binding events enhances sensitivity and reduces excess noise, but never below the bound set by Berg and Purcell.
\item Analytic results from the fluctuation dissipation theorem can be reproduced by careful numerical simulation, though small discrepancies remain to be understood.
\item Cells can push below the Berg--Purcell limit by factors $\sim 2$ with signal processing strategies that are more sophisticated than just averaging receptor occupancy.
\item These strategies only work away from equilibrium, providing a path to connect energy dissipation and signaling accuracy.
\item There are interesting generalizations to sensing time dependent concentrations, or the concentrations of multiple species by multiple receptors; results can be counterintuitive.
\end{itemize}
From all this work, the general conclusion is that Berg and Purcell got it right: Eq (\ref{BP}) defines a minimal noise level for sensing concentrations, up to factors of order unity that could also be seen as ambiguity in defining the size $a$ of the detector.

What the Berg--Purcell limit tells us is that the random arrival of molecules at their target site generates an effective concentration noise with variance
\begin{equation}
\sigma_c^2 = \left({{\delta c}\over c}\right)^2 c^2 \sim {c\over{D a\tau_{\rm avg}}}.
\end{equation}
This effective concentration noise will propagate through the genetic regulatory element, contributing to the variance in the output as
\begin{equation}
\sigma_{g,{\rm BP}}^2 (c) \sim {\bigg |} {{d\bar g (c)}\over{dc}}{\bigg |}^2 \sigma_c^2 .
\end{equation}
Putting this together with the counting noise from Eq (\ref{sigmag-count}) we have
\begin{eqnarray}
\sigma_g^2 (c) &=& \sigma_{g,{\rm count}}^2 (c)  + \sigma_{g,{\rm BP}}^2 (c) 
= {1\over {N_{\rm max}}} \bar g(c) + {\bigg |}{{d\bar g (c)}\over{dc}}{\bigg |}^2 {c\over{D a\tau_{\rm avg}}}\\
&=& {1\over {N_{\rm max}}}\left[ \bar g(c)   +  (c/c_0) {\bigg |}{{d\bar g (c)}\over{d(c/c_0)}}{\bigg |}^2\right],
\label{2partvar}
\end{eqnarray}
where the natural units of concentration are $c_0 = N_{\rm max}/Da\tau_{\rm avg}$.  From the numbers above, we see that real transcription factor concentrations are comparable to $c_0$, which is part of why the optimization of information transmission leads to interesting results.

Now we can make use of results from the previous lecture, \S3.1.  First we propagate the output noise $\sigma_g$ back through the input/output relation, as in Fig \ref{errorprop_fig}, to obtain an effective input noise
\begin{equation}
\sigma_c^{\rm eff} = {\bigg |}{{d\bar g (c)}\over{dc}} {\bigg |}^{-1} \sigma_g (c) 
= {{c_0}\over {h\sqrt{N_{\rm max}}}}
  {{(c/c_0)}\over {\bar g (c) [1-\bar g(c)]}} 
\left[  \bar g (c) + {{h^2}\over{(c/c_0)}} \bar g (c) [1-\bar g(c)]\right]^{1/2}.
\label{sigmac_final}
\end{equation}
Then we can work in the small noise limit to estimate the maximum information, from Eqs (\ref{Zdef}) and (\ref{Imax}),
\begin{equation}
I_{\rm max} = \log_2\left[ {1\over\sqrt{2\pi e}} \int_0^{c_{\rm max}} {{dc}\over{\sigma_c^{\rm eff}}}\right] \,\, {\rm bits},
\end{equation}
where we go back to conventional units and note explicitly that the transcription factor has some maximum concentration $c_{\rm max}$. Playing with this a bit one can see that
\begin{equation}
I_{\rm max} = {1\over 2} \log_2 \left[ {{N_{\rm max}}\over\sqrt{2\pi e}}\right]  + F(h,K/c_0; c_{\rm max}/c_0) .
\label{Fdef}
\end{equation}
Thus we can optimize the two parameters $h$ and $K$, and the answer will depend on the maximum concentration $c_{\rm max}$; again it is natural to express concentrations in units of $c_0$.

\begin{figure}[t]
\includegraphics[width=\linewidth]{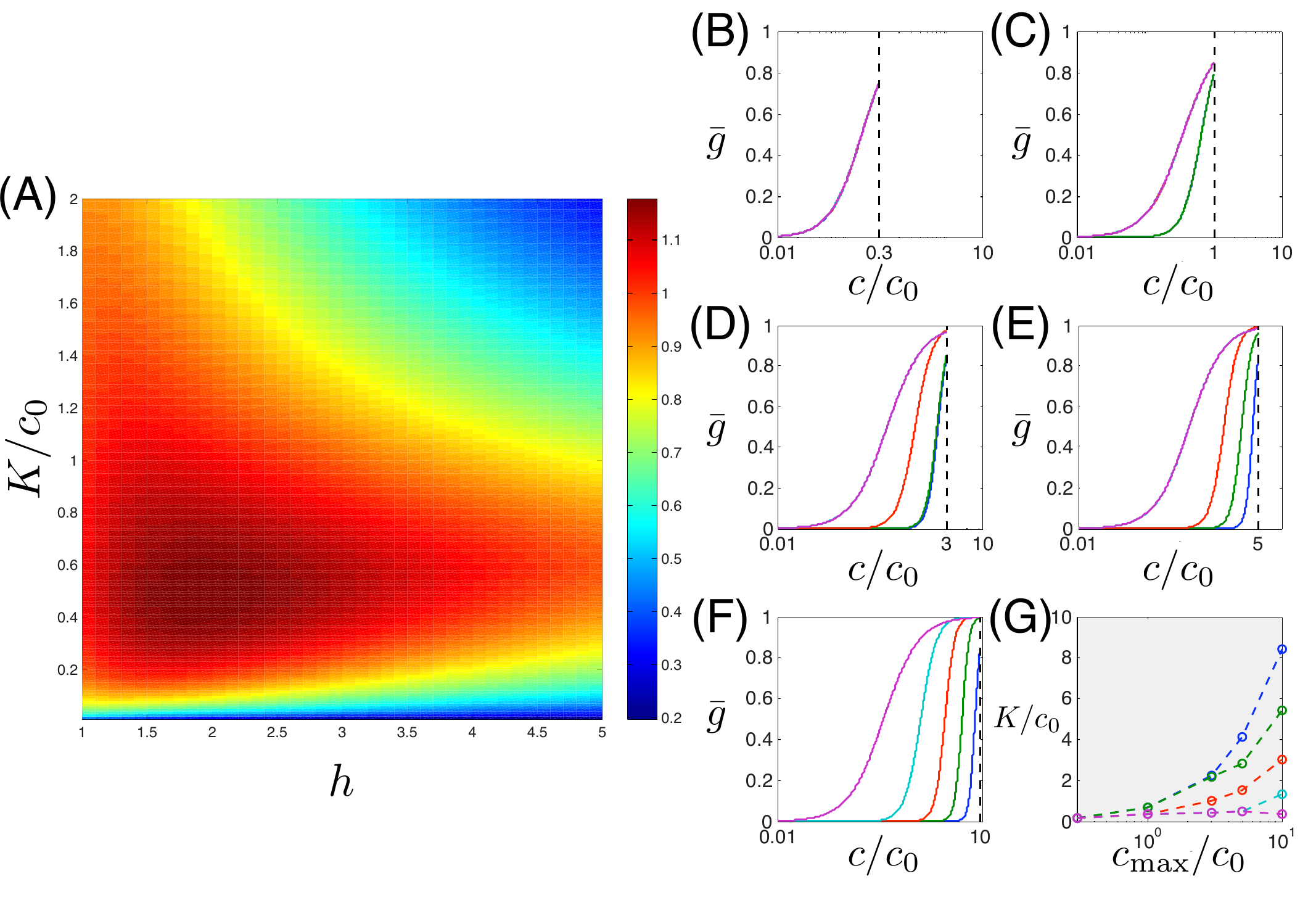}
\caption{Optimizing  information  about the concentration of a single transcription factor.   (A) A map of $F$ from Eq (\ref{Fdef}) as a function of $K$ and $h$, showing a single clear optimum at reasonable parameters ($c_{\rm max}/c_0 = 1$). (B--F) Optimal input/output relations $\bar g_{\rm i}(c)$ with five target genes, with increasing values of $c_{\rm max}/c_0$ indicated by dashed vertical lines.  At smaller $c_{\rm max}/c_0$ the optimal solutions are redundant, so fewer  distinct input/output relations are visible.  (G) Optimal values of $K/c_0$, showing a sequence of bifurcations as $c_{\rm max}/c_0$ increases and responses become distinct.  Redrawn from Ref \cite{tkacik+al_09}, with thanks to G Tka\v{c}ik and AM Walczak. \label{optgene1}}
\end{figure}

Figure \ref{optgene1}A shows the results of this optimization at $c_{\rm max}/c_0 = 1$ \cite{tkacik+al_09}.  Perhaps the most important conclusion is that the optimal parameters are perfectly sensible, not driven off to extreme values.  This happens because of the interplay between the two components of the noise, counting at the output and Berg--Purcell at the input.  

A natural generalization is to the case where the single transcription factor controls multiple genes, at expression levels $\{g_1,\, g_2,\, \cdots ,\, g_K\}$, but to keep things simple these target genes do not interact.  In Figure \ref{optgene1}B--G we see the results when there are $K=5$ targets and $c_{\rm max}/c_0$ changes across a dynamic range of $30\times$.  At small values of $c_{\rm max}/c_0$ the optimal solution is for all $K$ genes to have the same values of $K$ and $h$, so that they are completely redundant copies of one another (Fig \ref{optgene1}B).  I have always found this result fascinating because redundancy often is taken as prima facie evidence against any information theoretic optimization principle.  Here we see that redundancy actually is the result of such a principle, in the right regime.  I don't know whether this is a path to understanding the appearance of redundancy in biological signaling more broadly, but it surely is an object lesson.

As the maximum allowed concentration of the input transcription factor increases, the optimal strategy for information transmission changes and more of the target genes come to have distinguishable responses (Fig \ref{optgene1}B--F).  This happens through a series of bifurcations that we can visualize in a plot of $K/c_0$ vs $c_{\rm max}/c_0$ (Fig \ref{optgene1}G).   Successive bifurcations add distinct target gene responses at higher concentration (larger $K$) and with steeper responses (larger $h$), until the set of input/output relations ``tile'' the full dynamic range of inputs.

The results of Fig \ref{optgene1} are just in the case where a single transcription factor activates a set of non--interacting target genes.  I should admit that when we started thinking about these problems my colleagues and I thought that the path from these simple examples to something realistic would be quick, but we were wrong.  We did learn about some pieces of the problem.

\begin{figure}
\includegraphics[width=\linewidth]{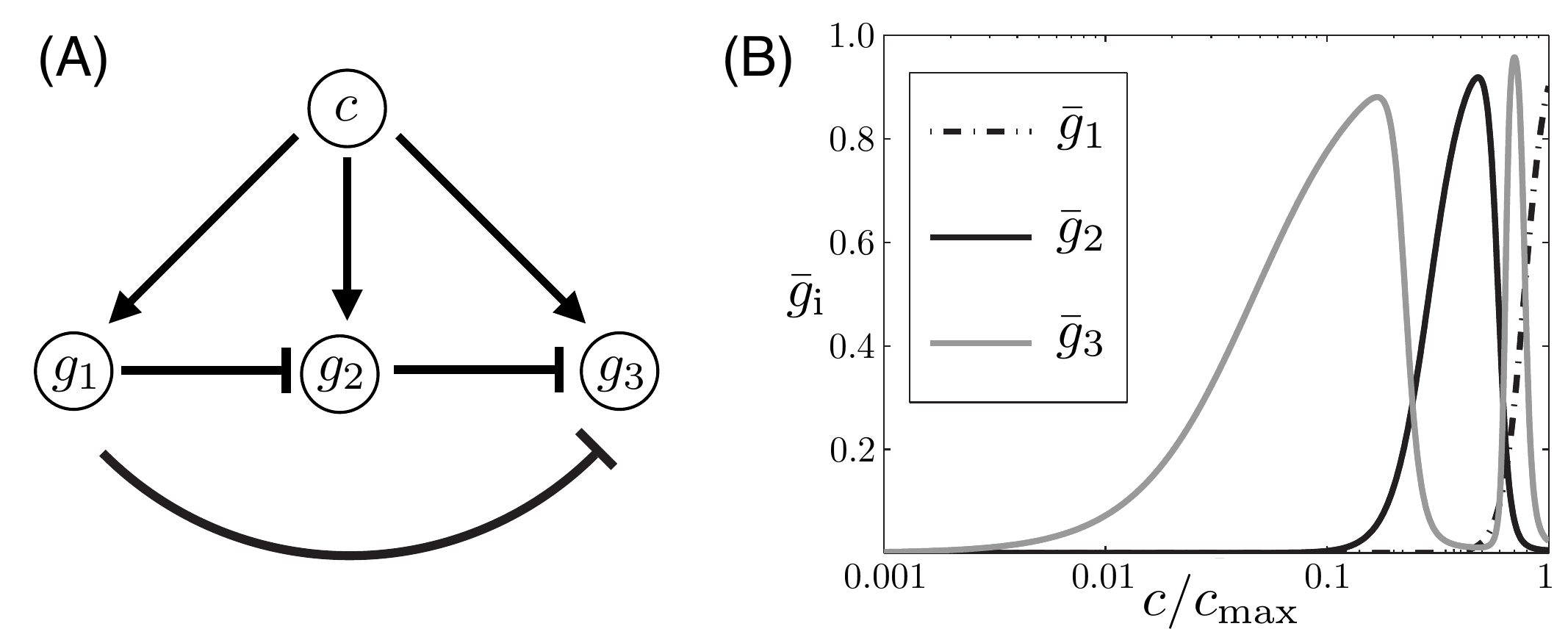}
\caption{Redundancy reduction. (A) A single transcription factor (concentration $c$) activates multiple genes $\{g_1, g_2, g_3\}$ that repress one another.  (B) Mean expression levels that optimize information transmission, with $c_{\rm max}/c_0 = 10$.  Redrawn from Ref  \cite{walczak+al_10}, with thanks to G Tka\v{c}ik and AM Walczak. \label{rep_red}}
\end{figure}

To begin, the pattern of responses from multiple targets in Fig \ref{optgene1} is redundant, because activation of the genes with large $K$ allows us to infer that the genes with smaller $K$ also are active.  This redundancy can be reduced, and information transmission enhanced, by repressive interactions among the targets, as shown in Fig \ref{rep_red} \cite{walczak+al_10}.  This parallels the discussion of lateral inhibition in the retina,  outlined in \S\ref{lec4:linearnets},  and produces profiles of (mean) expression level vs input concentration that  remind us of those shown by the gap genes.  These benefits of repression in enhancing efficiency are seen even when there are no feedback loops.  

The prototypical example of feedback is when a single target gene also can activate itself \cite{tkacik+al_12}.   This system has a bistable regime at sufficiently strong self--activation, but the optimal parameter values are  on the monostable side of this bifurcation.  As the bifurcation is approached, however, there is critical slowing down.  This emergent long time scale serves to reduce noise, but near criticality there is also a strong path for noise in the output to be injected back into the system.   The closeness of the optimum to the critical point thus depends on the maximal concentration of output molecules $N_{\rm max}$.

In many systems, including the fly embryo,  multiple cells or nuclei can exchange proteins or mRNA through diffusion.  This  generates a spatial averaging that can reduce noise, adding to the effects of temporal averaging.\footnote{This combination of spatial and temporal averaging seems to be necessary to understand how high levels of noise in the initiation of transcription  yield low levels of noise in the concentration of the gap gene proteins \cite{little+al_13}.}   Importantly, because diffusion itself is noisy, this effect can't push the final noise level below the Poisson  level of counting noise.  In the embryo, meaningful variations in input and output are laid out in space, and of course sufficiently strong diffusion will degrade these patterns.  The result is that there is an optimal diffusion constant that maximizes information transmission  \cite{sokolowski+tkacik_15}.

\begin{figure}
\includegraphics[width=\linewidth]{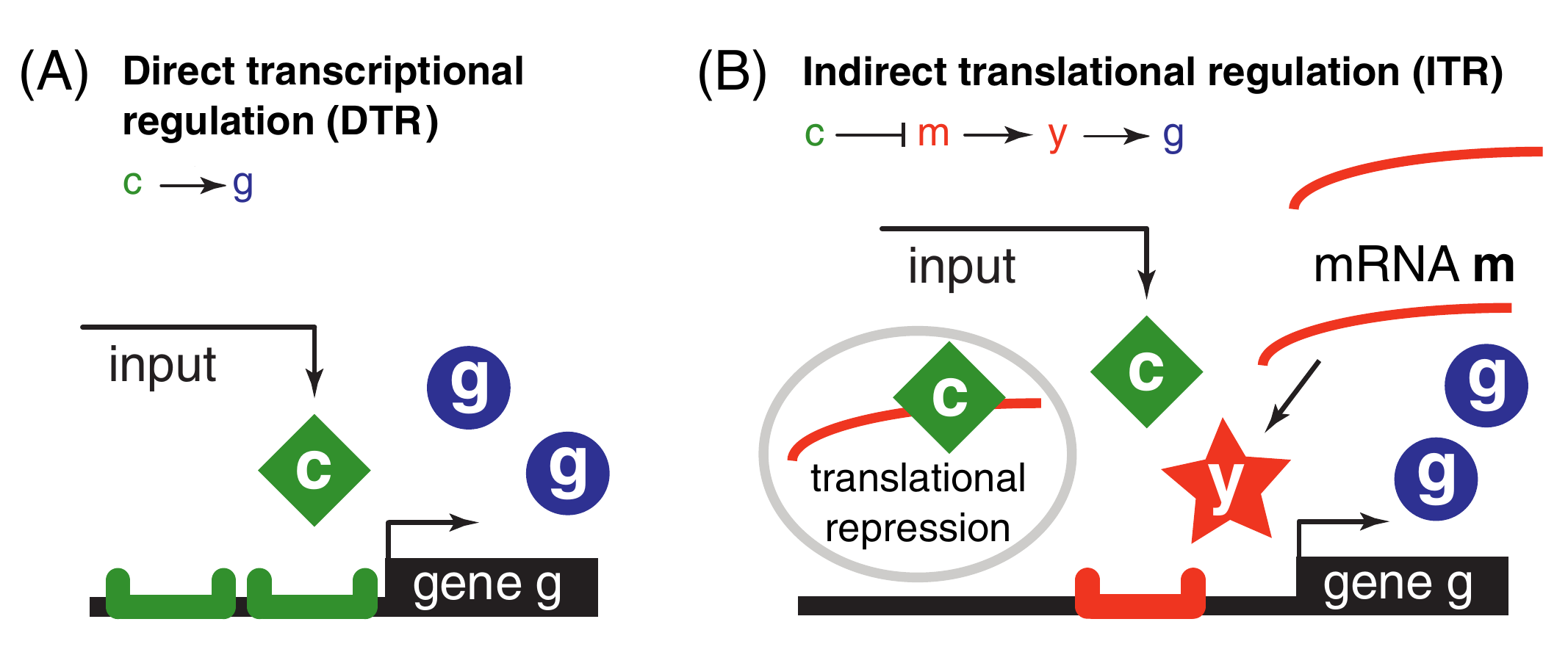}
\caption{Schematic  of direct transcriptional regulation (DTR) and indirect translational regulation (ITR). (A) In DTR, as above, activator TFs (green squares at concentration $c$) interact with   binding sites along the DNA to activate expression of the regulated gene $g$. (B) In ITR scenario, input molecules (green squares) bind to mRNAs $m$ of protein $y$ (red chain) to make the mRNA inaccessible for translation (gray oval). Translation proceeds from unbound mRNA molecules, giving rise to proteins $y$ (red stars). These proteins act as repressors   for gene $g$; the overall mapping from $c\rightarrow g$  is thus activating   in both scenarios. Redrawn from Ref  \cite{sokolowski+al_16}, with thanks to TR Sokolowski, G Tka\v{c}ik, and AM Walczak. \label{dualreg}}
\end{figure}

In Figures \ref{optgene1}B--F we see that even as the different target genes acquire distinct input/output relations, the optimal solutions still do not make much use of the lowest accessible concentrations; this regime is simply too noisy to allow reliable information transmission.  But suppose that the cell makes many mRNA molecules of some intermediate protein $y$, and that the protein $c$ binds to these mRNA and represses translation, as in Fig \ref{dualreg}; $y$ can then act indirectly as a transcriptional repressor so that the whole path $c\rightarrow g$ is activating.  Then the ``measurement'' of low concentrations is made in parallel at many sites, rather than at just one site along the DNA, and the response is largest and most reliable at the lowest concentrations.  This suggests that a molecule which functions both as a transcriptional activator and a translational repressor could make better use of its full dynamic range.  It requires a careful calculation to show that information transmission really can be  larger even when we constrain the total number of molecules, but it works \cite{sokolowski+al_16}.  One of the primary maternal morphogens in the fly embryo belongs to a whole class of proteins that function as such dual transcription/translation regulators \cite{dubnau+struhl_96,rivera-pomar+al_96,niessing+al_00,johnstone+lasko_01}, and our best estimates put this system in the regime where dual regulation enhances information transmission.  I found it remarkable that this almost baroque level of biological complexity can be derived as part of the solution to a fundamental physics problem faced by the cell.

Each of these pieces---repression to reduce redundancy, feedback to average over time, diffusion to average over space, and dual regulation---is an interesting problem by itself, and it was possible to make progress with a combination of analytic and (modest) numerical approaches.  But all of these things are happening at once in the fly embryo, even just in the gap gene network.  To put (most of) these pieces together requires a more sophisticated approach \cite{sokolowski+al_23}.

\subsection{The gap genes, once more}
\label{lec4-embryo}

\begin{figure}[t]
\includegraphics[width=\linewidth]{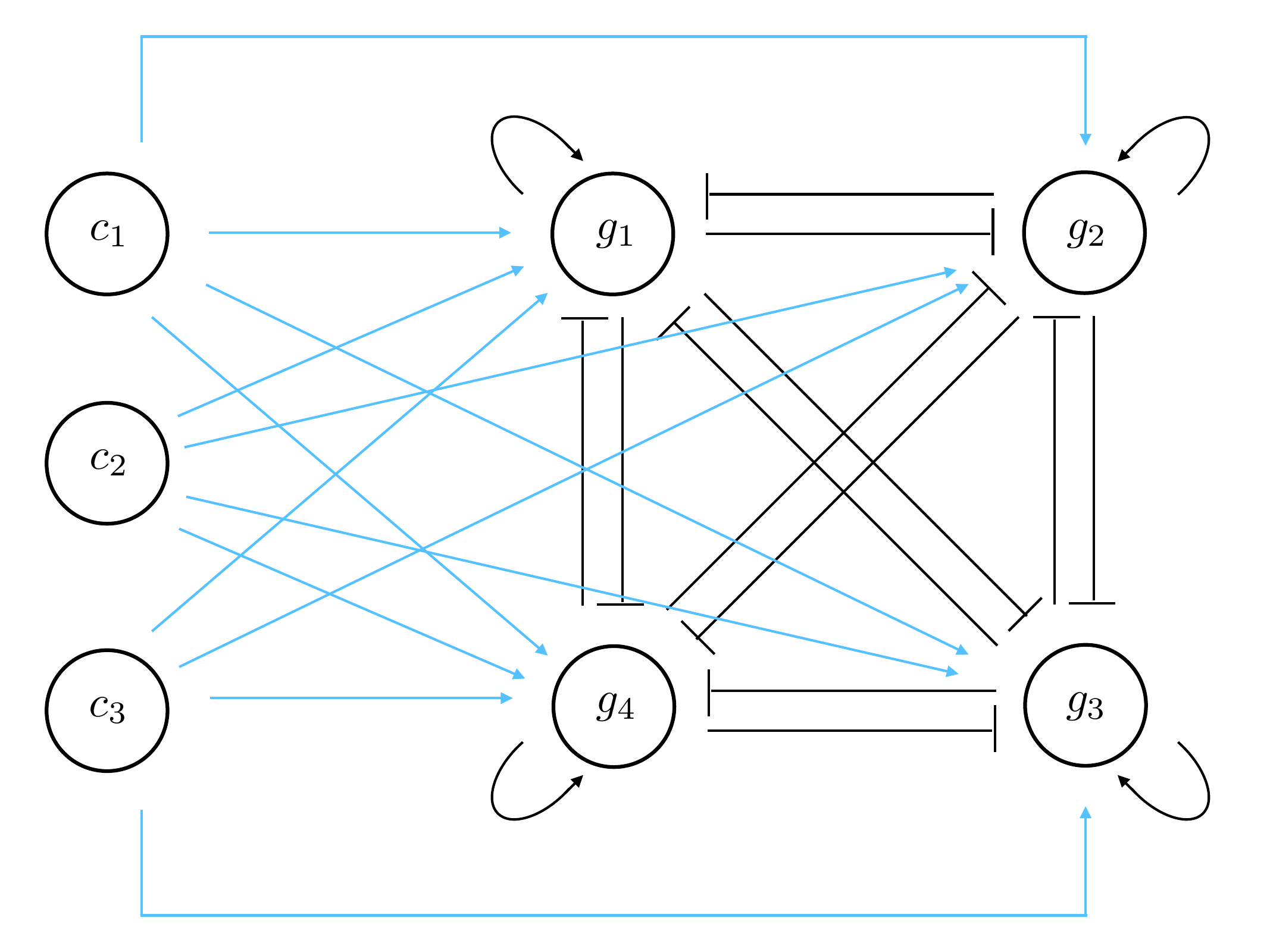}
\caption{The network of four gap genes $\{g_{\rm i}\}$ driven by three maternal inputs $\{c_\alpha\}$.  Regulatory interactions $c_\alpha \rightarrow g_{\rm i}$ are strictly feedforward (blue arrows).  In contrast,  interactions $g_{\rm i} \rightarrow g_{\rm j}$ can form feeback loops.  It is thought that $c_\alpha \rightarrow g_{\rm i}$ are largely activating while $g_{\rm i} \rightarrow g_{{\rm j}\neq {\rm i}}$ are largely repressive  (black blunt arrows), although there is evidence for self--activation $g_{\rm i} \rightarrow g_{\rm i}$ (curved black arrows).  The schematic includes this consensus, which proves to be a feature of the optimized network, though it is not imposed. \label{fig:network}}
\end{figure}

Let's plunge right in.  We have a collection of nuclei labeled ${\rm n} = 1,\, 2,\, \cdots ,\, N$ along the anterior--posterior axis of the embryo, with a distance $\Delta$ from one to the next.   Associated with each nucleus are the expression levels of the four gap genes $\{g_{\rm i}({\rm n})\}$, and these molecules can diffuse between neighbors with an effective diffusion constant $D$.  As before we will keep track of one concentration for each species, not worrying about the separate dynamics of mRNA and proteins.  We assume that each gap gene product has the same maximal synthesis rate $r_{\rm max}$ and the same decay time $\tau$; these are simplifications but also supported by experiment.  The gap genes are driven by maternal inputs $\{c_\alpha ({\rm n})\}$ with $\alpha = 1,\, 2,\, 3$, and we will assume that theese inputs are constant in time, which we know to be correct over the relevant time window for at least one of them \cite{gregor+al_07a}.  With these assumptions, the dynamics of the gap gene expression levels obey a generalization of the first equation that we wrote down in these lectures [Eq (\ref{AregB})]:
\begin{eqnarray}
{{dg_{\rm i}({\rm n},t)}\over {dt}} 
&=& 
r_{\rm max} f_{\rm i}\left( \{g_{\rm j} ({\rm n},t)\}; \{c_\alpha ({\rm n}) \}\right)
 - {1\over \tau}g_{\rm i}({\rm n},t) \nonumber\\
 &&\,\,\,\,\,\,\,\,\,\,
 + {D\over{\Delta^2}}\left[ g_{\rm i}({\rm n}+1,t) - 2 g_{\rm i}({\rm n},t) + g_{\rm i}({\rm n}-1,t)\right] + \eta_{\rm i}({\rm n}, t) ,
 \label{bigeq}
\end{eqnarray}
where $f_{\rm i}$ is a separate regulation function for each gap gene, and $\eta_{\rm i}({\rm n}, t)$ are noise terms.  Three static inputs are driving four interacting genes, as in Fig \ref{fig:network}, so there are $(3\times 4) + (4\times 4) = 28$ arrows, as noted in \S\ref{sec:introproblem}.

We take the maternal inputs $\{c_\alpha ({\rm n})\}$ as known, and fixed as we try to optimize the gap gene network itself. This certainly is fair for the input which is large at the anterior end, where we have accurate measurements in both live and fixed embryos establishing the peak absolute concentration and the approximate exponential decay with distance into the egg.  We will assume that the input which is large at the posterior end is just a mirror image, and that the terminal inputs are symmetric with the same peak concentration but more rapid decays.

In order to proceed we need a model for the regulation functions.  In particular we need to describe what happens as multiple regulatory arrows converge on a single target gene.  We have taken a simple view inspired by allostery in proteins 
\cite{monod+al_63,monod+al_65,perutz_90,phillips_20}.  In a single large protein molecule, binding of a small molecule at one point on the surface can influence the binding of molecules far away; a classic example is the cooperative binding of four oxygen molecules to hemoglobin in our blood.  Monod, Wyman, and Changeux (MWC) proposed that this happens because the protein can exist in two structures, and the small molecules have different binding energies to these two structures.  In this picture there is no direct interaction between the binding events; all interactions are mediated through the protein \cite{hopfield_73}.  In the simplest formulation these events occur at thermal equilibrium, even though one often is describing the activity of enzymes, ion channels, and other systems that evidently are not in equilibrium.  The idea is that the rate of the events that we are interested in is proportional to the occupancy of some state(s), and this occupancy is well approximated by estimates from equilibrium statistical mechanics.  There is a tradition of using such equilibrium arguments for transcriptional control as well \cite{bintu+al_05a,bintu+al_05b}, although the validity of this quasi--equilibrium view remains an open question \cite{zoller+al_22}.

\begin{figure}
\includegraphics[width=\linewidth]{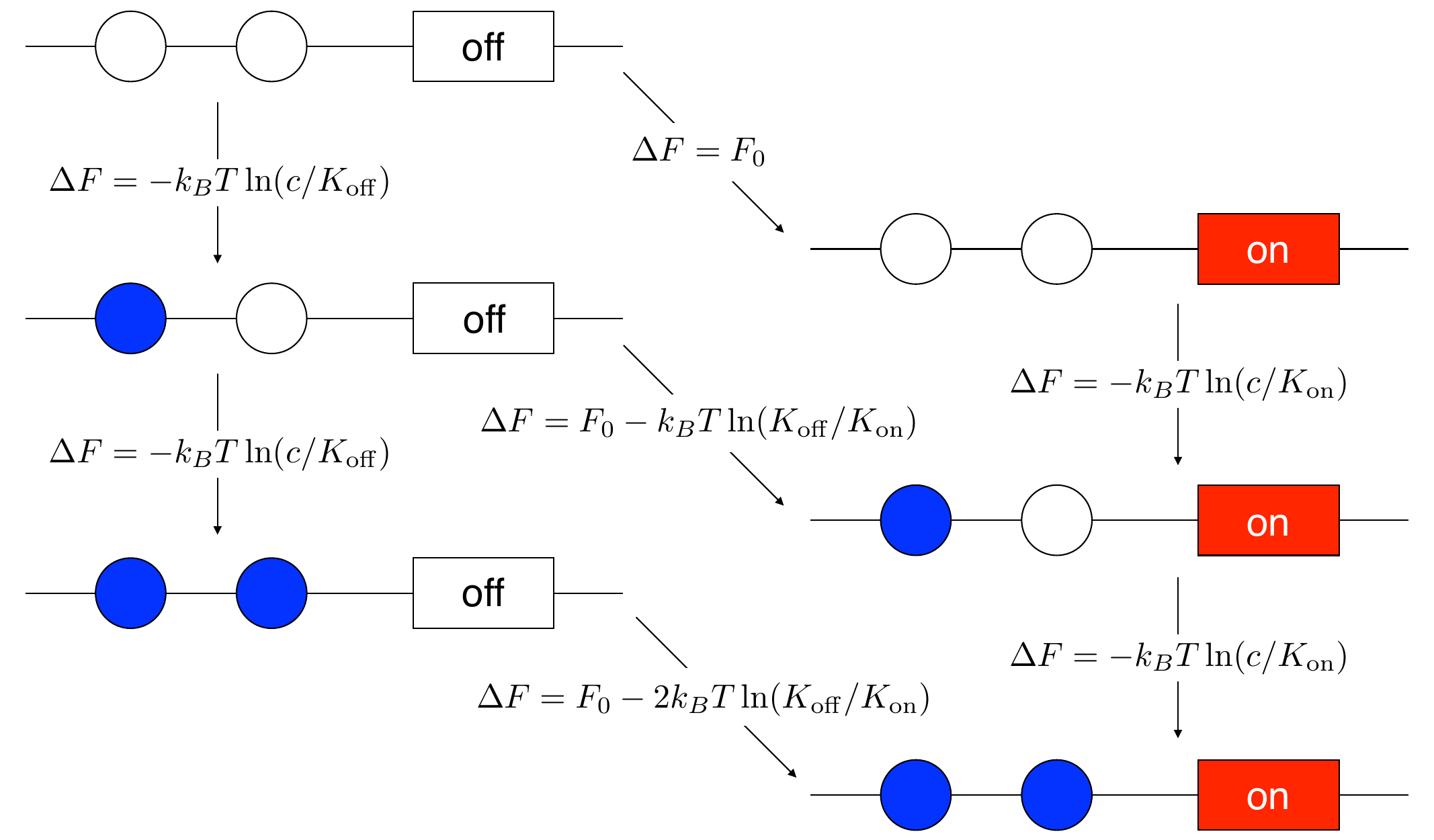}
\caption{The Monod--Wyman--Changeux (MWC) model for regulation.  Here there are two binding sites for the (blue) transcription factor (TF) protein, at concentration $c$. Binding at the two sites are independent events, but the binding constants $K_{\rm on}$ and $K_{\rm off}$ are different in the two states. If binding is stronger in the on state ($K_{\rm on} < K_{\rm off}$) then the free energy difference between off and on states increases as TFs bind.  \label{fig:MWC}}
\end{figure}

Figure \ref{fig:MWC} schematizes the MWC model for regulation of a gene by a single transcription factor (TF).  The system can be in ``on'' or ``off'' states, with transcription occurring in the on state, and there are two binding sites for the TF, which is at concentration $c$.  Assuming the system is in equilibrium we can write the probability of being in the on state in terms of the free energy differences as shown, to give
\begin{eqnarray}
P_{\rm on} &=& {{e^{-F_0/k_B T} (1+ c/K_{\rm on})^2}\over{ (1+ c/K_{\rm off})^2 + e^{-F_0/k_B T} (1+ c/K_{\rm on})^2}}\\
&=& {1\over{ 1 + \exp[- {\cal F}(c) ]}}\\
  {\cal F}(c)  &=& - {{F_0}\over{k_B T}} + 2  \ln\left({{1+c/K_{\rm on}}\over{1+c/K_{\rm off}}}\right) .\label{FMWC1}
\end{eqnarray}
Notice that if binding  is very tight in the on state and very weak in the off state, we should have $K_{\rm on} \ll c \ll K_{\rm off}$ and we can rewrite
\begin{equation}
P_{\rm on} = {{c^2}\over{c_{1/2}^2 + c^2}} ,
\end{equation}
with $c_{1/2} = K_{\rm on} e^{F_0/2k_B T}$.  We identify the probability of being in the on state with the regulation function $f$. Notice that factor of $2$ in Eq (\ref{FMWC1}) is counting the two binding sites.

One attractive feature of the MWC model is that it generalizes to having multiple TFs converge to regulate a single gene.  Consider that gene $\rm i$ has $H_{\rm ij}$ binding sites for the protein encoded by gene $\rm j$ and $H_{{\rm i}\alpha}$ sites for the maternal input $\alpha$. Then if we follow the same equilibrium statistical mechanics arguments as above, we will find
\begin{eqnarray}
f_{\rm i}\left( \{g_{\rm j} \}; \{c_\alpha  \}\right) &=& {1\over{1 +\exp\left[ - {\cal F}_i \left( \{g_{\rm j} \}; \{c_\alpha  \}\right)\right]}}\label{MWC_f1}\\
{\cal F}_{\rm i} \left( \{g_{\rm j} \}; \{c_\alpha  \}\right) &=& - {{F_{{\rm i}0}}\over{k_B T}} + \sum_{\rm j} H_{\rm ij}\ln \left(1 + {{g_{\rm j}}\over{K_{\rm ij}}}\right) + \sum_\alpha H_{{\rm i}\alpha}\ln \left(1 + {{c_\alpha}\over{K_{{\rm i}\alpha}}}\right) ,\label{MWC_f2}
\end{eqnarray}
where we work in the approximation that one of the two binding constants is large (very weak binding).  Note that by changing the sign of $H$ we can have TFs act as both activators or repressors.  This gives us a parameterization for the regulation functions in Eq (\ref{bigeq}), and as promised in the first lecture we have two parameters $K_{\rm ij}$ and $H_{\rm ij}$ for each pair of species that can interact.

The noise in Eq (\ref{bigeq}) has contributions from the input (Berg--Purcell) and output (counting) noise as in Eq (\ref{2partvar}).\footnote{To be fully realistic we can add a small extra noise  with constant fractional variance, which seems to be necessary to match the data \cite{sokolowski+al_23,swain+al_02}.} There is also noise attached to the diffusion terms, and one has to be careful that there is independent noise in the fluxes ${\rm n} \rightarrow {\rm n} \pm 1$, not independent noise added to each site; this insures that diffusion noise does not violate conservation of molecules.  All of these noise sources are white, as can be seen for example from the fact that variances in Eq (\ref{2partvar}) are inversely proportional to averaging times.  Spectral densities are set by the absolute numbers or concentrations of molecules, as above, and these are known quite well from experiment; for details see Ref \cite{sokolowski+al_23}.

Now that we have all the ingredients in Eq (\ref{bigeq}), we could just simulate.   But our goal is to optimize the parameters so as to maximize the information that expression levels $\{g_{\rm i}\}$ provide about position or cellular identity $\rm n$.  It would not be enough to have a single solution of these stochastic differential equations, we need to know about the whole {\em distribution} of solutions.  This quickly becomes intractable.  
Fortunately we know that in the real system noise levels are small and fluctuations are approximately Gaussian.  This means that we can linearize Eq (\ref{bigeq}) in the small fluctuations around the mean, and find closed equations for the time evolution of the means $\langle g_{\rm i} ({\rm n},t)\rangle$ and the covariance matrix $\langle \delta g_{\rm i}({\rm n},t)\delta g_{\rm j} ({\rm n},t)\rangle$.  Thus at a single setting of all the parameters $({\bf H}, {\bf K})$ we can do one (large) integration forward in time and find everything we need in order to evaluate the positional information at $t\sim 45\,{\rm min}$ into nuclear cycle 14.  We explore the 50+ dimensional parameter space by a version of simulated annealing \cite{sokolowski+al_23}.

\begin{figure}[t]
\includegraphics[width=\linewidth]{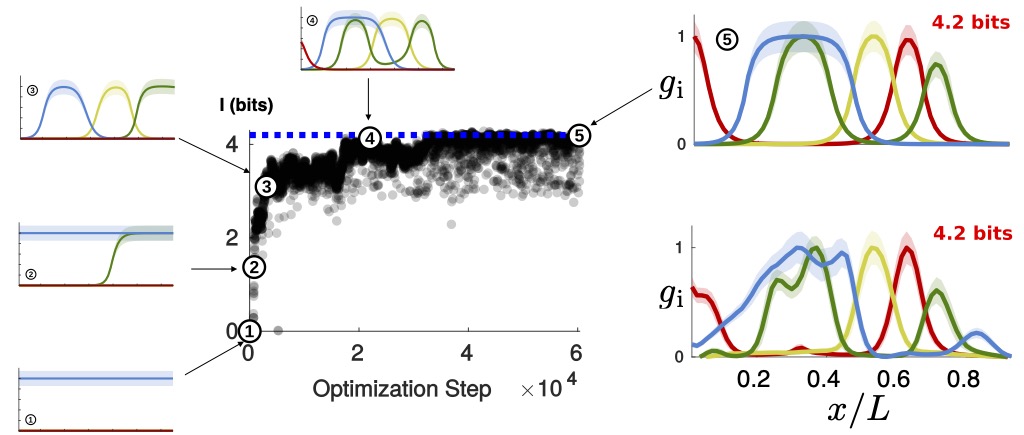}
\caption{Maximizing positional information in the gap gene network. Random initial parameter settings produce zero positional information, with all gap genes fully on or off across the length of the embryo (1).  Across ${\sim}20,000$ steps of optimization we find parameter settings that allow first one then two, three, and finally all four gap genes to be modulated by the dynamic range of maternal inputs ($2 \rightarrow 3 \rightarrow 4$).  The majority of the optimization run then tunes the parameters to maximize the positional information, arriving at a setting ($5$) that behaves much like the real network.  Redrawn from Ref  \cite{sokolowski+al_23}, with thanks to T Gregor, TR Sokolowksi, and G Tka\v{c}ik. \label{fig:optnet}}
\end{figure}

The essential results of the optimization process are shown in Fig \ref{fig:optnet}.  We start with some random setting of all the adjustable parameters, and typically this will leave some of the gap genes fully on and some fully off, uniformly across the entire embryo, so that there is zero positional information (point 1 in Fig \ref{fig:optnet}).  As we explore and anneal we find parameter settings that allow first one then two, three, and all four gaps genes to be driven on and off by the full dynamic range of the maternal inputs (points 2, 3, and 4). Much of the time spent in optimization is required to converge from these patterns onto something richer and more informative, finally arriving at an optimum (point 5).  This optimal network has spatial patterns of gene expression very similar to those of the real network, and the absolute magnitude of the positional information is close as well.

I want to emphasize that in deriving the patterns of gap gene expression that we see in the upper right of Fig \ref{fig:optnet}, there are no free parameters---all are determined by the optimization principle.  In fact if we set this principle aside and try to fit the model to the mean profiles, we don't do any better in matching the data.  More subtly, these best fits convey substantially less than the maximum information because they correspond to parameter settings with excess noise.  Thus in some way the optimization principle is getting us closer to the real network than standard model fitting.  Also, the small differences between the optimum and the real network must reflect a limitation in the class of models that we are considering, not a failure of optimization.
There are a few more interesting features of the solution(s):
\begin{itemize}
\item The positional information seen in real networks really is at the edge of what this class of models can support, as seen by random sampling of the parameter space.
\item Almost all of the possible interactions are realized in the optimal solution, and their signs agree with available data.
\item We have constrained the maximum molecule counts and concentrations; other local optima differ in the mean utilization of these resources.
\item We read out positional information at a moment in nuclear cycle 14 where the real expression levels are maximally informative, and do not constrain the dynamics, but the optimal networks have very slowly changing patterns, as with the real system.
\end{itemize}
We can also think of this as a ``laboratory'' in which to test alternative scenarios, making a change and then re--optimizing all the remaining parameters:
\begin{itemize}
\item We have taken the diffusion constant $D$ in Eq (\ref{bigeq}) as known, but there is a broad maximum of information near this measured value.
\item If we eliminate feedback loops we can still reach similar levels of positional information but only with unphysically large values of the $H$ parameters; within realistic settings feedback is essential.
\item If the same maximum number of molecules were spread across only three gap genes it would be impossible to achieve the same level of positional information; benefits of a fifth gap gene are minimal.
\item Although cross--regulation within the gap gene network provides some resilience, eliminating any one of the maternal inputs results in significantly less positional information.
\end{itemize}
I suspect we are just scratching the surface.  Perhaps the most intriguing observation is that different local optima have broadly consistent gap gene expression patterns, but with shifts and swaps similar to those seen in related species \cite{crombach+al_14,wotton+al_15,goltsev+al_04}.

\section{Conclusion}
\label{sec:conclusion}

This Summer School celebrates a special moment in the long history of interactions between physics and biology.  Just one generation back, physicists and biologists couldn't agree on much, but they could agree that searching for a theoretical physics of life was a waste of time.  Physicists saw biology as too messy, and biologists saw the physicists' search for simplicity and universality as a poor match to the evident complexity and diversity of life.    Much has changed.

I have emphasized that enormous progress in experiment has created new opportunities for theory.  In particular, experiments have revealed that many living systems exhibit behaviors with a precision and reproducibility far beyond what once was imagined.  In some cases this precision corresponds to functional behavior close to the limits of what is allowed by the laws of physics.  Thus, aspects of early embryonic development can take their place alongside classical examples such as photon counting in vision and molecule counting in bacterial chemotaxis.  

It is an old idea that evidence for near--optimal performance could be turned around and formulated as a theoretical principle from which aspects of mechanism and function can be derived.  This idea has a checkered history.  Many people have used optimization principles in the absence of any direct evidence for optimization, which is fine if the next experiments provide those direct tests.  More subtly, when we formulate an optimization principle we often must search among a class of possible mechanisms in order to find the optimum, and too often this class has been woefully oversimplified.  Again these simplifications are fine as starting points, but sometimes the more realistic calculations never come.  The message is that seemingly simple and clear principles of physical optimization become challenging when we try to use them in the realistic context of life's complexity.

I don't know if what I have described here will survive the next rounds of experiments.  But I am excited that we are  implementing physical notions of optimization in realistic settings.  We have asked how the embryo can best decode positional information contained in the {\em actual} expression levels of gap genes, and  this   makes detailed predictions for the distortion of the body plan in mutants; theory and experiment agree quantitatively, with no adjustable parameters (\S\ref{lec2-embryo}).  We have asked how the embryo can match the statistics of maternal inputs to the {\em measured} signal and noise characteristics of the gap gene network, and  the surprising  uniformity of positional information along the anterior--posterior axis emerges as a prediction of this optimization; the real embryo is within two percent of the optimum (\S\ref{lec3-embryo}).  Finally, we have asked how the architecture and parameters of a {\em reasonably realistic} model for the gap gene network can be tuned to maximize information transmission, and we find  networks very much like the real network emerging as a result, with no free parameters; this optimization principle predicts many features of the network that we did not constrain, and provides a framework for exploring the interplay of chance and necessity (\S\ref{lec4-embryo}).

I want to end by saying a few words about what we are not doing.  Much of the progress described here, and in other lectures at the Summer School, takes a phenomenological view, hoping to say something crisp about system level functions without digging into too much microscopic detail.  There is another community that is much more focused on molecules, and the divide between these communities can be quite stark.  To give one example, we probably know more about the functional dynamics of individual ion channels than about any other class of protein molecules, and we have a precise theoretical framework for how these molecular dynamics shape the electrical dynamics of single neurons.  There are $100+$ different kinds of channels encoded in the genome, and single neurons express a cocktail of perhaps seven different kinds.  But all of this molecular richness disappears rather abruptly once we start to talk about coding and computation in networks of neurons.  

As physicists we are used to the idea that macroscopic phenomena are described by coarse--grained models, and that in these models many microscopic details have disappeared (see Approach 3 in \S\ref{lec1-reactions}).  But in the examples we understand, from the inanimate world, we  can do this coarse--graining explicitly to see which (relevant) elements of the model survive to influence macroscopic dynamics and which (irrelevant) details are erased.  As far as I know nobody knows how to start with a realistic description of ion channels and coarse--grain to arrive at a model of neural networks.  Further, we do know that some molecular details must matter, because particular classes of cells in the brain will switch from using one kind of channel to another at crucial moments in development, and different types of cells use different combinations of channels.  We also know that not all macroscopic quantities are universal, and there is the danger that what is universal might not be relevant for the organism.

My concern about the gap between molecular and system level descriptions is not just the vague feeling that we are missing something.  In many cases, including the fly embryo, some of the most powerful new experimental tools allow direct and reliable manipulation at the molecular level.   The parameters of our network description, attached to the arrows in Fig \ref{fig:network}, are themselves encoded (in part) in the DNA sequence of transcription factor binding sites.  It now is possible to edit these sequences with single base pair precision, but we don't know how these manipulations relate to the circles and arrows, so it is hard to see how our current theories connect to an exploding set of new experiments.  In truth we don't even know if we can start with sequences and do a systematic calculation to arrive at something like Eqs (\ref{bigeq}, \ref{MWC_f1}, \ref{MWC_f2}).  A more complete physics of life will  build these bridges.

\section*{Acknowledgements}
Teaching in summer schools is a special pleasure, a stimulus to putting one's thoughts in order.  My thanks to the organizers---Anne--Florence Bitbol, Thierry Mora, Ilya Nemenman, and Aleksandra Walczak---for this opportunity, and for  years of friendship.  Thanks also to the students, who provided a warm and enthusiastic welcome. The ideas presented here owe much to many wonderful collaborators and friends, as may be seen from the references. But this is a school, and acknowledging these debts also is an opportunity to provide advice to the students.  

The idea that living systems approach fundamental physical limits to their performance has fascinated me from the first time I heard about photon counting in vision.   On arriving in Groningen as a postdoctoral fellow forty years ago (!), I had the good fortune to meet Rob de Ruyter van Steveninck, who was in the office next door.  The experiments he was doing on the fly visual system, combining measurements in the retina with measurements on motion--sensitive neurons deep in the brain, made it seem possible to connect general theoretical ideas about physical limits and optimization with detailed, quantitative experiments.   We could not have predicted that those early discussions would turn into decades of continuing collaboration and friendship.   We also have had the pleasure of being joined by many colleagues.  Lesson for theorists:  talk to the experimentalist next door, it literally can change your life.

When I moved to Princeton I wanted to take the ideas of signals, noise, and information that Rob and I had explored in neurons down to the molecular level; Sima Setayeshgar and I made some first theoretical efforts in this direction.  David Tank had moved at the same time, and was excited that new fluorescence techniques might make it possible to measure signals on this scale with enough precision to say something meaningful.   Eric Wieschaus joined our conversation, and while he was unclear on exactly what David and I wanted to do (we were unclear as well), he explained why we should do it in flies.  Eric ended by saying that if some physics student wanted to try, we should send the student to him to learn the relevant techniques.  Thomas Gregor was working with me on a theory project, but decided to take Eric up on his offer, and this led to spectacular new experiments.  In parallel, Curt Callan and I were working with another student, Ga\v{s}per Tka\v{c}ik, on purely theoretical questions about the optimization of information flow in transcriptional control. Luckily, Ga\v{s}per and Thomas sat in the same office, and they made many of the crucial theory/experiment connections on their own.  Again, we could not have predicted that these early discussions would form the basis for two decades of work, for collaborations with generations of new colleagues, and for valued friendships.  Lesson for theorists:  be ready for experimental developments which allow  speculative theoretical ideas to become concrete.

Lesson from both  stories: one experiment can test a theory or raise a specific theoretical question, but sometimes theory/experiment collaboration is a long term investment.

In addition to the handful of friends who were at the origins of these ideas, the specific things I discussed in these lectures involved collaborations with Naama Brenner, Julien Dubuis, Adrienne Fairhall, Dmitry Krotov, Geoff Lewen, Geoff Owen, Mariela Petkova, Marc Potters, Fred Rieke, Shiva Sinha, Thomas Sokolowski, and Aleksandra Walczak.  This bland listing doesn't do justice to all the fun we have had together.

\paragraph{Funding information.}
This work was supported in part by the US National Science Foundation through the Center for the Physics of Biological Function (PHY--1734030), and by Fellowships from the Simons Foundation and the John Simon Guggenheim Memorial Foundation.

\nolinenumbers


\begin{thebibliography}{99}
%
\bibitem{harrison_87}
ER Harrison, {\em Darkness at Night: A Riddle of the Universe} (Harvard University Press, Cambridge MA, 1987).
%
\bibitem{peebles_93}
PJE Peebles, {\em Principles of Physical Cosmology} (Princeton University Press, Princeton NJ, 1993).
%
\bibitem{anderson_84}
PW Anderson, {\em Basic Notions of Condensed Matter Physics} (Benjamin/Cummings, Menlo Park CA, 1984).
%
\bibitem{3man}
NW Timof\'eff--Ressovsky, KG Zimmer, and M Delbr\"uck, \"Uber die Natur der Genmutation und der Genstruktur 
{\em  Nachrichten von der Gesellschaft der Wissenschaften zu G\"ottingen. Neue band} {\bf 1,} 189 (1935).
%
\bibitem{sloan+fogel_11}
 PR Sloan and B Fogel, eds,  {\em Creating a Physical Biology. The Three--Man Paper and Early Molecular Biology} (The University of Chicago Press, Chicago IL, 2011).
 %
 \bibitem{schrodinger_44}
E Schr\"odinger, {\em What is Life?} (Cambridge University Press, Cambridge UK, 1944).
%
\bibitem{watson+al_14}
JD Watson,  TA Baker, SP Bell, A Gann, M Levine, and R Losick, {\em Molecular Biology of the Gene, 7th Edition} (Pearson, Boston MA, 2014).
%
\bibitem{alberts+al_02}
B Alberts, A Johnson, J Lewis, M Raff, K Roberts, and P Walter, {\em Molecular Biology of the Cell, 4th Edition} (Garland Science,  New York NY, 2002).
%
\bibitem{nusslein+wieschaus_80}
C N\"usslein--Volhard and E Wieschaus, Mutations affecting segment number and polarity in {\em Drosophila}. {\em Nature} {\bf 287,} 795 (1980). \url{https://doi.org/10.1038/287795a0}
%
\bibitem{petkova+al_19}
MD Petkova, G Tka\v{c}ik, W Bialek, EF Wieschaus, and T Gregor,  Optimal decoding of cellular identities in a genetic network.   {\em Cell} {\bf 176,} 844 (2019). \url{https://doi.org/10.1016/j.cell.2019.01.007}
%
\bibitem{efw+cnv_16}
E Wieschaus and C N\"usslein--Volhard,   The Heidelberg screen for pattern mutants of {\em Drosophila}: A personal account. {\em Annu Rev Cell Dev Biol} {\bf 32,} 1  (2016). \url{https://doi.org/10.1146/annurev-cellbio-113015-023138}
%
\bibitem{lawrence_92}
PA Lawrence,  {\em The Making of a Fly: The Genetics of Animal Design}.  (Blackwell, Oxford, 1992).
%
\bibitem{cahn_96}
RN Cahn, The eighteen parameters of the standard model in your everyday life.  {\em Rev Mod Phys} {\bf 68,} 952 (1996). \url{https://doi.org/10.1103/RevModPhys.68.951}
 %
\bibitem{bcs}
J Bardeen, LN Cooper, and JR Schrieffer, Theory of superconductivity.  {\em Phys Rev} {\bf 108,} 1175  (1957). \url{https://doi.org/10.1103/PhysRev.108.1175}
%
\bibitem{wilson_75}
KG Wilson, The renormalization group:  Critical phenomena and the Kondo problem.  {\em Rev Mod Phys} {\bf 47,} 773 (1975).
\url{https://doi.org/10.1103/RevModPhys.47.773}
%
\bibitem{parisi_88}
G Parisi, {\em Statistical Field Theory} (Addison--Wesley, Redwood City CA, 1988).
%
\bibitem{zinn-justin_89}
J Zinn--Justin, {\em Quantum Field Theory and Critical Phenomena} (Oxford University Press, Oxford UK and New York NY, 1989).
%
\bibitem{cardy_96}
J Cardy, {\em Scaling and Renormalization in Statistical Physics}
(Cambridge University Press, Cambridge, 1996).
%
\bibitem{laughlin_83}
RB Laughlin, Anomalous quantum Hall effect: An incompressible quantum fluid with fractionally charged excitations.  {\em Phys Rev Lett} {\bf 50,} 1395 (1983). \url{https://doi.org/10.1103/PhysRevLett.50.1395}
%
\bibitem{gross+wilczek_73}
DJ Gross and F Wilczek, Ultraviolet behavior of non--Abelian gauge theories.  {\em Phys Rev Lett} {\bf 30,} 1343 (1973). \url{https://doi.org/10.1103/PhysRevLett.30.1343}
%
\bibitem{politzer_73}
HD Politzer, Reliable perturbative results for strong interactions? {\em Phys Rev Lett} {\bf 30,} 1346 (1973). \url{https://doi.org/10.1103/PhysRevLett.30.1346}
%
\bibitem{callan+gross_73}
CG Callan, Jr and DJ Gross, Bjorken scaling in quantum field theory. {\em Phys Rev D} {\bf 8,} 4383 (1973). \url{https://doi.org/10.1103/PhysRevD.8.4383}
%
\bibitem{newman_04}
P Newman, Deep inelastic lepton--nucleon scattering at HERA. {\em Int J Mod Phys A} {\bf 19,} 1061 (2004). \url{https://doi.org/10.1142/S0217751X0401897X}
%
\bibitem{lecun+al_15}
Y LeCun, Y Bengio, and G Hinton, Deep learning. {\em Nature} {\bf 521,} 436 (2015). \url{https://doi.org/10.1038/nature14539}
%
\bibitem{vaswani+al_17}
A Vaswani, N Shazeer, N Parmar, J Uszkoreit, L Jones, AN Gomez, \L  ~Kaiser, I Polosukhin, Attention is all you need.  In {\em Advances in Neural Information Processing Systems} {\bf 30}, I Guyon et al, eds  (Curran Associates, 2017).  
\url{https://proceedings.neurips.cc/paper_files/paper/2017/file/3f5ee243547dee91fbd053c1c4a845aa-Paper.pdf}
%
\bibitem{brown+al_20}
TB Brown, J Kaplan, G Sastry, B Mann, N Ryder, P Dhariwal, A Neelakantan, A Askell,  S Agarwal,  M Subbiah, P Shyam, A Herbert--Voss, G Krueger, DM Ziegler, 
T Henighan, R Child,  J Wu,  A Ramesh, C Winter, C Hesse, M Chen, B Chess,
E Sigler, J Clark,  M Litwin, S Gray, C Berner,  S McCandlish, A Radford, I Sutskever, and  D Amodei, Language models are few--shot learners. {\em Advances in Neural Information Processing Systems} {\bf 33,} 1877 (2020). 
\url{https://proceedings.neurips.cc/paper\_files/paper/2020/file/1457c0d6bfcb4967418bfb8ac142f64a-Paper.pdf}
%
\bibitem{block_62}
HD Block, The perceptron: A model for brain functioning. I. {\em Rev Mod Phys} {\bf 34,} 123 (1962). \url{https://doi.org/10.1103/RevModPhys.34.123}
%
\bibitem{block+al_62}
HD Block, BW Knight, Jr, and F Rosenblatt, Analysis of a four-layer series-coupled perceptron. II. {\em Rev Mod Phys} {\bf 34,} 135 (1962). \url{https://doi.org/10.1103/RevModPhys.34.135}
%
\bibitem{hopfield_82}
JJ Hopfield, Neural networks and physical systems with emergent collective computational abilities. {\em Proc Natl Acad Sci (USA)}  {\bf 79,} 2554 (1982). \url{https://doi.org/10.1073/pnas.79.8.2554}
%
\bibitem{amit_89}
DJ Amit, {\em Modeling Brain Function: The World of Attractor Neural Networks} (Cambridge University Press, Cambridge UK, 1989).
%
\bibitem{hertz+al_91}
J Hertz, A Krogh, and RG Palmer, {\em Introduction to the Theory of Neural Computation} (Addison Wesley, Redwood City CA, 1991).
%
\bibitem{mehta+schwab_14}
P Mehta and DJ Schwab, An exact mapping between the variational renormalization group and deep learning. arXiv:1410.3831 [stat.ML] (2014). 
\url{https://doi.org/10.48550/arXiv.1410.3831}
%
\bibitem{mehta+al_19}
P Mehta, M Bukov, C--H Wang, AGR Day, C Richardson, CK Fisher, and DJ Schwab, A high-bias, low-variance introduction to machine learning for physicists. {\em Phys Repts} {\bf 810,} 1 (2019). \url{https://doi.org/10.1016/j.physrep.2019.03.001}
%
\bibitem{carleo+al_19}
G Carleo, I Cirac, K Cranmer, L Daudet, M Schuld, N Tishby, L Vogt--Maranto, and L Zdeborov\'a, Machine learning and the physical sciences. {\em Rev Mod Phys} {\bf 91,} 045002 (2019). \url{https://doi.org/10.1103/RevModPhys.91.045002}
%
\bibitem{kaplan+al_20}
J Kaplan, S McCandlish, T Henighan, TB Brown, B Chess, R Child, S Gray, A Radford, J Wu, and D Amodei, Scaling laws for neural language models. arXiv:2001.08361 [cs.LG] (2020).
\url{https://doi.org/10.48550/arXiv.2001.08361}
%
\bibitem{roberts+al_22}
DA Roberts and S Yaida, with B Hanin, {\em The Principles of Deep Learning Theory} (Cambridge University Press, Cambridge UK, 2022).  See also arXiv:2106.10165 [cs.LG] (2021). \url{https://doi.org/10.1017/9781009023405}
%
\bibitem{peliti_91}
L Peliti, ed, {\em Biologically Inspired Physics: Proceedings of a NATO Workshop, Carg\`ese 1990} (Plenum Press, New York, 1991).
%
\bibitem{nelson+al_89}
D Nelson, T Piran, and S Weinberg, eds, {\em Statistical Mechanics of Membranes and Surfaces} (World Scientific, Singapore, 1989).
%
\bibitem{ramaswamy_17}
S Ramaswamy, Active matter. {\em J Stat Mech} (2017) 054002. 
\url{https://doi.org/10.1088/1742-5468/aa6bc5}
%
\bibitem{desai+al_07}
MM Desai, DS Fisher, and AW Murray, The speed of evolution and maintenance of variation in asexual populations. {\em Curr Biol} {\bf 17,} 385 (2007). \url{https://doi.org/10.1016/j.cub.2007.01.072}
%
\bibitem{fisher_11}
DS Fisher, Leading the dog of selection by its mutational nose. {\em Proc Natl Acad Sci (USA)} {\bf 108,} 2633 (2011). \url{https://doi.org/10.1073/pnas.1100339108}
%
\bibitem{neher+hallatschek_12}
RA Neher and O Hallatschek, Genealogies of rapidly adapting populations. {\em Proc Natl Acad Sci (USA)} {\bf 110,} 437 (2012). \url{https://doi.org/10.1073/pnas.1213113110}
%
\bibitem{desai+al_13}
MM Desai, AM Walczak, and DS Fisher, Genetic diversity and the structure of Ggenealogies in rapidly adapting populations. {\em Genetics} {\bf 193,} 565 (2013). \url{https://doi.org/10.1534/genetics.112.147157}
%
\bibitem{fisher_13}
DS Fisher, Asexual evolution waves: Fluctuations and universality. {\em J Stat Mech} P01011 (2013). \url{https://doi.org/10.1088/1742-5468/2013/01/P01011}
%
\bibitem{levy+al_15}
S Levy, J Blundell, S Venkataram,  DA Petrov, DS Fisher, and G Sherlock, Quantitative evolutionary dynamics using high-resolution lineage tracking. {\em Nature} {\bf  519,} 181 (2015). \url{https://doi.org/10.1038/nature14279}
%
\bibitem{good+al_17}
BH Good, MJ McDonald, JE Barrick, RE Lenski, and MM Desai, The dynamics of molecular evolution over 60,000 generations. {\em Nature} {\bf 551,} 45 (2017). \url{https://doi.org/10.1038/nature24287}
%
\bibitem{ba+al_19}
AN Nguyen Ba, I Cvijovi\'c, JI Rojas Echenique, KR Lawrence, A Rego--Costa, X Liu, SF Levy, and MM Desai, High--resolution lineage tracking reveals travelling wave of adaptation in laboratory yeast. {\em Nature} {\bf 575,} 494 (2019).
\url{https://doi.org/10.1038/s41586-019-1749-3}
%
\bibitem{luksza+lassig_14}
M {\L}uksza and M L\"assig, A predictive fitness model for influenza. {\em Nature} {\bf 507,} 57 (2014). \url{https://doi.org/10.1038/nature13087}
%
\bibitem{neher+al_16}
RA Neher, T Bedford, RS Daniels, CA Russell, and BI Shraiman, Prediction, dynamics, and visualization of antigenic phenotypes of seasonal influenza viruses. {\em Proc Natl Acad Sci (USA)} {\bf 113,} E1701 (2016). \url{https://doi.org/10.1073/pnas.1525578113}
%
\bibitem{zanini+al_16}
F Zanini, J Brodin, L Thebo, C Lanz, G Bratt, J Albert, and RA Neher, Population genomics of intrapatient HIV-1 evolution. {\em eLife} {\bf 4,} e11282 (2016). \url{http://dx.doi.org/10.7554/eLife.11282.001}
%
\bibitem{transtrum+al_15}
MK Transtrum, BB Machta, KS Brown, BC Daniels, CR Myers, and JP Sethna, Perspective: Sloppiness and emergent theories in physics, biology, and beyond. {\em J Chem Phys} {\bf 143,} 010901 (2015). 
\url{https://doi.org/10.1063/1.4923066}
%
\bibitem{brown+al_04}
KS Brown, CC Hill, GA Calero, CR Myers, KH Less, JP Sethna, and RA Crione, The statistical mechanics of complex signaling networks: nerve growth factor signaling.  {\em Phys Biol} {\bf 1,} 184 (2004). 
\url{https://doi.org/10.1088/1478-3967/1/3/006}
%
\bibitem{rand+al_21}
DA Rand, A Raju, M S\'aez, F Corson, and ED Siggia, Geometry of gene regulatory dynamics. {\em Proc Natl Acad Sci (USA)} {\bf 118,} e2109729118 (2021). \url{https://doi.org/10.1073/pnas.2109729118}
%
\bibitem{hecht+al_42}
S Hecht, S Shlaer, and MH Pirenne, Energy, quanta and vision.  {\em J Gen Physiol} {\bf 25,} 819 (1942). \url{https://doi.org/10.1085/jgp.25.6.819}
%
\bibitem{barlow_52}
HB Barlow, The size of ommatidia in apposition eyes. {\em J Exp Biol} {\bf 29,} 667  (1952). \url{https://doi.org/10.1242/jeb.29.4.667}
%
\bibitem{berg+purcell_77}
HC Berg and  EM Purcell, Physics of chemoreception.  {\em Biohys J} {\bf   20,} 193 (1977). \url{https://doi.org/10.1016/S0006-3495(77)85544-6}
%
\bibitem{bialek_12}
W Bialek, {\em Biophysics: Searching for Principles} (Princeton University Press, Princeton, 2012).
%
\bibitem{nikolic+al_23}
M Nikoli\'c,  V  Antonetti,   F  Liu,  G  Muhaxheri,  MD Petkova,  M Scheeler,  EM Smith,  W  Bialek,  and T  Gregor.  Scale invariance in early embryonic development.  arXiv:2312.17684 [q--bio.MN] (2023). \url{https://doi.org/10.48550/arXiv.2312.17684}
%
\bibitem{shinomura+al_62}
O Shimomura, FH Johnson, and Y Saiga,  Extraction, purification and properties
of aequorin, a bioluminescent protein from the luminous hydromedusan, {\em Aequorea}. {\em J Cell Comp Physiol} {\bf 59,} 223 (1962). \url{https://doi.org/10.1002/jcp.1030590302}
%
\bibitem{johnson+al_62}
FH Johnson, O Shimomura, Y Saiga, LC Gershman, GT Reynolds, and JR Waters Jr, Quantum efficiency of {\em Cypridina} luminescence, with a note on that of {\em Aequorea.} {\em J Cell Comp Physiol} {\bf 60,}  85 (1962). \url{https://doi.org/10.1002/jcp.1030600111}
%
\bibitem{prasher+al_92}
DC Prasher, VK Eckenrode, WW Ward,  FG Prendergast, and MJ Cormier, Primary structure of the {\em Aequorea victoria} green fluorescent protein. {\em Gene} {\bf 111,} 229 (1992). \url{https://doi.org/10.1016/0378-1119(92)90691-H}
%
\bibitem{chalfie+al_94}
M Chalfie, Y Tu, G Euskirchen, WW Ward, and DC  Prasher,  Green fluorescent protein as a marker for gene expression. {\em Science} {\bf 263,} 802 (1994).
\url{https://doi.org/10.1126/science.8303295} 
%
\bibitem{tsien_09}
RY Tsien,  Constructing and exploiting the fluorescent protein paintbox.
{\em Angew Chem Int Ed}  {\bf 48,} 5612 (2009). 
\url{https://doi.org/10.1002/anie.200901916}
%
\bibitem{gregor+al_07a}
T Gregor, EF Wieschaus, AP McGregor, W Bialek, and DW Tank, Stability and nuclear dynamics of the Bicoid morphogen gradient.  {\em Cell} {\bf 130,}  141 (2007). \url{https://doi.org/10.1016/j.cell.2007.05.026}
%
\bibitem{dubuis+al_13a}
JO Dubuis, R Samanta, and T Gregor, Accurate measurements of dynamics and reproducibility in small genetic networks. {\em Mol Sys Bio} {\bf 9,} 639 (2013).  \url{https://doi.org/10.1038/msb.2012.72}
%
\bibitem{little+gregor_18}
SC Little and T Gregor, Single mRNA molecule detection in {\em Drosophila}. In {\em RNA Detection. Methods in Molecular Biology, vol 1649}, I  Gaspar, ed, pp  127--142 (Humana Press, New York, NY, 2018)
%
\bibitem{lubeck+cai_12}
E Lubeck and L Cai, Single--cell systems biology by super--resolution imaging and combinatorial labeling. {\em Nat Methods} {\bf 9,} 743 (2012).  
\url{https://doi.org/10.1038/nmeth.2069}
%
\bibitem{chen+al_15}
KH Chen, AN Boettiger, JR Moffitt, SS Wang, and X Zhuang, Spatially resolved, highly multiplexed RNA profiling in single cells.  {\em Science} {\bf 348,} aaa6090 (2015).   \url{https://doi.org/10.1126/science.aaa6090}
%
\bibitem{zoller+al_18}
B Zoller, SC Little, and T Gregor,  Diverse spatial expression patterns emerge from unified kinetics of transcriptional bursting. {\em Cell}  {\bf 175,} 835 (2018). \url{https://doi.org/10.1016/j.cell.2018.09.056}
%
\bibitem{larson+al_11}
DR Larson, D Zenklusen, B Wu, JA Chao, and RH Singer, Real-time observation of transcription initiation and elongation on an endogenous yeast gene. {\em Science} {\bf 332,} 475 (2011). \url{https://doi.org/10.1126/science.1202142}
%
\bibitem{lucas+al_13}
T Lucas, T Ferraro, B Roelens, JDLH Chanes, AM Walczak, M Coppey, and N Dostatni, Live imaging of Bicoid-dependent transcription in {\em Drosophila} embryos. {\em Curr Biol} {\bf 23,} 2135 (2013). \url{http://dx.doi.org/10.1016/j.cub.2013.08.053}
%
\bibitem{garcia+al_13}
HG Garcia, M Tikhonov, A Lin, and T Gregor, Quantitative imaging of transcription in living {\em Drosophila} embryos links polymerase activity to patterning. {\em Curr Biol} {\bf  23,} 2140 (2013). \url{https://doi.org/10.1016/j.cub.2013.08.054}
%
\bibitem{chen+al_23}
P--T Chen, B Zoller, M Levo, and T Gregor, Gene activity as the predictive indicator for transcriptional bursting dynamics. 	arXiv:2304.08770 [q--bio.MN] (2023). \url{https://doi.org/10.48550/arXiv.2304.08770}
%
\bibitem{chen+al_18}
H Chen, M Levo, L Barinov, M Fujioka, JB Jaynes, and T Gregor, Dynamic interplay between enhancer--promoter topology and gene activity.  {\em Nat Genet} {\bf 50,} 1296  (2018). 
\url{https://doi.org/10.1038/s41588-018-0175-z}
%
\bibitem{barinov+al_20}
L Barinov, S Ryabichko, W Bialek, and T Gregor,  Transcription-dependent spatial organization of a gene locus. arXiv:2012.15819  [q--bio.MN] (2020). \url{https://doi.org/10.48550/arXiv.2012.15819}
%
\bibitem{adrian_28}
ED Adrian,  {\em The Basis of Sensation: The Action of the Sense Organs.}  (WW Norton, New York, 1928).
%
\bibitem{hh_52}
AL Hodgkin and AF Huxley, A quantitative description of membrane current and its application to conduction and excitation in nerve.     {\em J Physiol (Lond)} {\bf 117,} 500 (1952). \url{https://doi.org/10.1113/jphysiol.1952.sp004764}
%
\bibitem{aidley_98}
DJ Aidley, {\em The Physiology of Excitable Cells,  Fourth Edition.}  (Cambridge University Press, Cambridge, 1998).
%
\bibitem{spikes}
F Rieke, D Warland, R de Ruyter van Steven\-inck, and W Bialek,
{\em Spikes: Exploring the Neural Code.}  (MIT Press, Cambridge, 1997).
%
\bibitem{segev+al_04} 
R Segev, J Goodhouse, J Puchalla, and MJ Berry II,  Recording spikes from a large fraction of the ganglion cells in a retinal patch. {\em Nat Neurosci} {\bf 7,} 1154 (2004). \url{https://doi.org/10.1038/nn1323}
%
\bibitem{litke_04}
AM Litke, N Bezayiff, EJ Chichilnisky, W Cunningham, W Dabrowski, AA Grillo, M Grivich, P Grybos, P Hottowy, S Kachiguine, et al,
%RS Kalmar, K Mathieson, D Petrusca, M Rahman, and A Sher,  
What does the eye tell the brain?: Development of a system for the large-scale recording of retinal output activity. {\em IEEE Trans Nucl Sci} {\bf 51,} 1434 (2004). \url{https://doi.org/10.1109/TNS.2004.832706}
%
\bibitem{marre+al_12}
O Marre, D Amodei, N Deshmukh, K Sadeghi, F Soo, TE Holy, and MJ Berry II,  Mapping a complete neural population in the retina. {\em J Neurosci} {\bf 32,} 14859 (2012).  \url{https://doi.org/10.1523/JNEUROSCI.0723-12.2012}
%
\bibitem{campbell+al_91}
PK Campbell, KE Jones, RJ Huber, KW Horch, and  RA Normann,
A silicon-based, three-dimensional neural interface: Manufacturing processes for an intracortical electrode array. {\em IEEE Trans Biomed Eng} {\bf 38,} 758 (1991). \url{https://doi.org/10.1109/10.83588}
%
\bibitem{jun+al_17}
JJ Jun, NA Steinmetz, JH Siegle, DJ Denman, M Bauza, B Barbarits, AK Lee, CA Anastassiou, A Andrei, \c{C} Aydin, et al,
%M Barbic, TJ Blanche, V Bonin, J Couto, B Dutta, SL Gratiy, DA Gutnisky, M H\"ausser, B Karsh, P Ledochowitsch, CM Lopez, C Mitelut, S Musa, M Okun, M Pachitariu, J Putzeys, PD Rich, C Rossant, W Sun, K Svoboda, M Carandini, KD Harris, C Koch, J O’Keefe, and TD Harris,  
Fully integrated silicon probes for high--density recording of neural activity. {\em Nature} {\bf 551,} 232 (2017). \url{https://doi.org/10.1038/nature24636}
%
\bibitem{chung+al_19}
JE Chung,  HR Joo, JL Fan, DF Liu, AH Barnett, S Chen, C Geaghan--Breiner, MP Karlsson, M Karlsson, KY Lee, et al,
%H Liang, JF Magland,  JA Pebbles, AC Tooker, LF Greengard, VM Tolosa, and LM Frank,  
High--density, long--lasting, and multi--region electrophysiological recordings using polymer electrode arrays. {\em Neuron} {\bf 101,} 21 (2019).
\url{https://doi.org/10.1016/j.neuron.2018.11.002}
%
\bibitem{gong_15}
Y Gong, The evolving capabilities of rhodopsin-based genetically encoded voltage indicators. {\em Curr Opin Chem Biol} {\bf 27,} 84 (2015). \url{https://doi.org/10.1016/j.cbpa.2015.05.006}
%
\bibitem{yang+st-pierre_16}
HH Yang and F St--Pierre, Genetically encoded voltage indicators: Opportunities and challenges. {\em J Neurosci} {\bf 36,} 9977 (2016). \url{https://doi.org/10.1523/JNEUROSCI.1095-16.2016}
%
\bibitem{platisa+al_22}
J Platisa, H Zeng, L Madisen, LB Cohen, VA Pieribone, and DA Storace, Voltage imaging in the olfactory bulb using transgenic mouse lines expressing the genetically encoded voltage indicator ArcLight. {\em Sci Rep} {\bf 12,} 1875 (2022). \url{https://doi.org/10.1038/s41598-021-04482-3}
%
\bibitem{evans+al_23}
SW Evans, D--Q Shi, M Chavarha, MH Pitt,  J Taxidis, B Madruga, JL Fan, F--J Hwang, SC van Keulen, C--M Suomivuori, et al,
%MM Pang, S Su, S Lee, YA Hao, G Zhang, D Jiang, L Pradhan, RH Roth, Y Liu, CC Dorian, AL Reese, A Negrean, A Losonczy, CD Makinson, S Wang, TR Clandinin, RO Dror, JB Ding, N Ji, P Golshani, LM Giocomo, G--Q Bi, and MZ Lin, 
A positively tuned voltage indicator for extended electrical recordings in the brain. {\em Nat Methods} {\bf 20,} 1104 (2023). \url{https://doi.org/10.1038/s41592-023-01913-z}
%
\bibitem{tian+al_09}
L Tian, SA Hires, T Mao, D Huber, ME Chiappe, SH Chalasani, L Petreanu, J Akerboom, SA McKinney, ER Schreiter, et al,
%CI Bargmann, V Jayaraman, K Svoboda, and  LL Looger,  
Imaging neural activity in worms, flies and mice with improved GCaMP calcium indicators. {\em Nat Methods} {\bf  6,} 875 (2009). \url{https://doi.org/10.1038/nmeth.1398}
%
\bibitem{tian+al_12}
L Tian, SA Hires, and LL Looger, Neural activity imaging with genetically encoded calcium indicators.  In {\em Imaging in Neuroscience}, F Helmchen and A Konnerth, eds, pp 647--656 (Cold Spring Harbor Laboratory Press, 2012). \url{https://doi.org/10.1101/pdb.top069609}
%
\bibitem{zhang+al_23}
YZhang, M R\'ozsa, Y Liang, D Bushey, Z Wei, J Zheng, D Reep, GJ Broussard, A Tsang, G Tsegaye, et al,
%S Narayan, CJ Obara, J--X Lim, R Patel, R Zhang, MB Ahrens, GC Turner, SS--H. Wang, WL Korff, ER Schreiter, K Svoboda, JP Hasseman, I Kolb, and  LL Looger, 
Fast and sensitive GCaMP calcium indicators for imaging neural populations. {\em Nature} {\bf 615,} 884 (2023). \url{https://doi.org/10.1038/s41586-023-05828-9}
%
\bibitem{denk+al_90}
W Denk, JH Strickler, and WW Webb,  Two--photon laser--scanning fluorescence microscopy. {\em Science} {\bf  248,} 73 (1990). \url{https://doi.org/10.1126/science.23210}
%
\bibitem{dombeck+al_07}
DA Dombeck, AN Khabbaz, F Collman, TL Adelman, and DW Tank, Imaging large--scale neural activity with cellular resolution in awake, mobile mice. {\em Neuron} {\bf 56,} 43 (2007). \url{https://doi.org/10.1016/j.neuron.2007.08.003}
%
\bibitem{song+al_17}
A Song,  AS Charles, SA Koay, JL Gauthier, SY Thiberge, JW Pillow, and DW Tank, Volumetric two-photon imaging of neurons using stereoscopy (vTwINS). {\em Nat Methods} {\bf 14,} 420 (2017). \url{https://doi.org/10.1038/nmeth.4226}
%
\bibitem{weisenburger+al_19}
S Weisenburger, F Tejera, J Demas, B Chen, J Manley, FT Sparks, FM Traub, T Daigle, H Zeng, A Losonczy, and  A Vaziri,
Volumetric Ca$^{2+}$ imaging in the mouse brain using hybrid multiplexed sculpted light microscopy. {\em Cell} {\bf 177,} 1050 (2019). \url{https://doi.org/10.1016/j.cell.2019.03.011}
%
\bibitem{demas+al_21}
J Demas, J Manley, F Tejera, K Barber, H Kim, FM Traub, B Chen, and A Vaziri, High-speed, cortex-wide volumetric recording of neuroactivity at cellular resolution using light beads microscopy. {\em Nat Methods} {\bf 18,} 1103 (2021).
\url{https://doi.org/10.1038/s41592-021-01239-8}
%
\bibitem{harvey+al_09}
CD Harvey, F Collman, DA Dombeck, and DW Tank, Intracellular dynamics of hippocampal place cells during virtual navigation. {\it Nature} {\bf 461,} 941  (2009). \url{https://doi.org/10.1038/nature08499}
%
\bibitem{ahrens+al_13}
MB Ahrens, MB Orger, DN Robson, JM Li, and PJ Keller, Whole--brain functional imaging at cellular resolution using light--sheet microscopy. {\em Nat Methods} {\bf 10,} 413 (2013). \url{https://doi.org/10.1038/nmeth.2434}
%
\bibitem{nguyen+al_16}
JP Nguyen, FB Shipley, AN Linder, GS Plummer, JW Shaevitz, and AM Leifer, Whole--brain calcium imaging with cellular resolution in freely behaving {\em Caenorhabditis elegans}  {\em Proc Natl Acad Sci (USA)} {\bf 113,} E1074 (2016). \url{https://doi.org/10.1073/pnas.1507110112}
%
\bibitem{nagel+al_03}
G Nagel, T Szellas, W Huhn, S Kateriya, N Adeishvili, P Berthold, D Ollig, P Hegemann, and E Bamberg, 
Channelrhodopsin--2, a directly light-gated cation--selective membrane channel. {\em Proc Natl Acad Sci (USA)} {\bf 100,} 13940 (2003). \url{https://doi.org/10.1073/pnas.1936192100} 
%
\bibitem{han+boyden_07}
X Han amd ES Boyden, Multiple--color optical activation, silencing, and desynchronization of neural activity, with single--spike temporal resolution. {\em PLoS One} {\bf 2,} e299 (2007). \url{https://doi.org/10.1371/journal.pone.0000299}
%
\bibitem{zhang+al_08}
F Zhang, M Prigge, F Beyri\`ere, SP Tsunoda, J Mattis, O Yizhar, P Hegemann, and  K Deisseroth,
Red--shifted optogenetic excitation: A tool for fast neural control derived from {\em Volvox carteri}. {\em Nat Neurosci} {\bf 11,} 631 (2008). \url{https://doi.org/10.1038/nn.2120}
%
\bibitem{randi+al_23}
F Randi, AK Sharma, S Dvali, and AM Leifer, Neural signal propagation atlas of {\em Caenorhabditis elegans}. {\em Nature} {\bf 623,} 406 2023). \url{https://doi.org/10.1038/s41586-023-06683-4}
%
\bibitem{meshulam+al_19}
L Meshulam, JL Gauthier, CD Brody, DW Tank, and W Bialek,  Coarse--graining, fixed points, and scaling in a large population of neurons.   {\em Phys Rev Lett} {\bf 123,} 178103 (2019). \url{https://doi.org/10.1103/PhysRevLett.123.178103}
%
\bibitem{meshulam+al_17}
L Meshulam, JL Gauthier, CD Brody, DW Tank, and W Bialek, Collective behavior of place and non--place neurons in the hippocampal network.     {\em Neuron} {\bf 96,} 1178 (2017). \url{https://doi.org/10.1016/j.neuron.2017.10.027}
%
\bibitem{svoboda+al_94}
K Svoboda, CF Schmidt, BJ Schnapp, and SM Block, Direct observation of kinesin stepping by optical trapping interferometry. {\em Nature} {\bf 365,} 721 (1993).
\url{https://doi.org/10.1038/365721a0}
%
\bibitem{abbondanzieri+al_05}
EA Abbondanzieri, WJ Greenleaf, JW Shaevitz, R Landick, and SM Block, Direct observation of base-pair stepping by RNA polymerase. {\em Nature} {\bf 438,} 460 (2005). \url{https://doi.org/10.1038/nature04268}
%
\bibitem{ueno+al_05}
H Ueno, T Suzuki, K Kinosita Jr, and M Yoshida, ATP--driven stepwise rotation of F$_0$F$_1$--ATP synthase. {\em Proc Natl Acad  Sci  (USA)} {\bf 102,}  1333 (2005). \url{https://doi.org/10.1073/pnas.0407857102}
%
\bibitem{cavagna+al_17}
A Cavagna, D Conti, C Creato, L Del Castello, I Giardina, TS Grigera, S Melillo, L Parisi, and M Viale, Dynamic scaling in natural swarms.  {\em Nat Phys} {\bf 13,} 914 (2017). \url{https://doi.org/10.1038/nphys4153}
%
\bibitem{cavagna+al_18}
 A Cavagna, I Giardina, and T Grigera, 
 The physics of flocking: Correlation as a compass from experiments to theory. 
 {\em Phys Repts} {\bf 728,} 1 (2018). \url{https://doi.org/10.1016/j.physrep.2017.11.003}
 %
 \bibitem{qin+al_20}
  B Qin, C Fei, AA Bridges, AA Mashruwala,  HA Stone, NS Wingreen, and  BL Bassler, Cell position fates and collective fountain flow in bacterial biofilms revealed by light--sheet microscopy.
{\em Science} {\bf 369,} 71 (2020).  \url{https://doi.org/10.1126/science.abb850}
 %
 \bibitem{copenhagen+al_21}
  K Copenhagen, {R Alert},  NS Wingreen, and  JW Shaevitz, Topological defects promote layer formation in {\em Myxococcus xanthus} colonies.
 {\em Nat Phys} {\bf 17,} 211 (2021). \url{https://doi.org/10.1038/s41567-020-01056-4}
%
\bibitem{warmflash+al_14}
A Warmflash, B Sorre, F Etoc, ED Siggia, and AH Brivanlou,
A method to recapitulate early embryonic spatial patterning in human embryonic stem cells.
{\em Nat Methods}  {\bf 11,} 847 (2014). \url{https://doi.org/10.1038/nmeth.3016}
%
\bibitem{shahbazi+al_19}
MN Shahbazi, ED  Siggia, and M Zernicka--Goetz,
Self--organization of stem cells into embryos: A window on early mammalian development.  {\em Science} {\bf 364,} 948 (2019). \url{https://doi.org/10.1126/science.aax0164}
%
\bibitem{decadal}
W Bialek, B Carragher, I Ciss\'e, MM Desai, OK Dudko, DI Goldman, J Kondev, PB Littlewood, AJ Liu, ME Moxon, JN Onuchic, MJ  Schnitzer, and CM Waterman, {\em Physics of Life}. (National Academies Press, Washington DC, 2022). \url{https://nap.edu/physicsoflife}
%
\bibitem{shlens_14}
J Shlens, A tutorial on principal component analysis. arXiv:1404.1100 [cs.LG] (2014). \url{https://doi.org/10.48550/arXiv.1404.1100}
%
\bibitem{roweis+saul_00}
ST Roweis and LK Saul, Nonlinear dimensionality reduction by locally linear embedding. {\em Science} {\bf 290,} 2323 (2000). \url{DOI: 10.1126/science.290.5500.2323}
%
\bibitem{gergen+al_86}
JP Gergen, D Coulter, and EF Wieschaus, Segmental pattern
and blastoderm cell identities. In {\em Gametogenesis and The Early
Embryo}, JG Gall, ed, pp 195--220 (Liss, New York, 1986).
%
\bibitem{dubuis+al_13b}
JO Dubuis, G Tka\v{c}ik, EF Wieschaus, T Gregor, and W Bialek, Positional information, in  bits.  {\em Proc Natl Acad Sci (USA)} {\bf 110,}  16301 (2013). \url{https://doi.org/10.1073/pnas.1315642110}
%
\bibitem{mcgough+al_23}
L McGough, H Casademunt, M Nikoli\' c, MD Petkova, T Gregor,  and W Bialek, Finding the last bits of positional information.  arXiv:2312.05963 [q--bio.MN] (2023).  \url{https://doi.org/10.48550/arXiv.2312.05963}
%
\bibitem{ptashne_92}
M Ptashne, {\em A Genetic Switch: Phage and Higher Organisms} (Blackwell, Cambridge MA, 1992).
%
\bibitem{pedone+al_96}
PV Pedone, R Ghirlando, GM Clore, AM Gronenbron, G Felsenfeld, and JG Omchinski,  The single Cys2--His2 zinc finger domain of the GAGA protein flanked by basic residues is sufficient for high--affinity specific DNA binding. {\em Proc Natl Acad Sci} USA {\bf 93,} 2822  (1996). \url{https://doi.org/10.1073/pnas.93.7.2822}
%
\bibitem{winston_99}
RL Winston, DP Millar, JM Gottesfeld,   and SB Kent,  Characterization of the DNA binding properties of the bHLH domain of Deadpan to single and tandem sites. {\em Biochemistry} {\bf 38,}  5138 (1999). \url{https://doi.org/10.1021/bi982856a}
%
\bibitem{gregor+al_07b}
T Gregor, DW Tank, EF Wieschaus, and W Bialek, Probing the limits to positional information.   {\em Cell} {\bf 130,} 153 (2007). \url{https://doi.org/10.1016/j.cell.2007.05.025}
%
\bibitem{bialek_02}
W Bialek, Thinking about the brain.  In  {\em Physics of Biomolecules
and Cells: Les Houches Session LXXV,} H Flyvbjerg, F J\"ulicher, P
Ormos, and F David, eds, pp 485--577 (EDP Sciences, Les Ulis;
Springer--Verlag, Berlin, 2002); arXiv:physics/0205030 (2002). 
\url{https://doi.org/10.48550/arXiv.physics/0205030}
%
\bibitem{field+rieke_02}
GD Field and F Rieke,  Nonlinear signal transfer from mouse rods to bipolar cells and implications for visual sensitivity. {\em Neuron} {\bf  34,} 773 (2002). \url{https://doi.org/10.1016/S0896-6273(02)00700-6}
%
\bibitem{rieke+al_91}
F Rieke, WG Owen, and W Bialek, Optimal filtering in the salamander retina.
{\em Advances in Neural Information Processing}   {\bf 3,}  377 (1991). \url{https://proceedings.neurips.cc/paper_files/paper/1990/file/019d385eb67632a7e958e23f24bd07d7-Paper.pdf}
%
\bibitem{hassenstein+reichardt_56}
S Hassenstein and W Reichardt, Systemtheoretische Analyse der Zeit-, Reihenfolgen- und Vorzeichenauswertung bei der Bewegungsperzeption des R\"usselk\"afers Chlorophanus. {\em Z Naturforsch} {\bf 11B,} 513 (1956).  \url{https://doi.org/10.1515/znb-1956-9-1004}
%
\bibitem{reichardt+poggio_76}
W Reichardt and T Poggio, Visual control of orientation behaviour in the fly: Part I. A quantitative analysis, {\em Q Rev Biophys} {\bf  9,} 311 (1976). \url{https://doi.org/10.1017/S0033583500002523}
%
\bibitem{adelson+bergen_85}
EH Adelson and JR Bergen, Spatiotemporal energy models for the perception of motion. {\em J Opt Soc Am A} {\bf  2,}  284 (1985). \url{https://doi.org/10.1364/JOSAA.2.000284}
%
\bibitem{santen+sperling_85}
JPH van Santen and G Sperling,  Elaborated Reichardt detectors. {\em J Opt Soc Am A} {\bf 2,}
300 (1985). \url{https://doi.org/10.1364/JOSAA.2.000300}
%
\bibitem{strausfeld_76}
NJ Strausfeld, {\em Atlas of an Insect Brain}  (Springer--Verlag, Berlin, 1976).
%
\bibitem{hausen_82}
K Hausen, Motion sensitive interneurons in the optomotor system of the fly. I. The horizontal cells: Structure and signals. {\em Biol Cybern} {\bf 45,} 143 (1982). \url{https://doi.org/10.1007/BF00335241}
%
\bibitem{stavenga+hardie_89}
DG Stavenga and RC Hardie, eds, {\em Facets of Vision}.  (Springer--Verlag, Berlin, 1989).
%
\bibitem{ruyter+laughlin_96a}
RR de Ruyter van Steveninck and SB Laughlin,  The rate of information transfer at graded-potential synapses. {\em Nature} {\bf  379,} 642. (1996). \url{https://doi.org/10.1038/379642a0}
%
\bibitem{ruyter+laughlin_96b}
RR de Ruyter van Steveninck and SB Laughlin,  Light adaptation and reliability in blowfly photoreceptors. {\em Int J Neural Syst} {\bf  7,} 437 (1996). \url{https://doi.org/10.1142/S0129065796000415}
%
\bibitem{bialek+al_91}
W Bialek, F Rieke, RR de Ruyter van Ste\-ven\-inck, and D Warland,  Reading a neural code.
{\em Science} {\bf 252,} 1854 (1991). \url{https://doi.org/10.1126/science.2063199}
%
\bibitem{ruyter+bialek_95}
R de Ruyter van Steveninck and W Bialek,  Reliability and statistical efficiency of a blowfly movement--sensitive neuron. {\em Phil Trans R Soc Lond} {\bf 348,} 321 (1995). \url{https://doi.org/10.1098/rstb.1995.0071}
%
\bibitem{potters+bialek_94}
M Potters and  W Bialek,  Statistical mechanics and visual signal processing.  {\em J Phys I France} {\bf 4}, 1755
(1994). \url{https://doi.org/10.1051/jp1:1994219}
%
\bibitem{sinha+al_21}
 SR Sinha, W Bialek, and RR de Ruyter van Steveninck,  Optimal local estimates of visual motion in a natural environment. {\em Phys Rev Lett} {\bf 126,} 018101 (2021). \url{https://doi.org/10.1103/PhysRevLett.126.018101}
%
\bibitem{fitzgerald+al_11}
JE Fitzgerald, AY Katsov, TR Clandinin, and MJ Schnitzer, Symmetries in stimulus statistics shape the form of visual motion estimators.  {\em Proc Natl Acad Sci (USA)} {\bf  108,} 12909 (2011). \url{https://doi.org/10.1073/pnas.1015680108}
%
\bibitem{behnia+al_14}
R Behnia, DA Clark, AG Carter, TR Clandinin, and C Desplan, Processing properties of ON and OFF pathways for {\em Drosophila} motion detection. {\em Nature} {\bf  512,} 427 (2014). \url{https://doi.org/10.1038/nature13427}
%
\bibitem{wolpert_69}
L Wolpert,  Positional information and the spatial pattern of cellular differentiation.  {\em J Theor Biol} {\bf 25,} 1 (1969). \url{https://doi.org/10.1016/S0022-5193(69)80016-0}
%
\bibitem{krotov+al_14}
 D Krotov, JO Dubuis, T Gregor, and W Bialek, Morphogenesis at criticality?  {\em Proc Natl Acad Sci (USA)} {\bf 111,} 3683  (2014). \url{https://doi.org/10.1073/pnas.1324186111}
 %
 \bibitem{thome+bialek_24}
 T Thome and W Bialek, A brief tutorial on information theory.   arXiv:2402.16556 [physics.bio--ph] (2024).
 \url{https://doi.org/10.48550/arXiv.2402.16556}
 %
\bibitem{shannon_48} CE  Shannon,  A mathematical theory of communication. {\em Bell Sys Tech J} {\bf 27,} 379  and  623  (1948).  \url{https://doi.org/10.1002/j.1538-7305.1948.tb01338.x} Reprinted in CE Shannon and W Weaver, {\em The Mathematical Theory of Communication} (University of Illinois Press, Urbana, 1949).
%
\bibitem{cover+thomas_91}
TM Cover and JA Thomas,  {\em Elements of Information Theory}  (Wiley, New York, 1991).
%
\bibitem{mezard+montanari_09}
M M\'ezard and A Montanari, {\em Information, Physics, and Computation.}  (Oxford University Press, Oxford and New York, 2009).
%
\bibitem{laughlin_81}
SB Laughlin, A simple coding procedure enhances a neuron's information capacity. {\em Z Naturforsch} {\bf 36c,} 910--912 (1981). \url{https://doi.org/10.1515/znc-1981-9-1040}
%
\bibitem{smirnakis+al_97}
S Smirnakis, MJ Berry II, DK Warland, W Bialek, and M Meister,  Adaptation of retinal processing to image contrast and spatial scale. {\em Nature} {\bf 386,} 69 (1997). \url{https://doi.org/10.1038/386069a0}
%
\bibitem{rieke+al_97}
F Rieke, D Warland, R de Ruyter van Steven\-inck, and W Bialek  {\em Spikes: Exploring the Neural Code} (MIT Press, Cambridge, 1997).
%
\bibitem{dayan+abbott_01}
LF Abbott and P Dayan, {\em Theoretical Neuroscience. Computational and Mathematical Modeling of Neural Systems} (MIT Press, Cambridge MA, 2001).
%
\bibitem{deboer+kuyper_68}
E de Boer and P Kuyper, Triggered correlation. {\em IEEE Trans Biomed Eng} {\bf 15,} 169 (1968).
\url{https://doi.org/10.1109/TBME.1968.4502561}
%
\bibitem{ruyter+bialek_88}
R de Ruyter van Steveninck and W Bialek, Real--time performance of a movement sensitive neuron in the blowfly visual system: Coding and information transfer in short spike sequences. {\em Proc R Soc Lond}  B {\bf 234,} 379--414 (1988). \url{https://doi.org/10.1098/rspb.1988.0055}
%
\bibitem{brenner+al_00}
N Brenner, W Bialek, and R de Ruyter van Steveninck, Adaptive rescaling optimizes information transmission.
 {\em Neuron} {\bf 26,} 695 (2000). \url{https://doi.org/10.1016/S0896-6273(00)81205-2}
%
\bibitem{bialek+ruyter_05}
W Bialek and R de Ruyter van Steveninck,  Features and dimensions: Motion estimation in fly vision. arXiv:q--bio/0505003 (2005). 
\url{https://doi.org/10.48550/arXiv.q-bio/0505003}
%
\bibitem{rust+al_05}
NC Rust, O Schwartz, JA Movshon, and EP Simoncelli, Spatiotemporal elements of Macaque V1 receptive fields. {\em Neuron} {\bf 46,} 945 (2005). \url{https://doi.org/10.1016/j.neuron.2005.05.021}
%
\bibitem{fairhall+al_01}
AL Fairhall, GD Lewen, W Bialek, and RR de Ruyter van  Steveninck, Efficiency and ambiguity in an adaptive neural code. {\em Nature}  {\bf 412,} 787 (2001). \url{https://doi.org/10.1038/35090500}
%
\bibitem{nagel+doupe_06}
KI Nagel and AJ Doupe, Temporal processing and adaptation in the songbird auditory forebrain. {\em Neuron} {\bf 51,} 845 (2006). \url{https://doi.org/10.1016/j.neuron.2006.08.030}
%
\bibitem{maravall+al_07}
M Maravall, RS Petersen, AL Fairhall, E Arabzadeh, and ME Diamond,  Shifts in coding properties and maintenance of information transmission during adaptation in barrel cortex. {\em PLoS Biol} {\bf 5,} e19 (2007). \url{https://doi.org/10.1371/journal.pbio.0050019}
%
\bibitem{dean+al_05}
I Dean, N Harper, and D McAlpine, Neural population coding of sound level adapts to stimulus statistics. {\em Nat Neurosci} {\bf 8,}  (2005). \url{https://doi.org/10.1038/nn1541}
%
\bibitem{wark+al_07}
B Wark, BN Lundstrom, and AL Fairhall, Sensory adaptation. {\em Curr Opin Neurobiol} {\bf 17,} 423 (2007).  \url{https://doi.org/10.1016/j.conb.2007.07.001}
%
\bibitem{kim+rieke_01}
KJ Kim and F Rieke,  Temporal contrast adaptation in the input and output signals of salamander retinal ganglion cells. {\em J Neurosci} {\bf 21,} 287 (2001). \url{https://doi.org/10.1523/JNEUROSCI.21-01-00287.2001}
%
\bibitem{kim+rieke_03}
KJ Kim and F Rieke, Slow Na$^+$ inactivation and variance adaptation in salamander retinal ganglion cells.  {\em J Neurosci} {\bf  23,} 1506 (2003). \url{https://doi.org/10.1523/JNEUROSCI.23-04-01506.2003}
%
\bibitem{tkacik+al_15}
G Tka\v{c}ik, JO Dubuis,  MD Petkova, and T Gregor, Positional information, positional error, and read--out precision in morphogenesis: a mathematical framework.  {\em Genetics} {\bf 199,} 39 (2015). \url{https://doi.org/10.1534/genetics.114.171850}
%
\bibitem{liu+al_13}
F Liu, AH Morrison, and T Gregor, Dynamic interpretation of maternal inputs by the {\em Drosophila} segmentation gene network.  {\em Proc Natl Acad Sci (USA)} {\bf 110,} 6724  (2013). \url{https://doi.org/10.1073/pnas.1220912110}
% 
\bibitem{arias+hayward_06}
AM Arias and P Hayward, Filtering transcriptional noise during development: Concepts and mechanisms. {\em Nat Rev Genet} {\bf 7,} 34--44 (2006). \url{https://doi.org/10.1038/nrg1750}
% 
\bibitem{lacalli_22}
TC Lacalli, Patterning, from conifers to consciousness: Turing’s theory and order from fluctuations.  {\em Front  Cell  Dev  Biol} {\bf 10,} 871950 (2022). \url{https://doi.org/10.3389/fcell.2022.871950}
%
\bibitem{bauer+al_21}
 M Bauer, MD Petkova, T Gregor, EF Wieschaus, and W Bialek,  Trading bits in the readout from a genetic network. {\em Proc Natl Acad Sci (USA)} {\bf 118,} e2109011118 (2021). \url{https://doi.org/10.1073/pnas.210901111}
%
\bibitem{bauer+bialek_23}
M Bauer and W Bialek, Information bottleneck in molecular sensing.    {\em PRX Life} {\bf 1,} 023005 (2023). \url{https://doi.org/10.1103/PRXLife.1.023005}
%
%\bibitem{waddington_57}
%CH Waddington, {\em The Strategy of the Genes. A Discussion of Some Aspects of Theoretical Biology.}  (Allen and Unwin, London, 1957).
%
\bibitem{tishby+al_99}
N Tishby, FC Pereira, and W Bialek, The information bottleneck method.
In {\em Proceedings of the 37th Annual Allerton Conference on Communication, Control
and Computing,} B Hajek and RS Sreenivas, eds,  pp 368--377
(University of Illinois, 1999); arXiv:physics/0004057 (2000). 
\url{https://doi.org/10.48550/arXiv.physics/0004057}
%
\bibitem{barlow_59}
HB Barlow, Sensory mechanisms, the reduction of redundancy, and intelligence.  In 
 {\em  Proceedings of the Symposium on the Mechanization of Thought Processes, Volume II}, DV Blake and AM Uttley, eds, pp 537--574 (HM Stationery Office, London, 1959).
 %
\bibitem{barlow_61}
HB Barlow, Possible principles underlying the transformation of sensory messages.  In  {\em Sensory Communication}, W Rosenblith, ed, pp 217--234 (MIT Press, Cambridge, 1961).
%
 \bibitem{schrodinger_25}
E Schr\"odinger, \"Uber das Verh\"altnis der Vierfarben- zur Dreifarbentheorie. {\em Sitzungsberichte der Akademie der Wissenschaften in Wien} Ila, mathematischnaturwissenschaftliche Klasse {\bf 134,} 417 (1925).  Translated with commentary by KK Niall, On the trichromatic and opponent--process theories: An article by E. Schr\"odinger. {\em Spat Vis} {\bf 3,} 79 (1988). \url{https://doi.org/10.1163/156856888X00050}
%
\bibitem{gegenfurter+sharpe_01}
KR Gegenfurtner and  LT Sharpe, eds, {\em Color Vision: From Genes to Perception} (Cambridge University Press, Cambridge, 2001).
%
\bibitem{buchsbaum+gottschalk_83}
G Buchsbaum and A Gottschalk, Trichromacy, opponent colours coding and optimum colour information transmission in the retina. {\em Proc R Soc Lond} B {\bf 220,} 89 (1983). \url{https://doi.org/10.1098/rspb.1983.0090}
%
\bibitem{ruderman+al_98}
DL Ruderman, TW Cronin, and C--C Chiao, Statistics of cone responses to natural images: Implications for visual coding. {\em J Opt Soc Am A} {\bf 15,} 2036 (1998). \url{https://doi.org/10.1364/JOSAA.15.002036}
%
\bibitem{atick+redlich_90}
JJ Atick and AN Redlich, Toward a theory of early visual processing.  {\em Neural Comput} {\bf 2,} 308 (1990). \url{https://doi.org/10.1162/neco.1990.2.3.308}
%
\bibitem{hateren_92}
JH van Hateren, Real and optimal neural images in early vision.  {\em Nature} {\bf 360,} 68 (1992). \url{https://doi.org/10.1038/360068a0}
%
\bibitem{ruderman+bialek_94}
DL Ruderman and W Bialek, Statistics of natural images: Scaling in the woods.  {\em Phys Rev Lett} {\bf 73}, 814 (1994). \url{https://doi.org/10.1103/PhysRevLett.73.814}
%
\bibitem{barlow_53}
HB Barlow, Summation and inhibition in the frog's retina. {\em J Physiol (Lond)} {\bf 119,} 69 (1953). \url{https://doi.org/10.1113/jphysiol.1953.sp004829}
%
\bibitem{kuffler_53}
SW Kuffler, Discharge patterns and functional organization of the mammalian retina. {\em J Neurophysiol} {\bf 16,} 37 (1953). \url{https://doi.org/10.1152/jn.1953.16.1.37}
%
\bibitem{hartline_69}
HK Hartline, Visual receptors and retinal interaction. In {\em Nobel Lectures, Physiology or Medicine 1963--1970.} (Elsevier Publishing, Amsterdam, 1972). \url{https://www.nobelprize.org/prizes/medicine/1967/hartline/lecture/}
%
\bibitem{shannon_49}
CE Shannon, Communication in the presence of noise. {\em Proc IRE}   {\bf 37,} 10 (1949). \url{https://doi.org/10.1109/JRPROC.1949.232969}
%
\bibitem{laughlin+ruyter_96}
SB Laughlin and RR de Ruyter van Steveninck, Measurements of signal transfer and noise suggest a new model for graded transmission at an adapting retinal synapse.  {\em J Physiol (Lond)} {\bf 494(P),} P19 (1996). \url{https://www.ncbi.nlm.nih.gov/pmc/articles/PMC1160695}
%
\bibitem{lundtsrom+al_08}
BN Lundstrom. MH Higgs, WJ Spain, and AL Fairhall, Fractional differentiation by neocortical pyramidal neurons. {\em Nat Neurosci} {\bf 11,} 1335 (2008). \url{https://doi.org/10.1038/nn.2212}
%
\bibitem{lundtsrom+al_10}
BL Lundstriom, AL Fairhall, and M Maravall, Multiple timescale encoding of slowly varying whisker
stimulus envelope in cortical and thalamic neurons {\em in vivo}. {\em J Neurosci} {\bf 30,} 5071 (2010). \url{https://doi.org/10.1523/JNEUROSCI.2193-09.2010}
%
\bibitem{thorson_74}
J Thorson and M Biederman--Thorson, Distributed relaxation processes in sensory adaptation. {\em Science} {\bf 183,} 161 (1974). \url{https://doi.org/10.1126/science.183.4121.161}
%
\bibitem{tkacik+bialek_09}
G Tka\v{c}ik and W Bialek,  Diffusion, dimensionality and noise in transcriptional regulation. {\em Phys Rev E} {\bf 79,} 051901 (2009). \url{https://doi.org/10.1103/PhysRevE.79.051901}
 %
\bibitem{bialek+setayeshgar_05}
W Bialek and S Setayeshgar, Physical limits to biochemical signaling.   {\em Proc Natl Acad Sci (USA)} {\bf 102,} 10040 (2005). \url{https://doi.org/10.1073/pnas.0504321102}
%
\bibitem{bialek+setayeshgar_08}
W Bialek and S Setayeshgar,  Cooperativity, sensitivity and noise in biochemical signaling.  {\em Phys Rev Lett} {\bf 100,} 258101 (2008). \url{https://doi.org/10.1103/PhysRevLett.100.258101}
%
\bibitem{vanzon+al_06}
JS van Zon, MJ Morelli, S Tanase--Nicola, and PR ten Wolde,  Diffusion of transcription factors can drastically enhance the noise in gene expresssion. {\em Biophys J} {\bf 91,} 4350 (2006). \url{https://doi.org/10.1529/biophysj.106.086157}
%
\bibitem{endres+wingreen_09}
RG Endres and NS Wingreen, Maximum likelihood and the single receptor.  {\em Phys Rev Lett} {\bf 103,} 158101 (2009). \url{https://doi.org/10.1103/PhysRevLett.103.158101}
%
\bibitem{mora+wingreen_10}
T  Mora and NS Wingreen, Limits of sensing temporal concentration changes by single cells. {\em Phys Rev Lett} {\bf 104,} 248101 (2010). \url{https://doi.org/10.1103/PhysRevLett.104.248101}
%
\bibitem{kaizu+al_14}
K Kaizu, W de Ronde, J Paijmans, K Takahashi, F  Tostevin, and PR  ten Wolde, The Berg--Purcell limit revisited. {\em Biophys J} {\bf 106,}  976 (2014). \url{https://doi.org/10.1016/j.bpj.2013.12.030}
%
\bibitem{mora_15}
T Mora, Physical limit to concentration sensing amid spurious ligands
{\em Phys Rev Lett} {\bf 115,} 038102 (2015). \url{https://doi.org/10.1103/PhysRevLett.115.038102}
%
\bibitem{carballo-pacheco+al_19}
M Carballo--Pacheco, J Desponds, T Gavrilchenko, A Mayer, R Prizak, G Reddy, I Nemenman, and T Mora,
Receptor crosstalk improves concentration sensing of multiple ligands.
{\em Phys Rev E} {\bf  99,} 022423 (2019). \url{https://doi.org/10.1103/PhysRevE.99.022423}
%
\bibitem{mora+nemenman_19}
T Mora and I Nemenman, Physical limit to concentration sensing in a changing environment. {\em  Phys Rev Lett} {\bf 123,} 198101 (2019). \url{https://doi.org/10.1103/PhysRevLett.123.198101}
%
\bibitem{wave+al_20}
V Ngampruetikorn, DJ Schwab, and GJ Stephens, Energy consumption and cooperation for optimal sensing. {\em Nat Commun} {\bf 11,} 975 (2020). \url{https://doi.org/10.1038/s41467-020-14806-y}
%
 \bibitem{tkacik+al_09}
 G Tka\v{c}ik, AM Walczak, and W Bialek,   Optimizing information flow in small genetic networks.  {\em Phys Rev E} {\bf 80,} 031920 (2009). \url{https://doi.org/10.1103/PhysRevE.80.031920}
 %
 \bibitem{walczak+al_10}
AM Walczak, G Tka\v{c}ik, and W Bialek,  Optimizing information flow in small genetic networks. II: Feed--forward interaction.  {\em Phys Rev E} {\bf 81,} 041905 (2010). \url{https://doi.org/10.1103/PhysRevE.81.041905}
%
 \bibitem{tkacik+al_12}
 G Tka\v{c}ik, AM Walczak, and  W Bialek, Optimizing information flow in small genetic networks. III. A self--interacting gene.   {\em Phys Rev E} {\bf 85,}  041903 (2012). \url{https://doi.org/10.1103/PhysRevE.85.041903}
%
 \bibitem{little+al_13}
SC Little, M Tikhonov, and T Gregor, Precise developmental gene expression arises from globally stochastic transcriptional activity. {\em Cell} {\bf 154,} 789 (2013). \url{https://doi.org/10.1016/j.cell.2013.07.025}
 %
 \bibitem{sokolowski+tkacik_15}
 TR Sokolowski  and G Tka\v{c}ik, Optimizing information flow in small genetic networks. IV.  Spatial coupling. {\em Phys Rev E} {\bf 91,} 062710 (2015). \url{https://doi.org/10.1103/PhysRevE.91.062710}
%
\bibitem{sokolowski+al_16}
TR Sokolowski, AM Walczak, W Bialek, and G Tka\v{c}ik, Extending the dynamic range of transcription factor action by translational regulation.  {\em Phys Rev E } {\bf 93,} 022404 (2016). \url{https://doi.org/10.1103/PhysRevE.93.022404}
%
\bibitem{dubnau+struhl_96}
J Dubnau and G Struhl, RNA recognition and translational regulation by a homeodomain protein. {\em Nature} {\bf 379,} 694 (1996). \url{https://doi.org/10.1038/379694a0}
%
\bibitem{rivera-pomar+al_96}
R Rivera-Pomar, D Niessing, U Schmidt--Ott, WJ Gehring, and  H J\"ackle,  RNA binding and translational suppression by bicoid. {\em Nature} {\bf 379,} 746 (1996). \url{https://doi.org/10.1038/379746a0}
%
\bibitem{niessing+al_00}
D Niessing, W Driever, F Sprengerm. H Taubert, and  H J\"ackle,  Homeodomain position 54 specifies transcriptional versus translational control by Bicoid. {\em Mol Cell} {\bf 5,} 395 (2000). \url{https://doi.org/10.1016/S1097-2765(00)80434-7}
%
\bibitem{johnstone+lasko_01}
O Johnstone and P Lasko, Translational regulation and RNA localization in {\em Drosophila} oocytes and embryos. {\em Annu Rev Genet} {\bf 35,} 365 (2001). \url{https://doi.org/10.1146/annurev.genet.35.102401.090756}
%
\bibitem{sokolowski+al_23}
TR Sokolowski, T Gregor, W Bialek, and G Tka\v{c}ik, Deriving a genetic regulatory network from an optimization principle.    arXiv:2302.05680 [physics.bio--ph] (2023). \url{https://doi.org/10.48550/arXiv.2302.05680}
%
\bibitem{monod+al_63}
J Monod, J--P Changeux, and F Jacob, Allosteric proteins and cellular control systems. {\em J Mol Biol} {\bf 6,} 306 (1963). \url{https://doi.org/10.1016/S0022-2836(63)80091-1}
%
\bibitem{monod+al_65}
J Monod, J Wyman, and J--P Changeux, On the nature of allosteric transitions: A plausible model. {\em J Mol Biol} {\bf 12,} 88 (1965). \url{https://doi.org/10.1016/S0022-2836(65)80285-6}
%
\bibitem{perutz_90}
MF Perutz, {\em Mechanisms of Cooperativity and Allosteric Regulation in Proteins} (Cambridge University Press, Cambridge, 1990).
%
\bibitem{phillips_20}
R Phillips, {\em The Molecular Switch: Signaling and Allostery} (Princeton University Press, Princeton NJ, 2020).
%
\bibitem{hopfield_73}
JJ Hopfield, Relation between structure, co--operativity and spectra
in a model of hemoglobin action.  {\em J Mol Biol} {\bf 77,} 207 (1973). \url{https://doi.org/10.1016/0022-2836(73)90332-X}
%
\bibitem{bintu+al_05a}
L Bintu, NE Buchler, HG Garcia, U Geraland, T Hwa, J Kondev, and R Phillips, Transcriptional regulation by the numbers: models. {\em Curr Opin Genet Dev} {\bf 15,} 116 (2005). \url{https://doi.org/10.1016/j.gde.2005.02.007}
%
\bibitem{bintu+al_05b}
L Bintu, NE Buchler, HG Garcia, U Gerland, T Hwa, J Kondev, T Kuhlman, and Rob Phillips, Transcriptional regulation by the numbers: applications. {\em Curr Opin Genet Dev} {\bf 15,} 125 (2005). \url{https://doi.org/10.1016/j.gde.2005.02.006}
%
\bibitem{zoller+al_22}
B Zoller, T Gregor, and G Tka\v{c}ik, Eukaryotic gene regulation at equilibrium, or non? {\em Curr Opin Syst Biol} {\bf 31,}  100435 (2022). 
 \url{https://doi.org/10.1016/j.coisb.2022.100435}
 %
 \bibitem{swain+al_02}
 PS Swain, MB Elowitz, and ED Siggia, Intrinsic and extrinsic contributions to stochasticity in gene expression. {\em Proc Natl Acad   Sci (USA)} {\bf 99,} 12795 (2002).
 \url{https://doi.org/10.1073/pnas.162041399}
%
\bibitem{crombach+al_14}
A Crombach, MA Garc\'ia--Solache, and J Jaeger,  Evolution of early development in dipterans: reverse-engineering the gap gene network in the moth midge {\em Clogmia albipunctata} (psychodidae). {\em Biosystems} {\bf 123,} 74 (2014).
\url{https://doi.org/10.1016/j.biosystems.2014.06.003}
%
\bibitem{wotton+al_15}
KR Wotton, E Jim\'enez--Guri, A Crombach, H Janssens, Anna Alcaine--Colet, S Lemke, U Schmidt--Ott, J Jaeger, Quantitative system drift compensates for altered maternal inputs to the gap gene network of the scuttle fly {\em Megaselia abdita}. {\em eLife} {\bf 4,}  04785 (2015). \url{ https://doi.org/10.7554/eLife.04785}
%
\bibitem{goltsev+al_04}
Y Goltsev, W Hsiong, G Lanzaro, and M Levine,  Different combinations of gap repressors for common stripes in {\em Anopheles} and {\em Drosophila embryos}. {\em Dev Biol} {\bf 275,} 435 (2004). \url{https://doi.org/10.1016/j.ydbio.2004.08.021}
%
\end{thebibliography}
\end{document}